\documentclass[aps,prd,reprint,superscriptaddress,showpacs,nofootinbib]{revtex4-1}
\usepackage{graphicx}
\usepackage{amsfonts}
\usepackage{amsmath}
\usepackage{mathrsfs}
\usepackage{epsfig}
\usepackage{slashed}
\usepackage{dcolumn}%

\usepackage{ulem}
\usepackage{color}
\usepackage{caption}
\usepackage{subcaption}
\usepackage{hyperref}

\def\nat{Nature\ }
\def\aap{Astron.\ Astrophys.\ }
\def\apj{Astrophys.\ J.\ }
\def\apjl{Astrophys.\ J.\ Lett.\ }

\def\aj{Astron.\ J.\ }
\def\mnras{Mon.\ Not.\ Roy.\ Astron.\ Soc.\ }
\def\physrep{Phys.\ Rept.\ }
\def\prd{Phys.\ Rev.\ D\ }

\def\jcap{J.\ Cosmol.\ Astropart.\ Phys.\ }

\def\kpc{\,\mathrm{kpc}}

\def\km{\,\mathrm{km}}
\def\TeV{\,\mathrm{TeV}}
\def\GeV{\,\mathrm{GeV}}

\def\GV{\,\mathrm{GV}}

\def\cm{\,\mathrm{cm}}
\def\m{\,\mathrm{m}}
\def\s{\,\mathrm{s}}
\def\p{\,\mathrm{p}}
\def\e{\,\mathrm{e}}
\def\He{\,\mathrm{He}}
\def\A{\,\mathrm{A}}
\def\sr{\,\mathrm{sr}}
\newcolumntype{p}{D{,}{\pm}{-1}}
\def\pbar{\,\bar{\text{p}}}
\def\pbarp{\,\bar{\text{p}}/\text{p}}
\def\psr{\,\mathrm{psr}}
\def\DM{\,\mathrm{DM}}

\def\phinuc{\phi_{\mathrm{nuc}}}
\def\phipbar{\phi_{\pbar}}
\def\phipos{\phi_{e^{+}}}
\def\cpos{c_{e^{+}}}
\def\pos{e^{+}}
\def\lep{e^{-}+e^{+}}
\def\Mdm{m_{\chi}}
\def\sigv{\langle \sigma v \rangle}
\def\etae{\eta_{e}}
\def\etamu{\eta_{\mu}}
\def\etatau{\eta_{\tau}}
\def\eebar{e^{-}e^{+}}
\def\mumubar{\mu \bar{\mu}}
\def\tautaubar{\tau \bar{\tau}}

\def\bgfra{\varepsilon_{\mathrm{bg}}}
\def\Aeff{A_{\mathrm{eff}}}
\def\otherfra{\varepsilon_{\mathrm{other}}}

\begin{document}


\title{A Simple and Natural Interpretations of the DAMPE Cosmic Ray Electron/Positron Spectrum within Two Sigma Deviations}

\author{Jia-Shu Niu}
\email{jsniu@itp.ac.cn}
\affiliation{Institute of Theoretical Physics, State Key Laboratory of Quantum Optics and Quantum Optics Devices, Shanxi University, Taiyuan, 030006, China}
\affiliation{CAS Key Laboratory of Theoretical Physics, Institute of Theoretical Physics, Chinese Academy of Sciences, Beijing, 100190, China}
\affiliation{School of Physical Sciences, University of Chinese Academy of Sciences, No.~19A Yuquan Road, Beijing 100049, China}

\author{Tianjun Li}%
\email{tli@itp.ac.cn}
\affiliation{CAS Key Laboratory of Theoretical Physics, Institute of Theoretical Physics, Chinese Academy of Sciences, Beijing, 100190, China}
\affiliation{School of Physical Sciences, University of Chinese Academy of Sciences, No.~19A Yuquan Road, Beijing 100049, China}

\author{Fang-Zhou Xu}%
\affiliation{Institute of Modern Physics and Center for High Energy Physics, Tsinghua University, Beijing 100084, China}

\date{\today}

\begin{abstract}
  The DArk Matter Particle Explorer (DAMPE) experiment has recently announced the first results
for the measurement of total electron plus positron fluxes between 25 GeV and 4.6 TeV. A spectral break
at about 0.9 TeV and a tentative peak excess around 1.4 TeV have been found. However, it is very difficult 
to reproduce both the peak signal and the smooth background including spectral break simultaneously. 
We point out that the numbers of events in the two energy ranges (bins) close to the 1.4 TeV excess have 
$1\sigma$ deficits. With the basic physics principles such as simplicity and naturalness, we consider
the  $-2\sigma$, $+2\sigma$, and $-1\sigma$ deviations due to statistical fluctuations for the 
1229.3~GeV bin, 1411.4~GeV bin,  and 1620.5~GeV bin. Interestingly, we show that 
all the DAMPE data can be explained consistently via both the continuous distributed pulsar and 
dark matter interpretations, which have $\chi^{2} \simeq 17.2 $ and $\chi^{2}
\simeq 13.9$ (for all the 38 points in DAMPE electron/positron spectrum with 3 of them revised), respectively. 
These results are different from the previous analyses by neglecting the 1.4 TeV excess.
 At the same time, we do a similar global fitting on the newly released CALET lepton data, which could also be interpreted by such configurations.
Moreover, we present a $U(1)_D$ dark matter model with Breit-Wigner mechanism, which can provide
the proper dark matter annihilation cross section and escape the CMB constraint.
Furthermore, we suggest a few ways to test our proposal.

\end{abstract}


                              
\maketitle


\section{Introduction}
\label{intro}
Because of the strong radiative cooling via synchrotron
and inverse Compton scattering (ICS) processes, 
the TeV electrons can only travel a short distance of about 
 a few kpc in the Milky Way. Therefore,
the nearby Cosmic Ray (CR) sources such as pulsars~\citep{Shen1970,Harding1987,Aharonian1995,Chi1996,Zhang2001} and dark matter (DM)~\citep{Bergstrom2000,Bertone2005,Bergstrom2009} 
can be probed via the high energy electrons and positrons. 
The spectra of the cosmic ray electrons and positrons (CREs)  
have been measured up to TeV energy scales by the ground-based and space-borne experiments,
for example, HESS~\citep{Aharonian2008,Aharonian2009}, VERITAS~\citep{Staszak2015,Holder2017}, FermiLAT~\citep{Abdollahi2017,Meehan2017}, AMS-02~\citep{AMS02_lepton_sum,AMS02_lepton,AMS02_fraction01,AMS02_fraction02}, and CALET~\citep{CALET2017,CALET2018}. 
In particular, the excesses of the electrons~\citep{Chang2008,Aharonian2008,Abdo2009,Aguilar2014,Aguilar2014} and positrons~\citep{Adriani2009,Ackermann2012,Aguilar2013,Accardo2014} have
been discovered as well.

Recently, the DArk Matter Particle Explorer (DAMPE), which is a new generation space-borne 
experiment to measure CRs and was launched in December 2015,
has announced the first results of high energy CR electron plus positron
($e^{-}+e^{+}$) flux from 25 GeV to 4.6 TeV with unprecedentedly high quality~\citep{DAMPE2017}. 
The energy resolution of the DAMPE is better than 1.5\% at TeV energies,
and the hadron rejection power is about $10^5$. Thus, DAMPE is able to 
reveal (fine) structures of the electron and positron fluxes.
The main DAMPE spectrum can be fitted by a smoothly broken power-law model with
a spectral break around 0.9 TeV, which confirms the previous results by
HESS experiment~\citep{Aharonian2008,Aharonian2009}. And there exists a tentative peak-like flux excess around 1.4 TeV.
Thus, the DAMPE results have stimulated the extensive studies~\citep{Gu2017,Fang2017,Fan2017,Yuan2017_dampe,Duan2017,Gu2017a,Chao2017,Tang2017,Zu2017,Liu2017,Cao2017,Athron2017,Chao2017a,Gao2017, Niu2017_dampe, Jin2017_dampe,Huang2017,Duan2017a,Cao2017a,Ghorbani2017,Nomura2017,Gu2017b,Li2017,Chen2017,Yang2017,Ding2017,Ge2017,Liu2017,Okada2017,Sui2017,Cao2017b,Han2017, Niu2017_dampe, Cholis2017, Fowlie:2017fya}.
The spectral break can be explained by the broad distributed pulsars, pulsar wind nebulae (PWNe), 
supernova remnants (SNRs)~\citep{Fang2017,Yuan2017_dampe}, and by the  
dark matter annihilation and decay in the galaxy halo~\citep{Yuan2017_dampe,Niu2017_dampe,Athron2017,Jin2017_dampe}. 
Also, the tentative peak is always interpreted by local pulsars, PWNe, 
and SNRs~\citep{Fang2017,Yuan2017_dampe,Cholis2017}), and by the DM sub-halos, clumps, and
 mini-spikes~\citep{Fan2017,Duan2017,Gu2017,Athron2017,Liu2017,Cao2017,Jin2017_dampe,Huang2017,Yang2017,Ge2017}. 
 Another important interpretation ascribes observed CR spectrum puzzling features to a nearly 2-3 Myr Super Nova \citep{KNS2015,KNS2018}, which could naturally explain not only proton to helium ratio, positron and anti-proton fluxes, but also plateau in the cosmic ray dipole anisotropy.

However, one can easily show that it is impossible to explain both the spectral break and the
tentative peak simultaneously~\citep{Fang2017,Yuan2017_dampe,Cholis2017,Jin2017_dampe,Huang2017}). 
In addition, we have 74, 93, and 33 events for three continuous bins or energy ranges
[1148.2, 1318.3]~GeV, [1318.3, 1513.6]~GeV, and [1513.6, 1737.8]~GeV, respectively,
which for simplicity we shall call 1229.3~GeV bin, 1411.4~GeV bin, and 1620.5~GeV bin~\citep{DAMPE2017}.
The  number of events and fluxes for these bins are given in Table~\ref{tab:tuned_results}.
From Figure 2 of the DAMPE's paper~\citep{DAMPE2017}, it is obvious that the 1411.4~GeV bin 
has a little bit more than $3\sigma$
excess, while the 1229.3~GeV bin and 1620.5~GeV bin have about $1\sigma$ deficits.
Therefore, it is very difficult to explain the events in these three bins, especially
the first two, no matter by the pulsar or dark matter interpretations.

\begin{table*}[htb]
\begin{center}
\begin{tabular}{c | c|c|c|c|c|c}
\hline\hline
Energy Bins (GeV)  &$N$ (original) & $\Phi(\lep) \pm \sigma_{\mathrm{stat}} \pm \sigma_{\mathrm{sys}}$ (original) &$N$ (revised) & $\Phi(\lep) \pm \sigma_{\mathrm{stat}} \pm \sigma_{\mathrm{sys}}$ (revised) &$\Delta N$   &$\Delta N_{2\sigma}$    \\
$[1148.2,~1318.3]$ &74 &$(4.38 \pm 0.53 \pm 0.14) \times 10^{-8}$ &92 &$(5.44 \pm 0.48 \pm 0.14) \times 10^{-8}$ &+18 &$\pm$18  \\
$[1318.3,~1513.6]$ &93 &$(4.99 \pm 0.53 \pm 0.17) \times 10^{-8}$ &73 &$(3.92 \pm 0.60 \pm 0.17) \times 10^{-8}$ &-20 &$\pm$20  \\
$[1513.6,~1737.8]$ &33 &$(1.52 \pm 0.28 \pm 0.06) \times 10^{-8}$ &39 &$(1.80 \pm 0.26 \pm 0.06) \times 10^{-8}$ &+6  &$\pm$12  \\

\hline\hline
\end{tabular}
\end{center}
\caption{The original and revised numbers of events and fluxes, $\Delta N$, and $\Delta N_{2\sigma}$ for 
the 1229.3~GeV bin, 1411.4~GeV bin, and 1620.5~GeV bin.
Here, $\Delta N$ and $\Delta N_{2\sigma}$ are the adjusted numbers of events and 
the numbers of events for $2\sigma$ deviations from statistical fluctuations.
Thus, we should require $|\Delta N| \le |\Delta N_{2\sigma}|$. }
\label{tab:tuned_results}
\end{table*}

From the theoretical physics point of view, we would like to explain nature with basic
principles such as simplicity and naturalness, or say truth and beauty! 
In the words of Sir Isaac Newton,  
``Truth is ever to be found in the simplicity, and not in the multiplicity and confusion of things.''
Therefore, to explain all the DAMPE data via a simple and natural way, we propose that
the excess in the 1411.4~GeV bin and the deficits in the 1229.3~GeV bin and 1620.5~GeV bin
arise from the $+2\sigma$, $-2\sigma$, and $-1\sigma$ deviations due to
statistical fluctuations, which happened frequently in collider experiments. Remarkably,
we can indeed explain all the DAMPE data consistently via the pulsar and dark matter interpretations,
which have $\chi^{2} \simeq 17.2 $ and $\chi^{2} \simeq 13.9$ (for all the 38 points in DAMPE electron/positron spectrum with 3 of them revised), respectively. 
Our results are different from the previous analyses by neglecting the 1.4 TeV excess~\citep{Niu2017_dampe}.
 As a comparison, the newly released CALET lepton data is used to do a similar global fitting, which could give us some more supports on the origin of the lepton excess.
In addition, we present a $U(1)_D$ dark matter model with Breit-Wigner mechanism, which can provide
the proper dark matter annihilation cross section and escape the CMB constraint.
Furthermore, we suggest a few ways to test our proposal as well as the 1.4 TeV excess.

\section{Statistical Fluctuations}
\label{sec:1}
In the DAMPE's paper~\citep{DAMPE2017}, the numbers of events and 
the CRE fluxes with $1\sigma$ statistical and systematic errors have been given in its Table 1. 
To evaluate the
uncertainties for numbers of the events, we need to understand their relations.
The relation between the number of events and fluxes in each energy bin is~\citep{DAMPE2017,AMS02_lepton_sum}
\begin{equation}
\Phi(\lep) = \frac{N(E) \cdot (1 - \bgfra(E))}{\Aeff(E) \cdot T \cdot \Delta E} \cdot \otherfra(E),
\end{equation}
where $N$ is the number of ($\lep$) events, $\Aeff$ is the effective detector acceptance, 
$T$ is the operating time, $\Delta E$ is the energy range of the bin, 
$\bgfra$ is the background fraction of the events, and $\otherfra$ represents the effects caused 
by other mechanisms which were not given in the Table 1 of Ref.~\citep{DAMPE2017}.

Taking $T = 530$~days and $\otherfra = 1.3$, we can reproduce the corresponding results 
in the 1229.3~GeV bin, 1411.4~GeV bin, and 1620.5~GeV bin
 within the uncertainty $< 0.1 \%$. Consequently, we use the formula
\begin{equation}
\Phi(\lep) = \frac{N(E) \cdot (1 - \bgfra(E))}{\Aeff(E) \cdot T \cdot \Delta E} \cdot 1.3
\end{equation}
in this letter to calculate the fluxes in these bins.

We calculate the $2\sigma$ deviations for the number of events ($\Delta N_{2\sigma}$) 
from the flux statistical fluctuations as follows
\begin{equation}
\Delta N_{2\sigma} = \frac{\Delta \Phi(\lep)_{2 \sigma_{\mathrm{stat}}}}{\Phi(\lep)} \cdot N~.
\end{equation}
Thus, for the 1229.3~GeV bin, 1411.4~GeV bin, and 1620.5~GeV bin,
we obtain $\Delta N_{2\sigma} = \pm 18,~\pm 20,~\pm 12$, respectively. 
Assume  $-2\sigma$, $+2\sigma$, and $-1\sigma$ deviations for
these bins from statistical fluctuations, 
we have $\Delta N = + 18,~- 20,~+ 6$, respectively. 
Therefore, the revised numbers of events for the 1229.3~GeV bin, 1411.4~GeV bin, and 1620.5~GeV bin,
 are 92, 73, and 39, respectively.

Furthermore, we reestimate the statistical uncertainties in these bins based on the revised 
numbers of events via the formula
\begin{equation}
\Delta N_{1\sigma } \simeq \frac{1}{\sqrt{N}} ~,
\label{Eq-N1S}
\end{equation}
and then calculate the corresponding fluxes and their statistical uncertainties.
The systematical uncertainties are assumed to be invariant. All the detailed information for
 these three bins are given in Table~\ref{tab:tuned_results}. By the way, as a cross check,
with Eq.~(\ref{Eq-N1S}), we have reproduced similar $1\sigma$ statistical uncertainties 
of the original fluxes
in the DAMPE's paper~\citep{DAMPE2017}.

\section{Fitting Procedure}
\label{sec2}

In CR theory, the CR electrons are considered to be accelerated during the acceleration of CR nuclei at the sources, e.g. SNRs. On the other hand, the CR positrons are produced as secondary particles from CR nuclei interaction with the interstellar medium (ISM) \citep{Adriani2009,AMS2013,Barwick1997,AMS01}. From the observed spectra of positrons and electrons \citep{AMS02_fraction01,AMS02_fraction02,AMS02_lepton,AMS02_lepton_sum}, we can conclude that there should exist some extra sources producing electron-positron pairs. As we stated in the first section, these extra sources could be astrophysical sources or DM annihilation or decay. As a result, the CREs data contains (i) the primary electrons; (ii) the secondary electrons; (iii) the secondary positrons; (iv) the extra source of electron-positron pairs. If we want to study the properties of the extra source, we should deduct the primary electrons and secondary electrons/positrons first.

The primary electrons are always assumed to have a power-law injection and the secondary electrons/positrons are determined mainly by the CR proton and helium nucleus interact with ISM. Consequently, we should do global fitting to all these related data simultaneously which can avoid the bias of choosing the lepton background parameters.

The public code  {\sc dragon} \footnote{https://github.com/cosmicrays/DRAGON} \citep{Evoli2008} was used to do numerical calculations. Some custom modifications are performed in the original code, such as the possibility to use specie-dependent injection spectra, which is not allowed by default in {\sc dragon}.

In view of some discrepancies when fitting  the new data \citep{Johannesson2016}, we use a factor $c_{\He}$ to re-scale the helium-4 abundance (which has a default value of $7.199 \times 10^4$), which helps us to get a better global fitting.

\subsection{Background}

In this work, we use the widely used  diffusion-reacceleration model which can give a consistent fitting results to the AMS-02 nuclei data (see for e.g., \citep{Niu2017,Yuan2017,Yuan2018,Zhu2018}). In the whole propagation region, a uniform diffusion coefficient ($D_{xx} = D_{yy}= D_{zz} = D_0\beta \left( R/R_0 \right)^{\delta}$) is employed to describe the propagation .

The hardening of the nuclei spectra at $\sim 300 \GeV$ (which has been observed by ATIC-2 \citep{ATIC2006}, CREAM \citep{CREAM2010}, PAMELA \citep{PAMELA2011}, and AMS-02 \citep{AMS02_proton,AMS02_helium}) is considered by adding breaks in the primary source injections. At the same time, considering the observed significant difference in the slopes of proton and helium (of about $\sim 0.1$ \citep{Adriani2011,AMS02_proton,AMS02_helium}), we use separate primary source spectra settings for proton and helium. In summary, for nuclei primary source injections, each of them has 2 breaks at rigidity $R_{\A1}$ and $R_{\A2}$. The corresponding slopes are $\nu_{\A1}$ ($R \le R_{\A1}$), $\nu_{\A2}$ ($R_{\A1} < R \le R_{\A2}$) and $\nu_{\A3}$ ($R > R_{\A3}$).

For CR electrons primary source, we use one break $R_{e}$ for electron primary source, and the corresponding slopes are $\nu_{e1}$ ($R \le R_{e}$) and $\nu_{e2}$ (($R > R_{e}$)).

In order to  take into account the uncertainties when calculating the secondary CR particles' fluxes, we employ parameters $c_{\pbar}$ and $\cpos$ to re-scale the calculated secondary flux to fit the data \citep{Tan1983,Duperray2003,Kappl2014,diMauro2014,Lin2015}. Note that the above mentioned uncertainties may not be simply represented by a constant factor, but most probably are  energy dependent \citep{Delahaye2009,Mori2009}. Here we expect that a constant factor could be a simple assumption. 

The force-field approximation \citep{Gleeson1968} is used to describe the effects of solar modulation effects. $\phinuc$, $\phipbar$ and $\phipos$ are used to modulate the local interstellar spectra of nuclei (proton and helium), anti-proton and positrons respectively, which based on the charge-sign dependence of solar modulation. On the other hand, Because the DAMPE lepton data $\gtrsim 20 \GeV$, we did not consider the modulation effects on electrons (or leptons).

\subsection{Extra Sources}

We consider  both pulsar and DM scenarios to generate the CRE excesses in the observed spectrum 
by the DAMPE experiment. 
For the pulsar scenario, a continuous distributed pulsar background was used~\citep{Lin2015,Niu2017_dampe}. 
The injection spectrum of such sources is  assumed to be a power law with an exponential cutoff
\begin{equation}
  q_e^{\psr}(p) = N_{\psr}(R/\mathrm{10\GeV})^{-\nu_{\psr}} \exp{(-R/R_\mathrm{c})},
  \label{eq:psr_injection}
\end{equation}
where $N_{\psr}$ is the normalization factor, $\nu_{\psr}$ is the spectral index, and 
$R_\mathrm{c}$ is the cutoff rigidity. 
For the DM scenario, we employ the Einasto profile~\citep{Navarro2004,Merritt2006,Einasto2009,Navarro2010}
\begin{equation}
\rho(r)=\rho_\odot \exp
\left[
-\left( \frac{2}{\alpha}\right)
\left(\frac{r^{\alpha}-r_\odot^{\alpha}}{r_{s}^{\alpha}} \right)
\right] ,
\end{equation}
with $\alpha\approx 0.17$, $r_{s}\approx 20 \kpc$, and $\rho_\odot \approx 0.39 \GeV \cm ^{-3}$ is 
the local DM relic density \citep{Catena2010,Weber2010,Salucci2010,Pato2010,Iocco2011}. And the source term, which
 we use to add the CRE particles from the annihilations of the Majorana DM particles, is
\begin{equation}
\label{eq:dm_source}
Q(\boldsymbol{r},p)=\frac{\rho(\boldsymbol{r})^2}{2 m_{\chi}^2}\langle \sigma v \rangle 
\sum_{f} \eta_{f} \frac{dN^{(f)}}{dp} ,
\end{equation}
where $\langle \sigma v \rangle$ is the velocity-averaged DM annihilation cross section multiplied by DM relative velocity (referred as cross section), $\rho(\boldsymbol{r})$ is the DM density distribution, and $dN^{(f)}/dp$ is the injection energy spectrum  of CREs from DM annihilating into the Standard Model (SM) final states via leptonic channels 
$f{\bar f}$ ($\eebar$, $\mumubar$, and $\tautaubar$) with $\eta_{f}$ ($\etae$, $\etamu$, and $\etatau$) 
the corresponding branching fractions. Here, we normalized $\eta_{f}$ as $\etae + \etamu + \etatau = 1$.

The parameters related to the extra source of the leptons for pulsar scenario is $(N_{\psr}, \nu_{\psr}, R_{c})$, and for DM scenario is $(\Mdm, \sigv, \etae, \etamu, \etatau)$.

\subsection{Data Sets and Parameters}

As in Ref.~\citep{Niu2017_dampe}, we perform a global fitting on the data set including the proton fluxes from AMS-02 and CREAM \citep{AMS02_proton,CREAM2010} helium flux from AMS-02 and CREAM\footnote{The CREAM data was used as the supplement of the AMS-02 data because it is more compatible with the AMS-02 data when $R \gtrsim 1 \TeV$.} \citep{AMS02_helium,CREAM2010}, $\pbarp$ ratio from AMS-02 \citep{AMS02_pbar_proton}, positrons flux from AMS-02 \citep{AMS02_lepton}, and CRE flux from DAMPE \citep{DAMPE2017}, which could account for the primary electrons, the secondary leptons, and the extra leptons in a self-consistent way \footnote{The errors used in our global fitting are the quadratic summation over statistical and systematic errors.}. Moreover, the employed AMS-02 positron flux is used to calibrate the positron contribution in the DAMPE CRE flux in energy region $\lesssim 300 \GeV$\footnote{The systematics between AMS-02 and DAMPE are dealt with by employing a re-scale factor $\cpos$ on positron flux.}. The framework of the fitting procedure is the same as our previous work \citep{Niu2017_dampe,Niu2017}, where the details can be found.

Considering the systematics between different CREs spectra observed by different experiments (see in Fig. \ref{fig:lepton_psr} and \ref{fig:lepton_dm}), we take the newly released CREs spectrum from CALET \citep{CALET2018} as a comparison to do global fitting as that on DAMPE CREs spectrum.
Because both of these experiments are implemented in space and have a similar ability to obtain CREs data, a reasonable explanation on lepton excess should explain both of them simultaneously.
In this case, DAMPE and CALET could be considered as the maximum and minimum CREs flux examples.

Altogether, the data set in our global fitting is 
 \begin{align*}
D = &\{D^{\text{AMS-02}}_{\p}, D^{\text{AMS-02}}_{\He},  D^{\text{AMS-02}}_{\pbarp}, D^{\text{CREAM}}_{\p}, \\
&D^{\text{CREAM}}_{\He}, D^{\text{AMS-02}}_{\pos}, D^{\text{DAMPE}}_{\lep} / D^{\text{CALET}}_{\lep} \}~.
\end{align*}

The parameter sets for pulsar scenario is 
\begin{align*}
\boldsymbol{\theta}_{\psr} =  &\{ D_{0}, \delta, z_{h}, v_{A}, | N_{\p}, R_{\p1}, R_{\p2}, \nu_{\p1}, \nu_{\p2}, \nu_{\p3}, \\
&  R_{\He1},  R_{\He2}, \nu_{\He1}, \nu_{\He2}, \nu_{\He3}, |  \phinuc, \phipbar, c_{\He}, c_{\pbar},  | \\
& N_{\e}, R_{\e1}, \nu_{\e1}, \nu_{\e2}, |   \\
& N_{\psr}, \nu_{\psr}, R_{c}, | \\
&  \phipos, c_{\pos} \}~,
\end{align*}
for DM scenario is 
\begin{align*}
\boldsymbol{\theta}_{\DM} =  &\{ D_{0}, \delta, z_{h}, v_{A}, | N_{\p}, R_{\p1}, R_{\p2}, \nu_{\p1}, \nu_{\p2}, \nu_{\p3}, \\
&  R_{\He1},  R_{\He2}, \nu_{\He1}, \nu_{\He2}, \nu_{\He3}, |  \phinuc, \phipbar, c_{\He}, c_{\pbar}, | \\
& N_{\e}, R_{\e1}, \nu_{\e1}, \nu_{\e2}, |   \\
& \Mdm, \sigv, \eta_{e}, \eta_{\mu}, \eta_{\tau},  | \\
& \phipos, c_{\pos} \}~.
\end{align*}

Note that, most of these 2 scenarios' parameters in the set $\boldsymbol{\theta}_{\psr}$ and $\boldsymbol{\theta}_{\DM}$ is the same with each other except those which account the extra sources of lepton.

\section{Results}
\label{sec3}

When the Markov Chains have reached their equilibrium state, we take the samples of the parameters as their posterior probability distribution functions. 
The best-fit values, statistical mean values, standard deviations and allowed intervals at $95 \%$ CL for parameters in set $\boldsymbol{\theta}_{\psr}$ and $\boldsymbol{\theta}_{\DM}$ for DAMPE and CALET   are shown in Appendix, Table \ref{tab:params_psr_dampe} (DAMPE, pulsar scenario), Table \ref{tab:params_psr_calet} (CALET, pulsar scenario),  Table \ref{tab:params_dm_dampe} (DAMPE, DM scenario), and Table \ref{tab:params_dm_calet} (CALET, DM scenario), respectively. For best fit results of the global fitting, we got $\chi^{2}/d.o.f = 243.13/299 $ (DAMPE, pulsar scenario), $\chi^{2}/d.o.f = 229.98/301 $ (CALET, pulsar scenario), $\chi^{2}/d.o.f = 262.94/297 $ (DAMPE, DM scenario), and $\chi^{2}/d.o.f = 265.03/299 $ (CALET, DM scenario).

\subsection{Background}

The best-fitting results and the corresponding residuals of the proton flux, helium flux and $\pbarp$ ratio for pulsar scenario is showed in Appendix, Fig. \ref{fig:nuclei_psr_results}, for DM scenario is showed in Appendix, Fig. \ref{fig:nuclei_dm_results}. In these figures, we can see that the nuclei data is perfectly reproduced, which would provide a good precondition for the  fitting on the lepton data.

Although the main purpose of this work does not focus on the CR propagation models, we would like to emphasize some points here: (i) As shown in Appendix, Table \ref{tab:params_psr_dampe}, Table \ref{tab:params_psr_calet}, Table \ref{tab:params_dm_dampe}, and Table \ref{tab:params_dm_calet}, we got larger best-fit values of $D_{0}$ and $z_{h}$ than previous works \citep{Niu2017}. This mainly because the newly released AMS-02 nuclei spectra favor large values of $D_{0}$ and $z_{h}$.\footnote{A similar global fitting result can be found in Ref. \citep{Niu2018}, which use a different numerical tool {\sc galprop} to do calculation.} Moreover, the employed 2 breaks in nuclei primary source injection strengthen the classical degeneracy between $D_{0}$ and $z_{h}$ based on the data set we used in this work (without B/C), which both got larger best-fit values in this work. (ii) The employed 2 breaks in the nuclei primary source injection accounted for the observed hardening in the observed spectra, other than use only one break and let $\delta$ compromise the different slopes in high energy regions, which lead to a smaller value of $\delta$ and fitting uncertainties on $\delta$ ($\lesssim 0.01$) than previous works.

\subsection{Extra Sources}

\begin{figure*}[!htbp]
  \centering
  \includegraphics[width=0.8\textwidth]{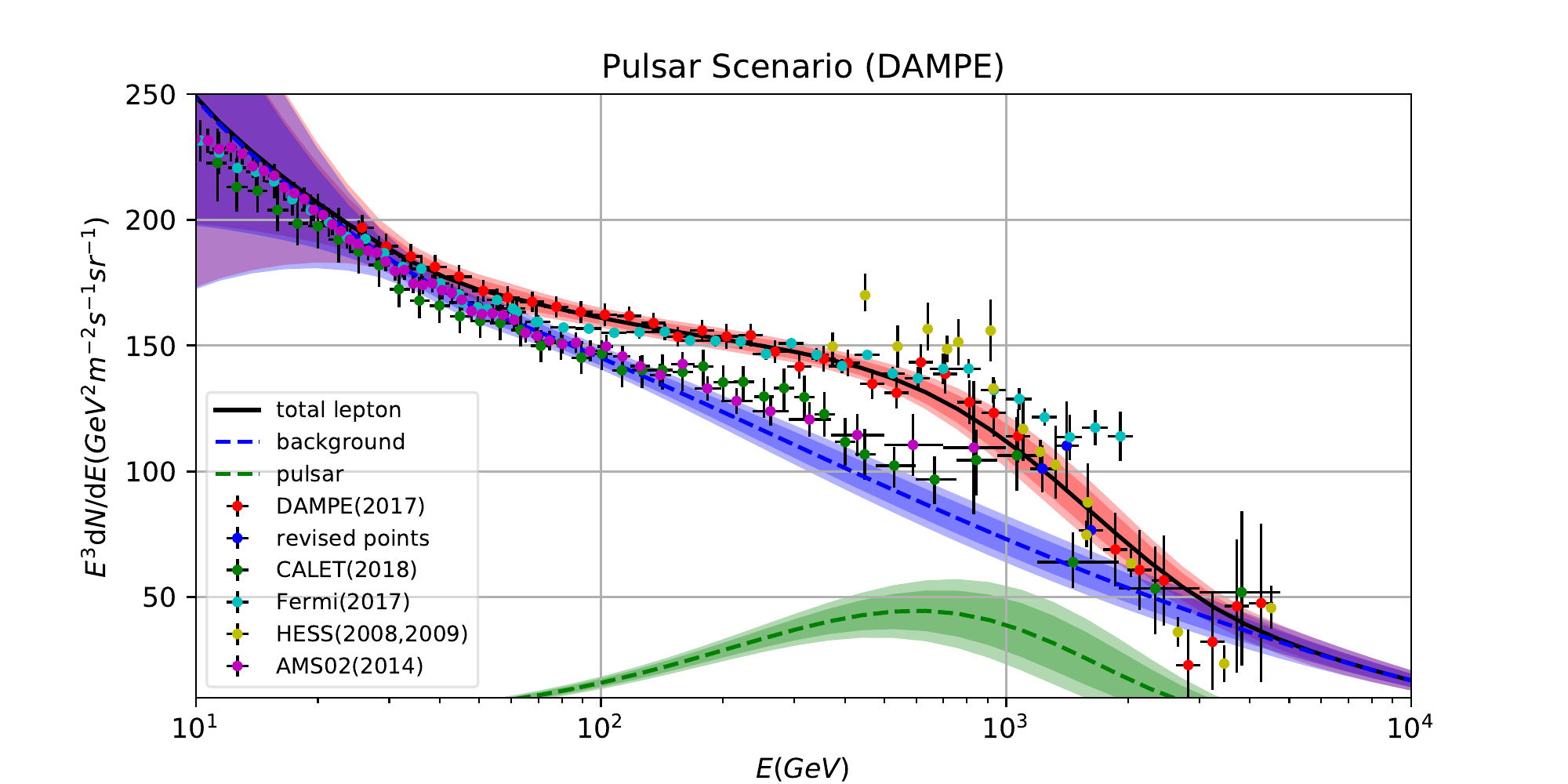}
  \includegraphics[width=0.8\textwidth]{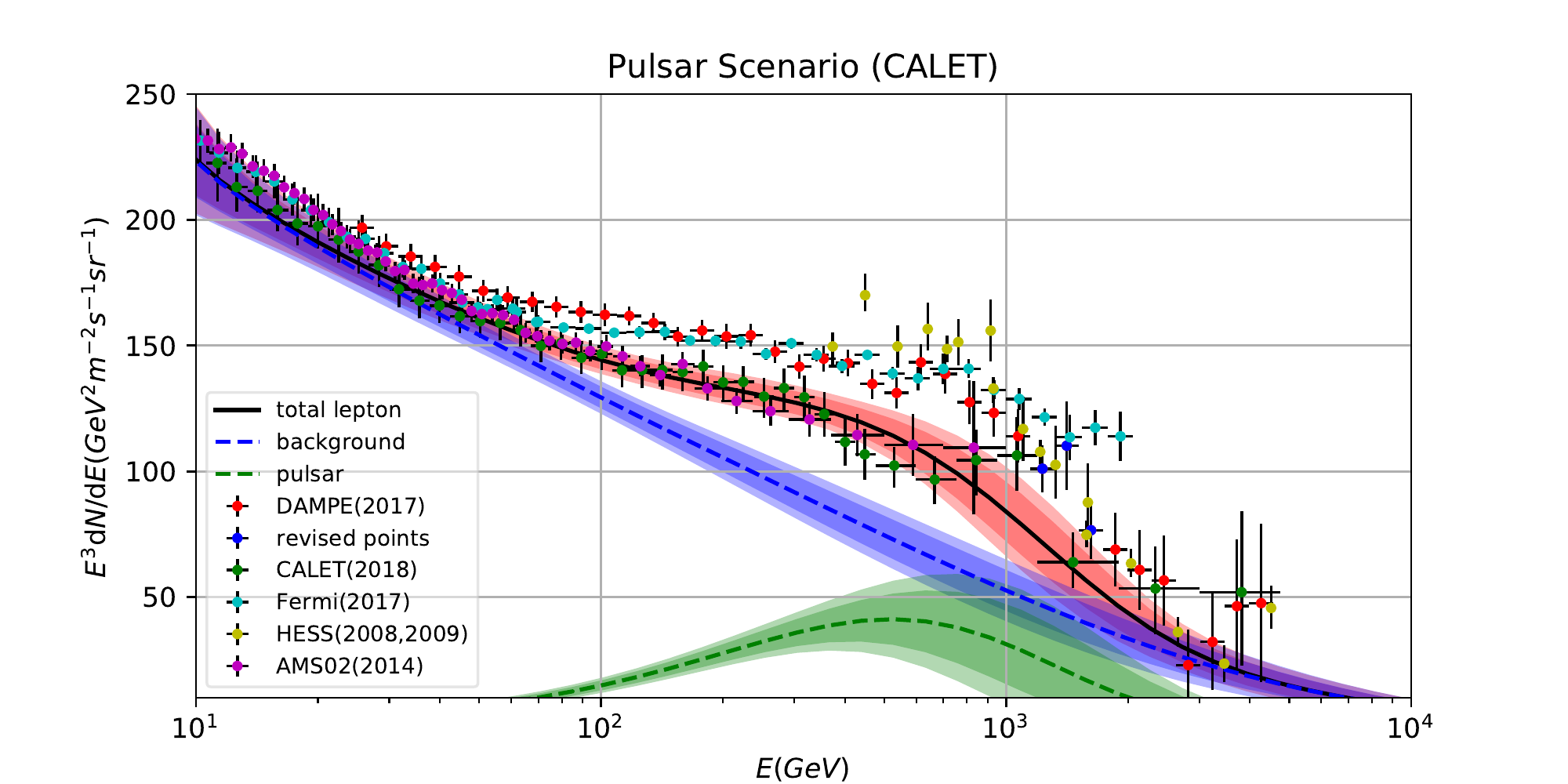}
  \caption{The global fitting results of the DAMPE and CALET lepton flux for pulsar scenario. The $2\sigma$ (deep color) and $3\sigma$ (light color) bounds of total fitted results (red), contribution from background (blue) and  pulsar (green) are also shown in the figure. And we have $\chi^{2} \simeq 17.2$ (DAMPE) and $\chi^{2} \simeq 16.1$ (CALET). (Data sources: DAMPE \citep{DAMPE2017}, CALET \citep{CALET2018}, Fermi \citep{Fermi2017}, HESS \citep{Aharonian2008,Aharonian2009} and AMS-02 \citep{AMS02_lepton_sum}.)}
  \label{fig:lepton_psr}
\end{figure*}

\begin{figure*}[!htbp]
  \centering
  \includegraphics[width=0.8\textwidth]{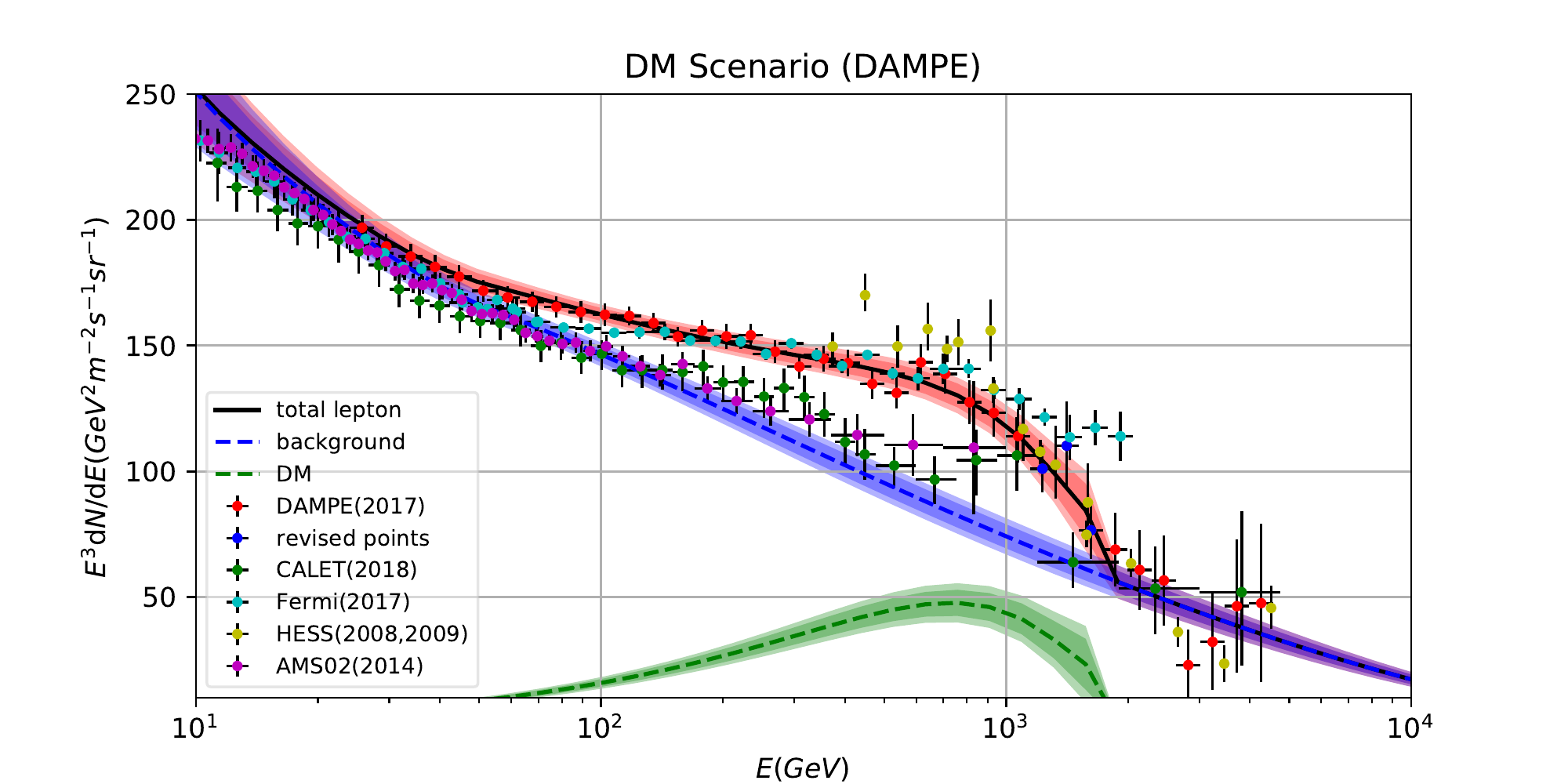}
  \includegraphics[width=0.8\textwidth]{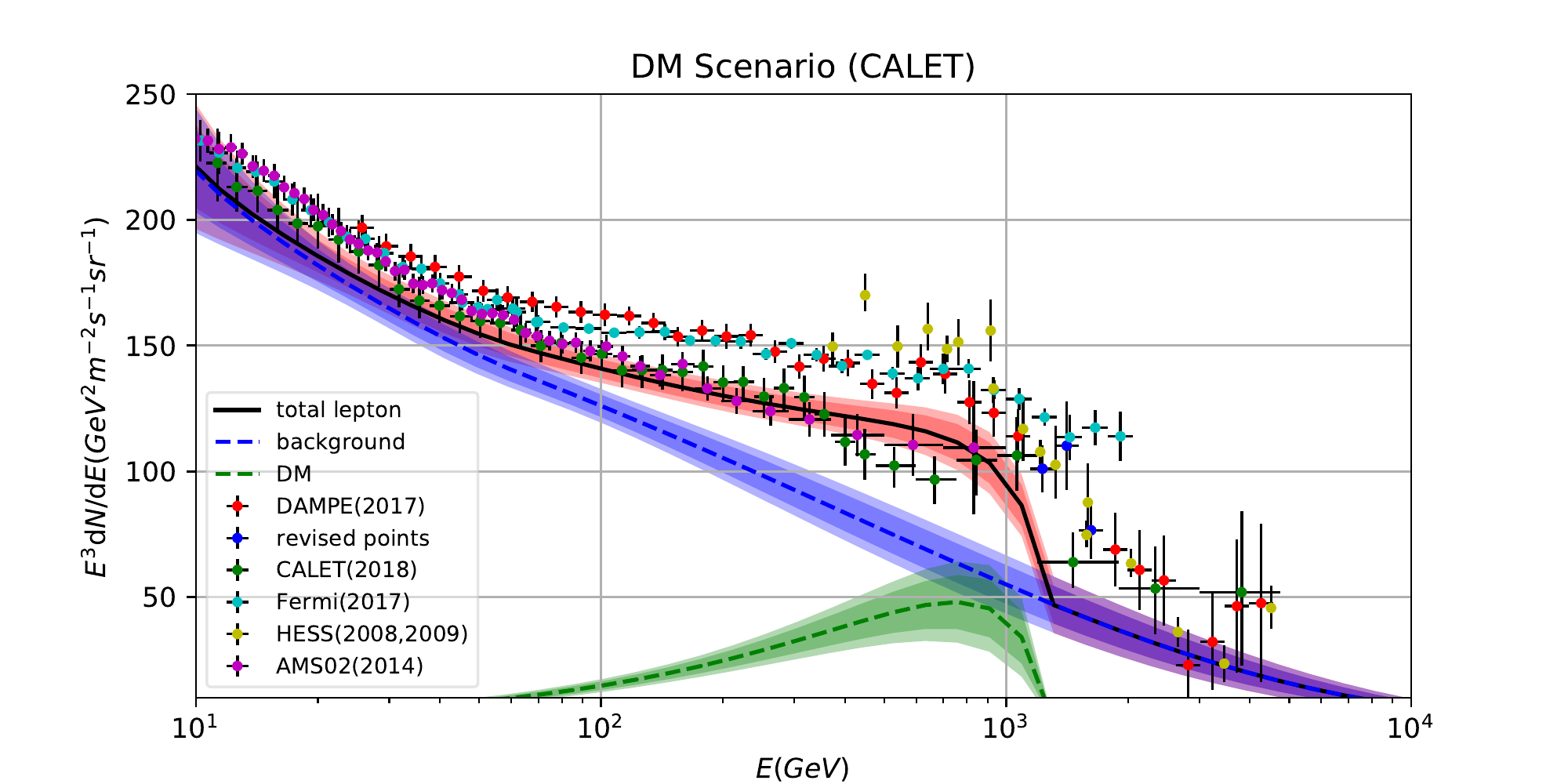}
  \caption{The global fitting results of the DAMPE and CALET lepton flux for DM scenario. The $2\sigma$ (deep color) and $3\sigma$ (light color) bounds of total fitted results (red), contribution from background (blue) and  DM (green) are also shown in the figure. And we have $\chi^{2} \simeq 13.9$ (DAMPE) and $\chi^{2} \simeq 25.7$ (CALET).}
  \label{fig:lepton_dm}
\end{figure*}

\begin{figure*}[!htbp]
  \centering
  \includegraphics[width=0.48\textwidth]{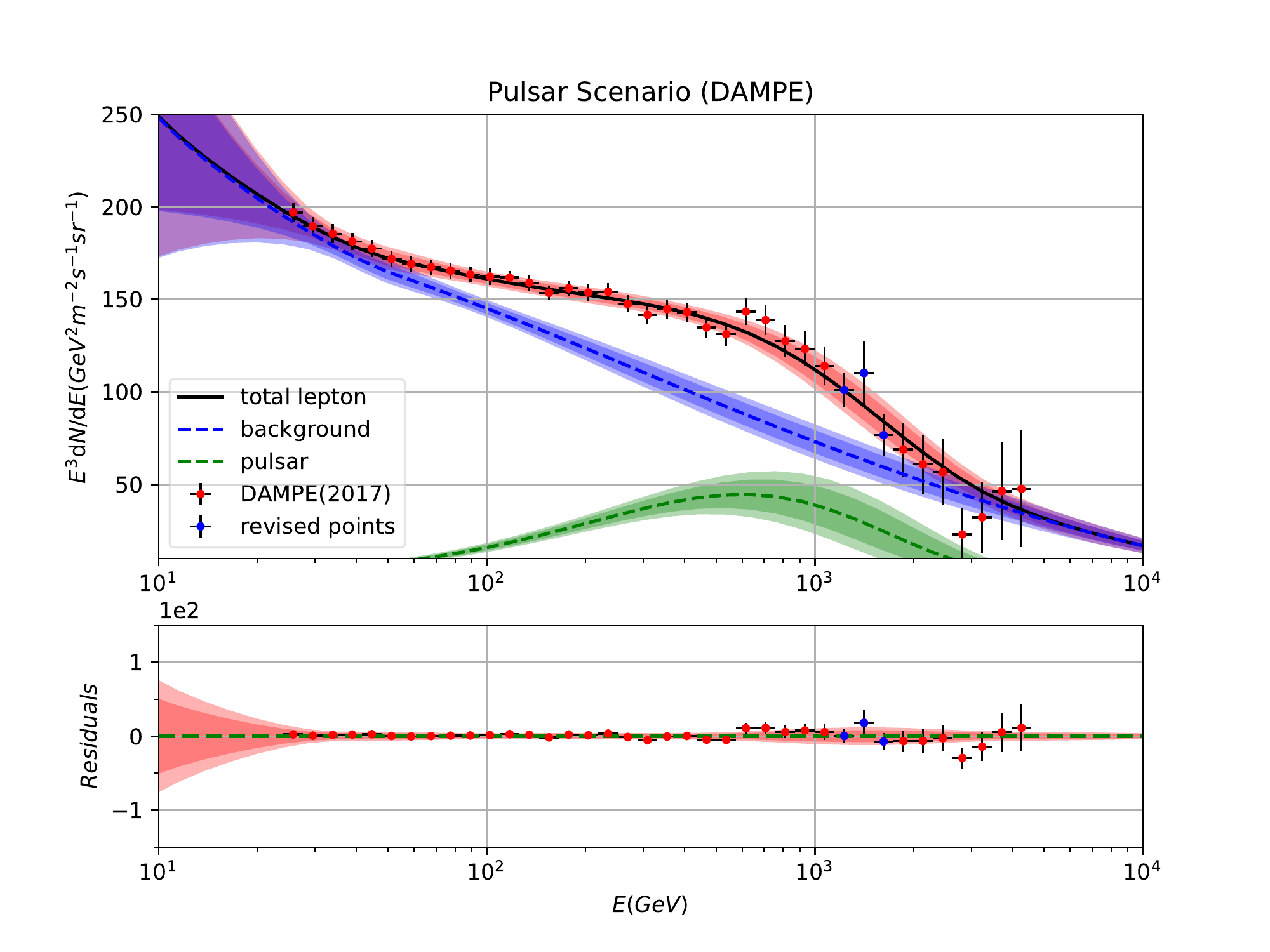}
  \includegraphics[width=0.48\textwidth]{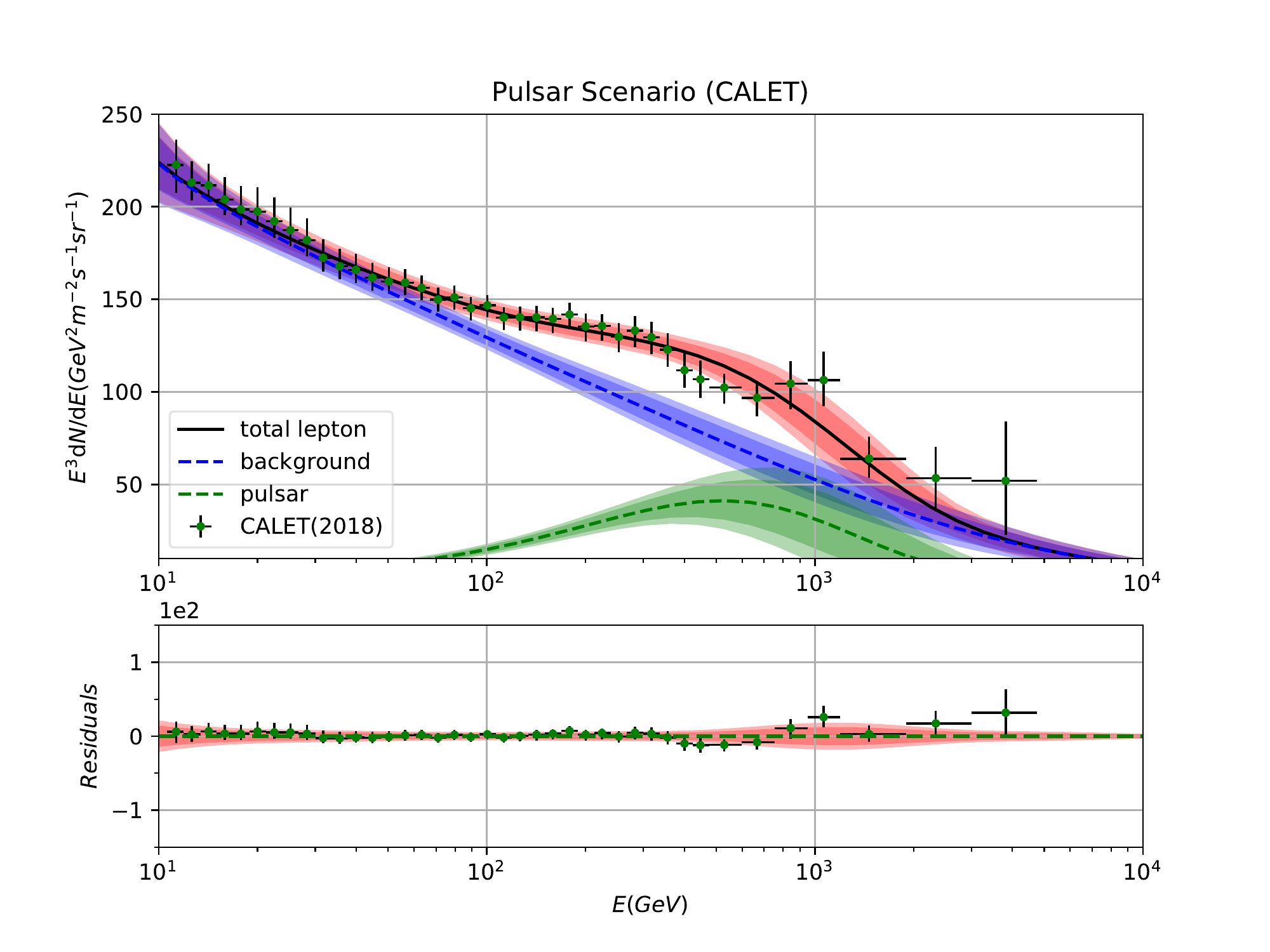}
  \includegraphics[width=0.48\textwidth]{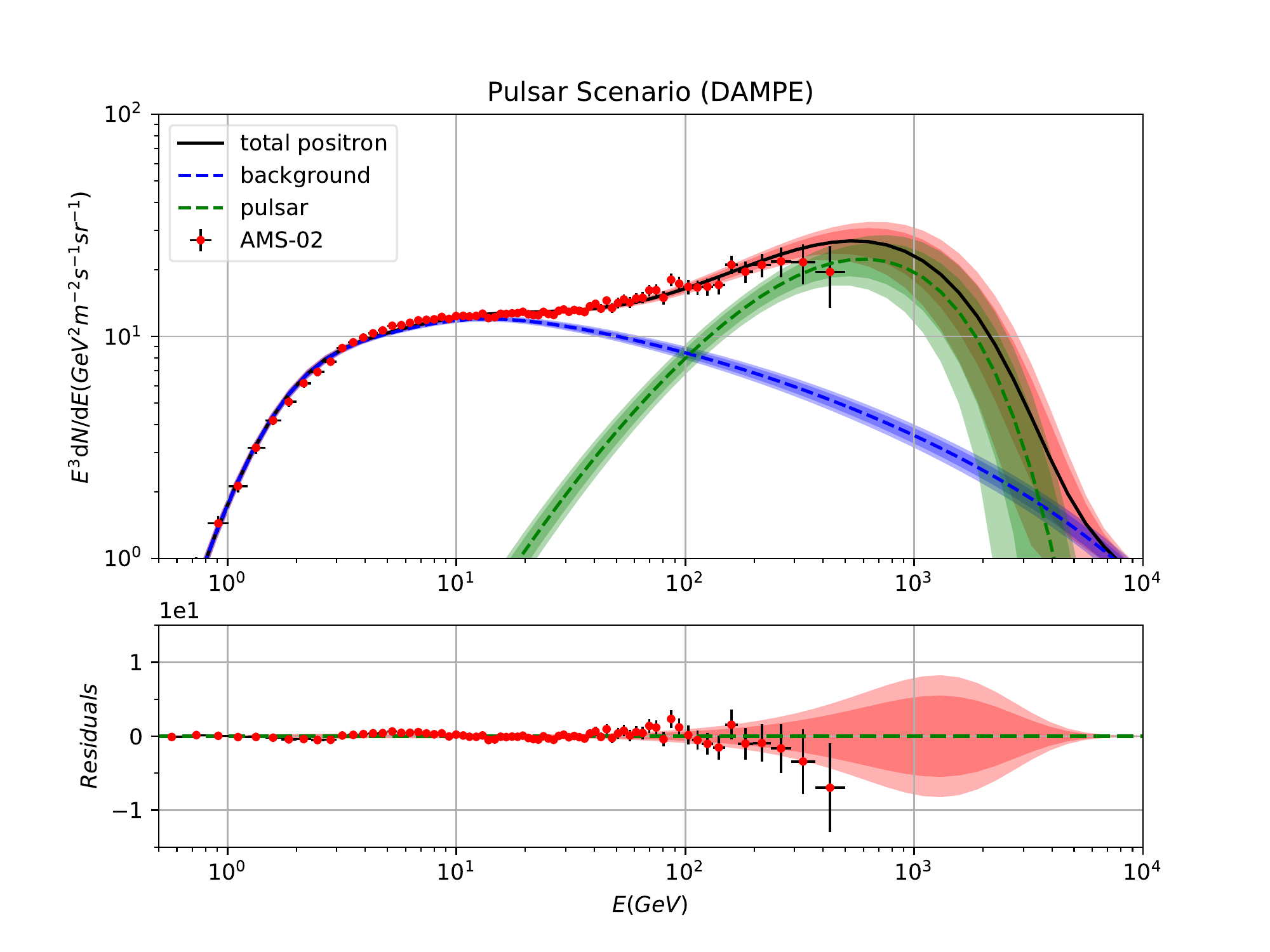}
  \includegraphics[width=0.48\textwidth]{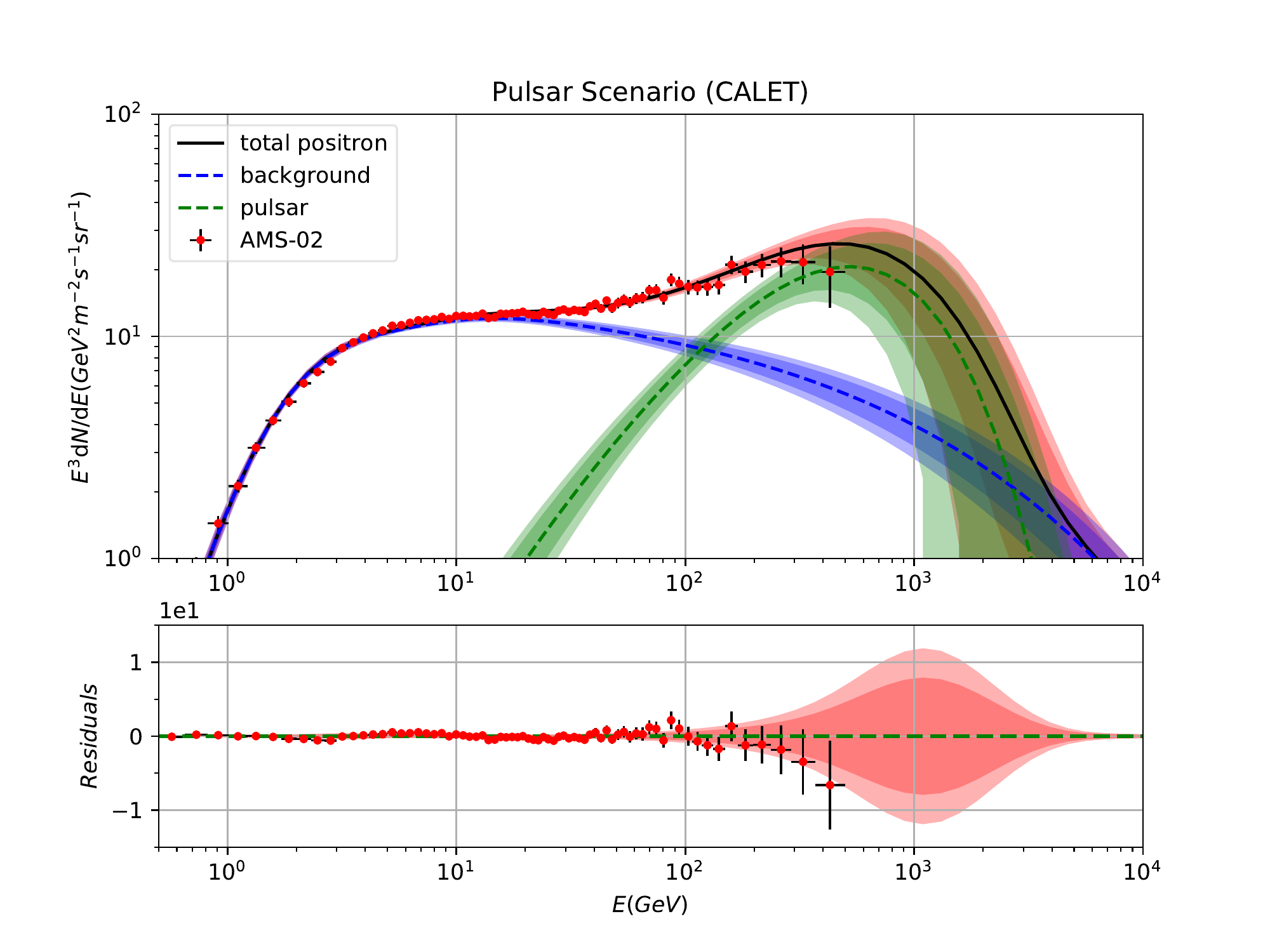}
  \caption{The global fitting results and the corresponding residuals to the lepton (DAMPE and CALET) and positron (AMS-02) flux for pulsar scenario. The $2\sigma$ (deep color) and $3\sigma$ (light color) bounds of total fitted results (red), contribution from background (blue) and  pulsar (green) are also shown in the figure. Different from Fig. \ref{fig:lepton_psr} and Fig. \ref{fig:lepton_dm}, all the experiment data which did not participate in the global fitting has not been not represented.}
\label{fig:lepton_pos_psr}
\end{figure*}

\begin{figure*}[!htbp]
  \centering
  \includegraphics[width=0.48\textwidth]{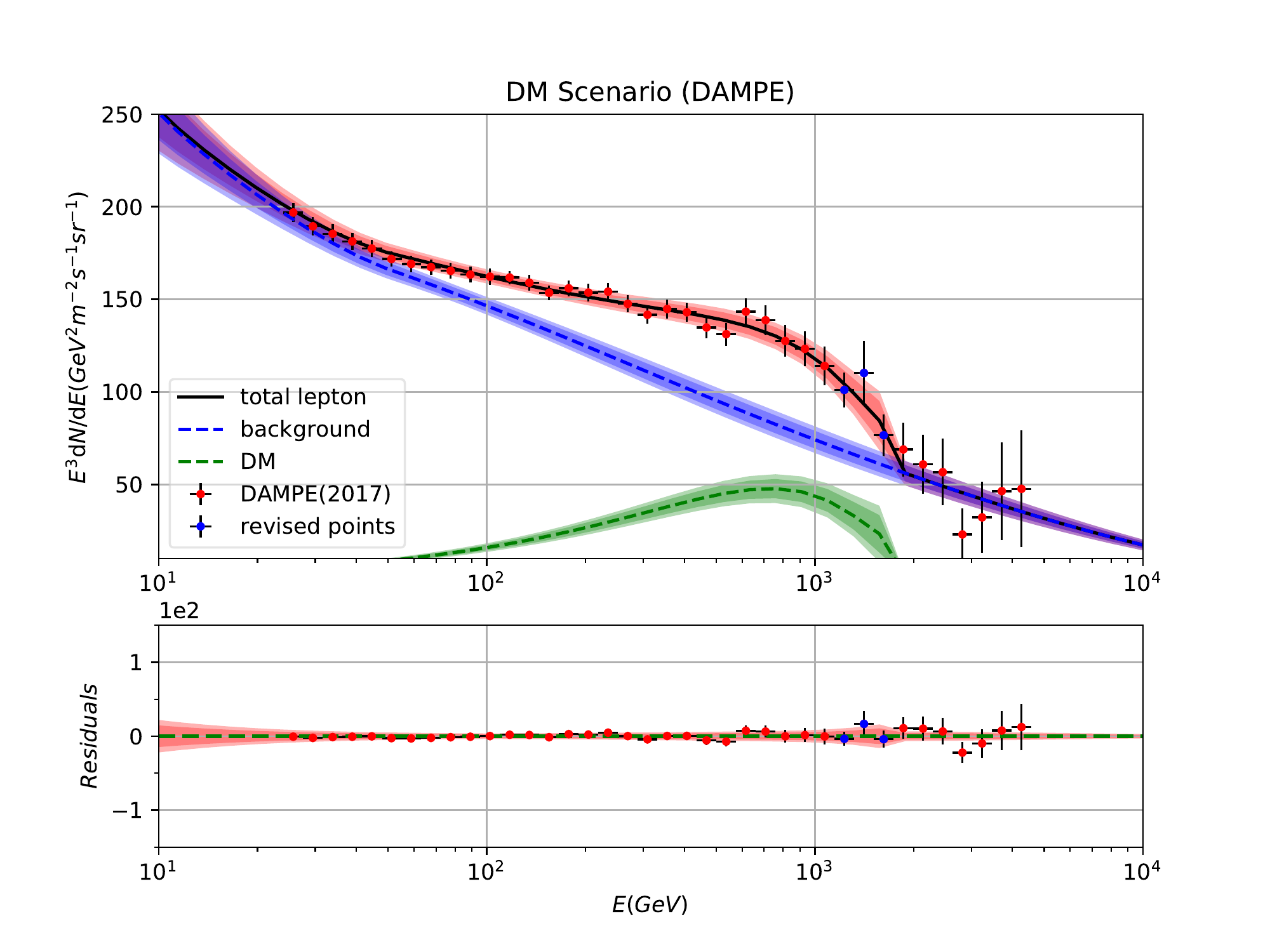}
  \includegraphics[width=0.48\textwidth]{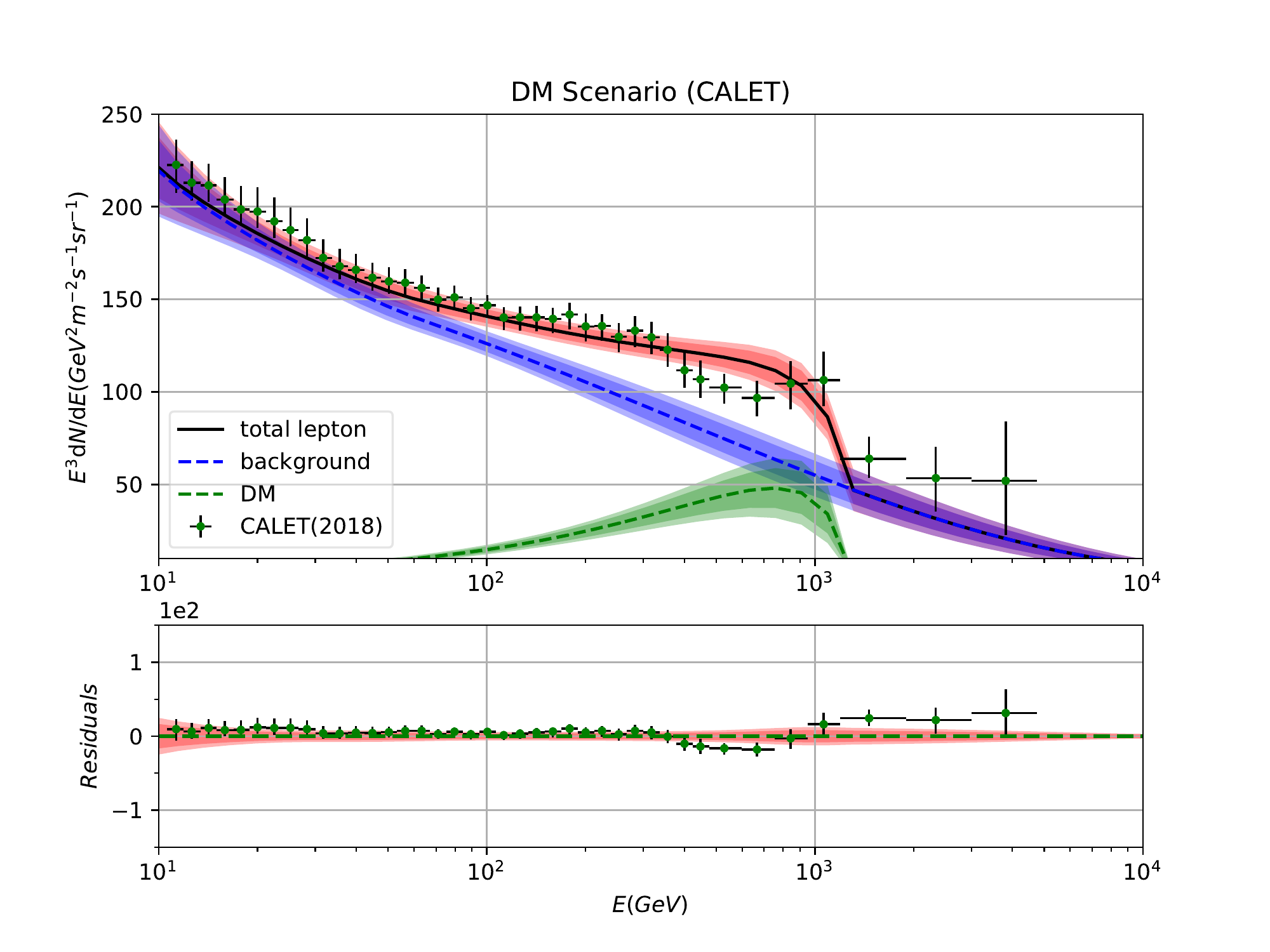}
  \includegraphics[width=0.48\textwidth]{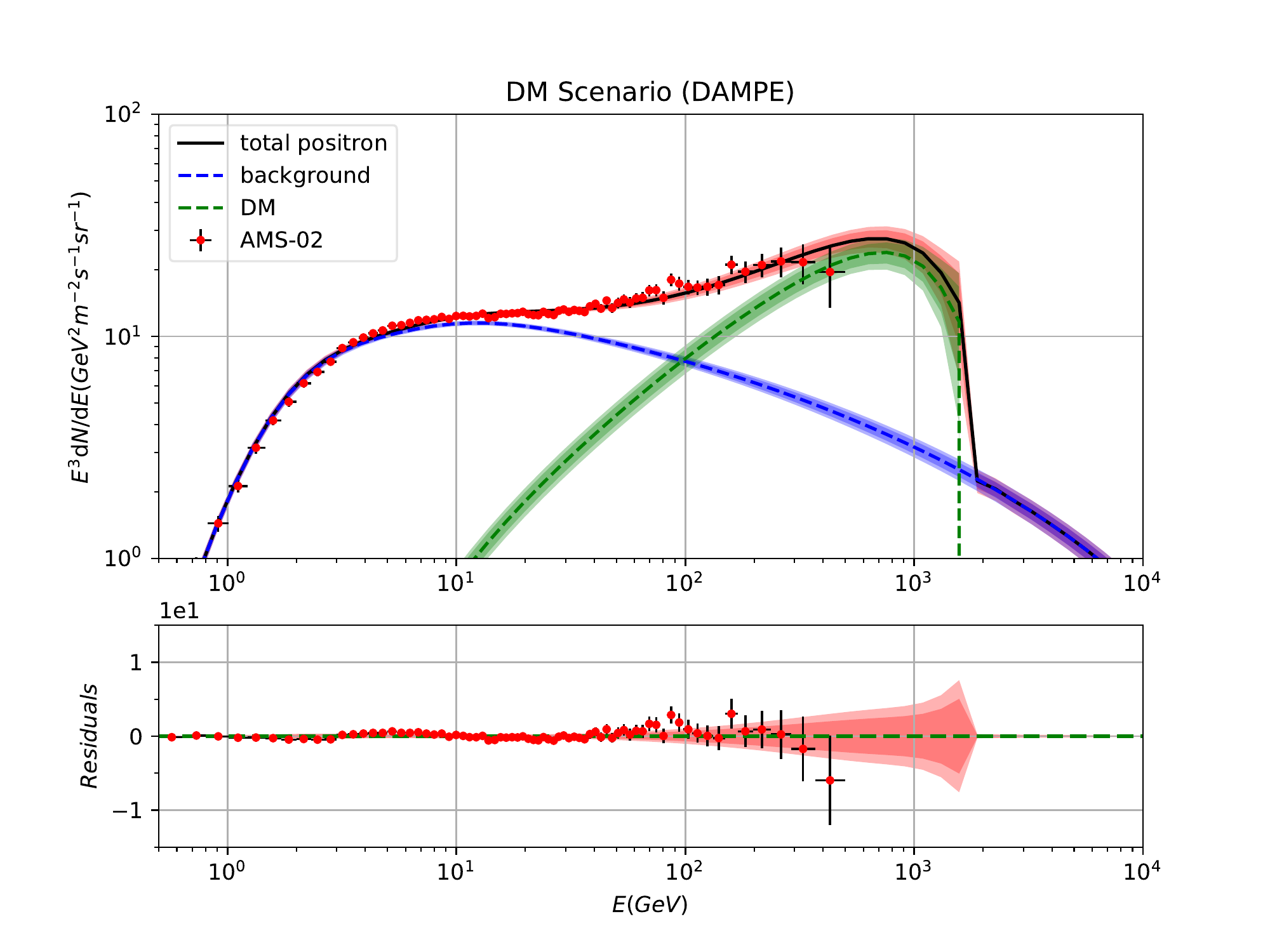}
  \includegraphics[width=0.48\textwidth]{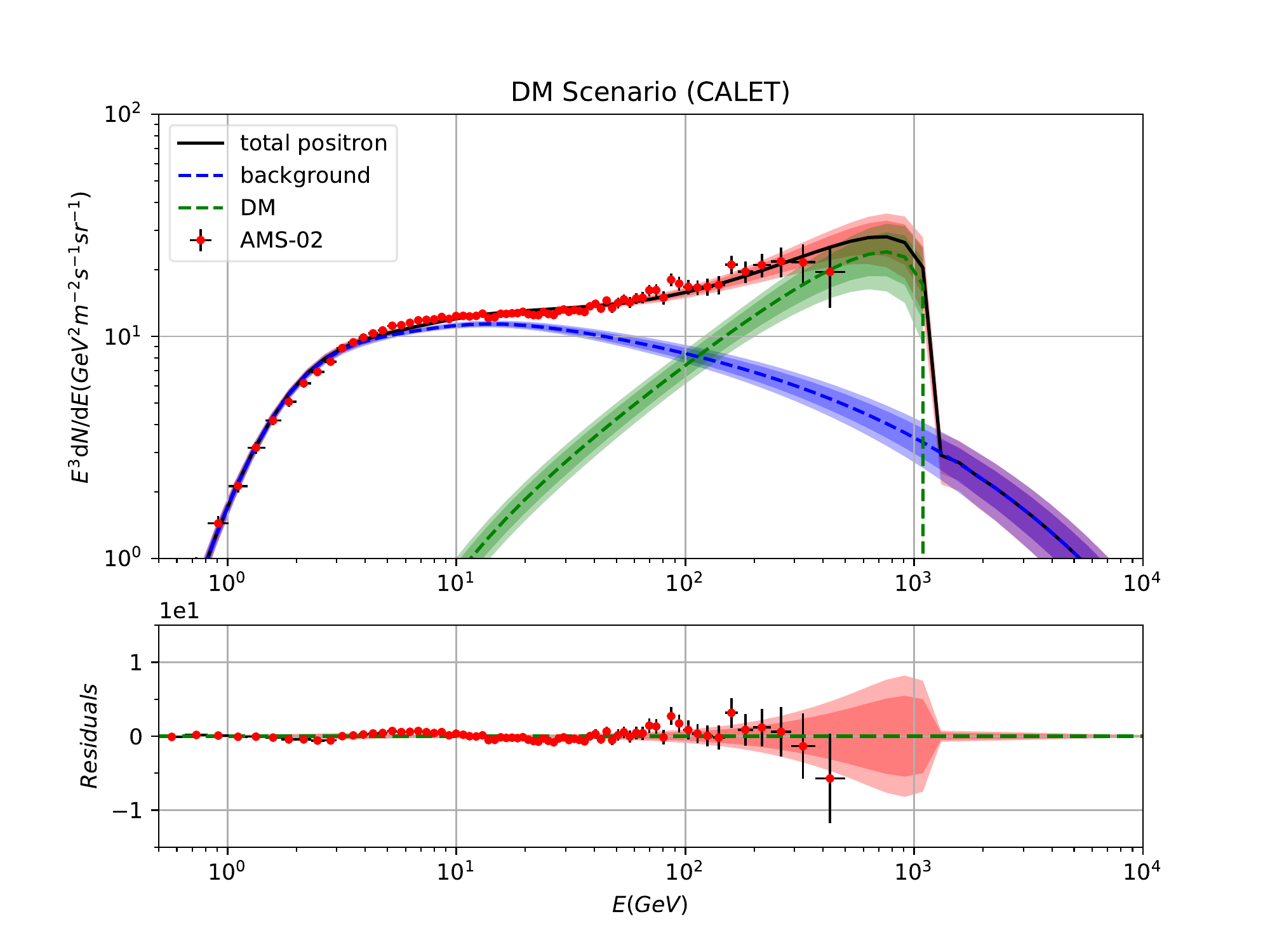}
  \caption{The same as Fig. \ref{fig:lepton_pos_psr}, but for DM scenario.}
\label{fig:lepton_pos_dm}
\end{figure*}

The fitting results of the pulsar and DM scenario on DAMPE and CALET CREs spectrum are given in Figs. \ref{fig:lepton_psr} and \ref{fig:lepton_dm} respectively, which also shows some  CREs spectrum from other experiments. Cleaner fitting results are shown in Figs. \ref{fig:lepton_pos_psr} and \ref{fig:lepton_pos_dm}.\footnote{Here, one should note that both proton and positron flux from AMS-02 are well fitted in such configurations, but the coincidence of the proton and positron flux in this work is caused by chance.}  From these figures, we can conclude that both scenarios could provide a excellent fittings to the DAMPE CREs spectrum within $3\sigma$ fitting deviation, which do not need to employ extra local sources. At the same time, the CALET CREs spectrum could also be fitted by both scenarios, although the fitting result is not that good because of a suspected bump at about 0.9 - 1 $\TeV$.

For the best fit result on the DAMPE and CALET CREs spectrum, we got $\chi^{2} \simeq 17.2 $ (DAMPE, pulsar scenario), $\chi^{2} \simeq 16.1 $ (CALET, pulsar scenario), $\chi^{2} \simeq 13.9 $ (DAMPE, DM scenario) and $\chi^{2} \simeq 25.7 $ (CALET, DM scenario). This indicate that CALET CREs data disfavors the DM scenario because of the defective fit on the suspected bump at about 0.9 - 1 $\TeV$. This need more events accumulated in the future.

The detailed results of the constraints on the parameters could be found in Appendix, Table \ref{tab:params_psr_dampe}, Table \ref{tab:params_psr_calet}, Table \ref{tab:params_dm_dampe}, and Table \ref{tab:params_dm_calet}. Most of the parameters are slightly different between DAMPE and CALET fitting because of the systematics between them. One point we want to mention is that in the fitting results of CALET CREs data, DM scenario, $\etae \simeq 0.930$, $\etamu \simeq 0.042 $, and $\etatau \simeq 0.028 $, which is largely different from the DAMPE results.

In the following of this work, we will focus on the analysis of the DAMPE results, whose method could be extended to deal with the CALET results without difficulties.

For the pulsar scenario, the fitting results give $\nu_{\psr} \simeq 0.62$, which is obviously different 
from the fitting results in previous works (see for e.g., \citep{Profumo2012}). In standard pulsar models, 
the injection spectrum indices of CREs from pulsars are always 
in the range $\nu_{\psr} \in [1.0,2.4]$ \citep{Reynolds1988,Thompson1994,Fierro1995}. As a result, 
 more attention should be paid in future researches. This may indicate: 
(i) there is something wrong or inaccuracy with the classical pulsar CRE injection model; 
(ii) the CRE excess is not contributed primarily by pulsars.
Moreover, the cut-off is $R_c \simeq 692$~GV.
In the previous work~\citep{Niu2017_dampe} where  the 1.4 TeV peak excess was neglected,
we obtained that the spectral index of the injection is
$\nu_{\psr} \simeq 0.65$ and the cut-off is $R_c \simeq 650$~GV.  Thus, there exist about $+5\%$ and $-5\%$ deviations
for $\nu_{\psr}$ and $R_c$, respectively.

For the DM scenario, we obtain $\sigv \simeq 4.07 \times 10^{-23} \cm^{2} \s^{-1}$ and $\Mdm \simeq 1884 \GeV$. 
The value of $\sigv$ is about 3 orders larger than that of thermal DM \citep{Jungman1996}. To explain
this discrepancy, we will present a concrete model in the next section. Moreover, 
we have $\etae \simeq 0.465$, $\etamu \simeq 0.510 $, and $\etatau \simeq 0.025 $. So the DM annihilation into
$\tautaubar$ is highly suppressed, which provides some hints to construct an appropriate DM model.
In our previous work \citep{Niu2017_dampe} where  the 1.4 TeV peak excess was neglected,
we have $\sigv \simeq 1.48 \times 10^{-23} \cm^{2} \s^{-1}$, $\Mdm \simeq 1208 \GeV$, $\etae \simeq \etamu \simeq 0.5$,
while $\etatau$ is highly suppressed. Thus, we have similar results on branching fractions,
but different DM masses and annihilation cross sections.

\section{Model Building}
\label{sec4}
Because we have $\etae \sim 0.465$, $\etamu \sim 0.510 $, and $\etatau \sim 0.025 $,
the constraints from the Fermi-LAT observations of 
dwarf spheroidal galaxies~\cite{Ackermann:2011wa, GeringerSameth:2011iw, Tsai:2012cs, Ackermann:2015zua, Li:2015kag, Profumo:2017obk} 
can be avoided~\cite{Yuan2017_dampe}. 
To escape the constraints from the Planck observations of CMB anisotropies \cite{Ade:2015xua}, we employ
the Breit-Wigner mechanism \cite{Feldman:2008xs, Ibe:2008ye, Guo:2009aj, Bi:2009uj, Bi:2011qm, Hisano:2011dc, Bai:2017fav, Xiang:2017jou}.  
We consider the dark $U(1)_D$ model where the SM fermions and Higgs fields
are neutral under it. We introduce one SM singlet Higgs field $S$, one chiral fermionic dark matter particle
$\chi$, and three pairs of the vector-like particles (${\widehat{XE}}_i, {\widehat{XE}}_i^c)$, whose quantum numbers under
the $SU(3)_C\times SU(2)_L \times U(1)_Y \times U(1)_D$ are
\begin{eqnarray}
& S: (\mathbf{1}, \mathbf{1}, \mathbf{0}, \mathbf{2})~,~~~
\chi: (\mathbf{1}, \mathbf{1}, \mathbf{0}, \mathbf{-1}) \nonumber \\
& {\widehat{XE}}_i: (\mathbf{1}, \mathbf{1}, \mathbf{-1}, \mathbf{-2})~,~~~
{\widehat{XE}}^c_i: (\mathbf{1}, \mathbf{1}, \mathbf{1}, \mathbf{2}) ~.~\,
\end{eqnarray}
The relevant Lagrangian is 
\begin{eqnarray}
-{\cal L} &=& -m_S^2 |S|^2 + \frac{\lambda}{2} |S|^4 + \left(M^V_{ij} {\widehat{XE}}^c_i {\widehat{XE}}_j
\right. \nonumber \\  && \left. + y_{ij} S {\widehat E}_i^c {\widehat{XE}}_j +y S  \chi \chi + {\rm H.C.}\right) ~,~\,
\end{eqnarray}
where ${\widehat E}_i^c$ are the right-handed charged leptons.
For simplicity, we choose $M^V_{ij} = M^V_{i} \delta_{ij}$ and $y_{ij} = y_i \delta_{ij}$.
After $S$ acquires a Vacuum Expectation Value (VEV), the $U(1)_D$ gauge symmetry is broken
down to a $Z_2$ symmetry under which $\chi$ is odd. Thus, $\chi$ is a DM matter candidate.
For simplicity, we assume that the mass of $U(1)_D$ gauge boson is about twice of $\chi$ mass, 
{\it i.e.}, $M_{Z'} \simeq 2 m_{\chi}$, while the Higgs field $S$ and vector-like particles
are heavier than $M_{Z'}$.
Moreover,  ${\widehat E}_i^c$ and ${\widehat{XE}}_i^c$ will be mixed due to
the $M^V_{i} {\widehat{XE}}^c_i {\widehat{XE}}_i$ and $y_{i} S {\widehat E}_i^c {\widehat{XE}}_i$ terms, and we obtain 
the mass eigenstates $E_i^c$ and $XE_i^c$ by neglecting the tiny charged lepton masses
\begin{eqnarray}
\left(
\begin{array}{c}
E_i^c \\
XE_i^c
\end{array} \right)=
\left(
\begin{array}{cc}
\cos\theta_i & \sin\theta_i \\
-\sin\theta_i & \cos\theta_i
\end{array}
\right)
\left(
\begin{array}{c}
{\widehat E}_i^{c} \\
{\widehat{XE}}_i^{c \prime}
\end{array} \right)
~,~\,
\end{eqnarray}
where $\tan\theta_i = -y\langle S \rangle/M^V_i$.

Neglecting the charged lepton masses again, we obtain
\begin{eqnarray}
\sigma v = \sum_{i=1}^3\frac{g'^4\sin^2\theta_i}{6\pi}\frac{s-m_\chi^2}{(s-m_{Z'}^2)^2+(m_{Z'}\Gamma_{Z'})^2}~,~\,
\end{eqnarray}
where $m_\chi = y \langle S \rangle$, and $g'$ and $M_{Z'}$ are the gauge coupling and gauge boson mass
for $U(1)_D$ gauge symmetry.

For $m_{Z'}\simeq 2m_\chi$, $Z'$ decays dominantly into leptons, and the decay width is 
\begin{eqnarray}
\Gamma_{Z'}=\sum_{i=1}^3 \frac{g'^2\sin^2\theta_i}{6\pi} m_{Z'}~.~\,
\end{eqnarray}

To explain the DM best fit results, we choose
\begin{eqnarray}
&& g' \simeq 0.028,~ m_\chi \simeq 1884~\mbox{GeV},~
\frac{m_{Z'}-2m_\chi}{m_{Z'}} \simeq 3.0\times 10^{-6},~\nonumber \\
&& \sin\theta_e \simeq 0.21~,~~~\sin\theta_\mu \simeq 0.22~,~~~
\sin\theta_\tau \simeq 0.05~.~\,
\end{eqnarray}
And then we obtain $\langle\sigma v\rangle \simeq 4.07\times 10^{-23} \mbox{cm}^3\mbox{s}^{-1}$,  
and $\eta_e :\eta_\mu : \eta_\tau \simeq 0.465 : 0.510 : 0.025$. Of course,
there exists fine-tuning between $m_{Z'}$ and $m_\chi$ , which deserves further study.
For some solutions, see Ref.~\cite{Bai:2017fav}.

\section{Discussions and Conclusion}
\label{sec5}
First, we would like to point out that
if the numbers of events in the 1229.3~GeV bin and 1411.4~GeV bin are exchanged, 
we can also explain the DAMPE's data similarly. Of course,
the most important question is how to test our proposal that
there exists statistical fluctuations in the 1229.3~GeV bin, 1411.4~GeV bin, and 1620.5~GeV bin. 
For the data analyses, we suggest that one chooses different energy ranges to study the data again.
For example, we can shift the energy ranges by $\pm 50$~GeV and $\pm 100$~GeV 
for the high energy bins, and then study the corrsponding events and fluxes.
In the future, DAMPE will provide us more accurate spectrum data reaching up 
to $\sim 10 \TeV$, which can give us a unprecedented opportunity to study the origin and 
propagation of CREs. We predict that the CRE spectrum would be more continuous. In particular,
the peak excess in the 1411.4~GeV bin as well as the deficits in the 1229.3~GeV bin and 1620.5~GeV bin 
will all decrease! Moreover,
if the 1.4 TeV peak signal was proved to be correct, we do need a local source of high energy CREs. 
Other experiment is needed as a cross check if such signal arises from DM annihilation, 
for example, our recent work~\citep{Niu2017_DAV} proposed a novel scenario to probe 
the interaction between DM particles 
and electrons for the DM mass range $5 \GeV \lesssim \Mdm \lesssim 10 \TeV$.

In summary, with the simplicity and naturalness physics principle, we proposed that there exists
the  $-2\sigma$, $+2\sigma$, and $-1\sigma$ deviations due to statistical fluctuations for the 
1229.3~GeV bin, 1411.4~GeV bin,  and 1620.5~GeV bin of the DAMPE data. Interestingly, we showed that 
all the DAMPE data can be explained consistently via both the pulsar and 
dark matter interpretations, which have $\chi^{2} \simeq 17.2 $ and $\chi^{2}
\simeq 13.9$ (for all the 38 points in DAMPE electron/positron spectrum with 3 of them revised), respectively. 
These results are different from the previous analyses by neglecting the 1.4 TeV excess. 
 At the same time, we employed the newly released CALET CREs spectrum to do a similar global fitting, which cold also be fitted by continuous distributed pulsar and DM scenarios.
Moreover, we presented a $U(1)_D$ dark matter model with Breit-Wigner mechanism, which can provide
the proper dark matter annihilation cross section and escape the CMB constraint.
Furthermore, we suggested a few ways to test our proposal.

\begin{acknowledgements}
We would like to thank Xiao-Jun Bi and Yi-Zhong Fan for helpful discussions, and 
thank D. Maurin, et al. \cite{Maurin2014} for collecting the database and associated online tools  
for charged cosmic-ray measurements.
This research was supported in part by the Projects 11475238 and 11875062 (from National  
Science Foundation of China), and the Projects 11747601 (from Key Research Program of Frontier Sciences, Chinese Academy of Sciences). 
The calculation in this paper are supported by HPC Cluster of SKLTP/ITP-CAS.
\end{acknowledgements}



\begin{thebibliography}{120}%
\makeatletter
\providecommand \@ifxundefined [1]{%
 \@ifx{#1\undefined}
}%
\providecommand \@ifnum [1]{%
 \ifnum #1\expandafter \@firstoftwo
 \else \expandafter \@secondoftwo
 \fi
}%
\providecommand \@ifx [1]{%
 \ifx #1\expandafter \@firstoftwo
 \else \expandafter \@secondoftwo
 \fi
}%
\providecommand \natexlab [1]{#1}%
\providecommand \enquote  [1]{``#1''}%
\providecommand \bibnamefont  [1]{#1}%
\providecommand \bibfnamefont [1]{#1}%
\providecommand \citenamefont [1]{#1}%
\providecommand \href@noop [0]{\@secondoftwo}%
\providecommand \href [0]{\begingroup \@sanitize@url \@href}%
\providecommand \@href[1]{\@@startlink{#1}\@@href}%
\providecommand \@@href[1]{\endgroup#1\@@endlink}%
\providecommand \@sanitize@url [0]{\catcode `\\12\catcode `\$12\catcode
  `\&12\catcode `\#12\catcode `\^12\catcode `\_12\catcode `\%12\relax}%
\providecommand \@@startlink[1]{}%
\providecommand \@@endlink[0]{}%
\providecommand \url  [0]{\begingroup\@sanitize@url \@url }%
\providecommand \@url [1]{\endgroup\@href {#1}{\urlprefix }}%
\providecommand \urlprefix  [0]{URL }%
\providecommand \Eprint [0]{\href }%
\providecommand \doibase [0]{http://dx.doi.org/}%
\providecommand \selectlanguage [0]{\@gobble}%
\providecommand \bibinfo  [0]{\@secondoftwo}%
\providecommand \bibfield  [0]{\@secondoftwo}%
\providecommand \translation [1]{[#1]}%
\providecommand \BibitemOpen [0]{}%
\providecommand \bibitemStop [0]{}%
\providecommand \bibitemNoStop [0]{.\EOS\space}%
\providecommand \EOS [0]{\spacefactor3000\relax}%
\providecommand \BibitemShut  [1]{\csname bibitem#1\endcsname}%
\let\auto@bib@innerbib\@empty
\bibitem [{\citenamefont {{Shen}}(1970)}]{Shen1970}%
  \BibitemOpen
  \bibfield  {author} {\bibinfo {author} {\bibfnamefont {C.~S.}\ \bibnamefont
  {{Shen}}},\ }\href {\doibase 10.1086/180650} {\bibfield  {journal} {\bibinfo
  {journal} {\apjl}\ }\textbf {\bibinfo {volume} {162}},\ \bibinfo {pages}
  {L181} (\bibinfo {year} {1970})}\BibitemShut {NoStop}%
\bibitem [{\citenamefont {{Harding}}\ and\ \citenamefont
  {{Ramaty}}(1987)}]{Harding1987}%
  \BibitemOpen
  \bibfield  {author} {\bibinfo {author} {\bibfnamefont {A.~K.}\ \bibnamefont
  {{Harding}}}\ and\ \bibinfo {author} {\bibfnamefont {R.}~\bibnamefont
  {{Ramaty}}},\ }\href@noop {} {\bibfield  {journal} {\bibinfo  {journal}
  {International Cosmic Ray Conference}\ }\textbf {\bibinfo {volume} {2}},\
  \bibinfo {pages} {92} (\bibinfo {year} {1987})}\BibitemShut {NoStop}%
\bibitem [{\citenamefont {{Aharonian}}\ \emph {et~al.}(1995)\citenamefont
  {{Aharonian}}, \citenamefont {{Atoyan}},\ and\ \citenamefont
  {{Voelk}}}]{Aharonian1995}%
  \BibitemOpen
  \bibfield  {author} {\bibinfo {author} {\bibfnamefont {F.~A.}\ \bibnamefont
  {{Aharonian}}}, \bibinfo {author} {\bibfnamefont {A.~M.}\ \bibnamefont
  {{Atoyan}}}, \ and\ \bibinfo {author} {\bibfnamefont {H.~J.}\ \bibnamefont
  {{Voelk}}},\ }\href@noop {} {\bibfield  {journal} {\bibinfo  {journal}
  {\aap}\ }\textbf {\bibinfo {volume} {294}},\ \bibinfo {pages} {L41} (\bibinfo
  {year} {1995})}\BibitemShut {NoStop}%
\bibitem [{\citenamefont {{Chi}}\ \emph {et~al.}(1996)\citenamefont {{Chi}},
  \citenamefont {{Cheng}},\ and\ \citenamefont {{Young}}}]{Chi1996}%
  \BibitemOpen
  \bibfield  {author} {\bibinfo {author} {\bibfnamefont {X.}~\bibnamefont
  {{Chi}}}, \bibinfo {author} {\bibfnamefont {K.~S.}\ \bibnamefont {{Cheng}}},
  \ and\ \bibinfo {author} {\bibfnamefont {E.~C.~M.}\ \bibnamefont {{Young}}},\
  }\href {\doibase 10.1086/309943} {\bibfield  {journal} {\bibinfo  {journal}
  {\apjl}\ }\textbf {\bibinfo {volume} {459}},\ \bibinfo {pages} {L83}
  (\bibinfo {year} {1996})}\BibitemShut {NoStop}%
\bibitem [{\citenamefont {{Zhang}}\ and\ \citenamefont
  {{Cheng}}(2001)}]{Zhang2001}%
  \BibitemOpen
  \bibfield  {author} {\bibinfo {author} {\bibfnamefont {L.}~\bibnamefont
  {{Zhang}}}\ and\ \bibinfo {author} {\bibfnamefont {K.~S.}\ \bibnamefont
  {{Cheng}}},\ }\href {\doibase 10.1051/0004-6361:20010021} {\bibfield
  {journal} {\bibinfo  {journal} {\aap}\ }\textbf {\bibinfo {volume} {368}},\
  \bibinfo {pages} {1063} (\bibinfo {year} {2001})}\BibitemShut {NoStop}%
\bibitem [{\citenamefont {{Bergstr{\"o}m}}(2000)}]{Bergstrom2000}%
  \BibitemOpen
  \bibfield  {author} {\bibinfo {author} {\bibfnamefont {L.}~\bibnamefont
  {{Bergstr{\"o}m}}},\ }\href {\doibase 10.1088/0034-4885/63/5/2r3} {\bibfield
  {journal} {\bibinfo  {journal} {Reports on Progress in Physics}\ }\textbf
  {\bibinfo {volume} {63}},\ \bibinfo {pages} {793} (\bibinfo {year} {2000})},\
  \Eprint {http://arxiv.org/abs/hep-ph/0002126} {hep-ph/0002126} \BibitemShut
  {NoStop}%
\bibitem [{\citenamefont {{Bertone}}\ \emph {et~al.}(2005)\citenamefont
  {{Bertone}}, \citenamefont {{Hooper}},\ and\ \citenamefont
  {{Silk}}}]{Bertone2005}%
  \BibitemOpen
  \bibfield  {author} {\bibinfo {author} {\bibfnamefont {G.}~\bibnamefont
  {{Bertone}}}, \bibinfo {author} {\bibfnamefont {D.}~\bibnamefont {{Hooper}}},
  \ and\ \bibinfo {author} {\bibfnamefont {J.}~\bibnamefont {{Silk}}},\ }\href
  {\doibase 10.1016/j.physrep.2004.08.031} {\bibfield  {journal} {\bibinfo
  {journal} {\physrep}\ }\textbf {\bibinfo {volume} {405}},\ \bibinfo {pages}
  {279} (\bibinfo {year} {2005})},\ \Eprint
  {http://arxiv.org/abs/hep-ph/0404175} {hep-ph/0404175} \BibitemShut {NoStop}%
\bibitem [{\citenamefont {{Bergstr{\"o}m}}\ \emph {et~al.}(2009)\citenamefont
  {{Bergstr{\"o}m}}, \citenamefont {{Edsj{\"o}}},\ and\ \citenamefont
  {{Zaharijas}}}]{Bergstrom2009}%
  \BibitemOpen
  \bibfield  {author} {\bibinfo {author} {\bibfnamefont {L.}~\bibnamefont
  {{Bergstr{\"o}m}}}, \bibinfo {author} {\bibfnamefont {J.}~\bibnamefont
  {{Edsj{\"o}}}}, \ and\ \bibinfo {author} {\bibfnamefont {G.}~\bibnamefont
  {{Zaharijas}}},\ }\href {\doibase 10.1103/PhysRevLett.103.031103} {\bibfield
  {journal} {\bibinfo  {journal} {Physical Review Letters}\ }\textbf {\bibinfo
  {volume} {103}},\ \bibinfo {eid} {031103} (\bibinfo {year} {2009})},\ \Eprint
  {http://arxiv.org/abs/0905.0333} {arXiv:0905.0333 [astro-ph.HE]} \BibitemShut
  {NoStop}%
\bibitem [{\citenamefont {{Aharonian}}\ \emph {et~al.}(2008)\citenamefont
  {{Aharonian}}, \citenamefont {{Akhperjanian}}, \citenamefont {{Barres de
  Almeida}}, \citenamefont {{Bazer-Bachi}}, \citenamefont {{Becherini}},
  \citenamefont {{Behera}}, \citenamefont {{Benbow}}, \citenamefont
  {{Bernl{\"o}hr}}, \citenamefont {{Boisson}}, \citenamefont {{Bochow}},\ and\
  \citenamefont {et~al.}}]{Aharonian2008}%
  \BibitemOpen
  \bibfield  {author} {\bibinfo {author} {\bibfnamefont {F.}~\bibnamefont
  {{Aharonian}}}, \bibinfo {author} {\bibfnamefont {A.~G.}\ \bibnamefont
  {{Akhperjanian}}}, \bibinfo {author} {\bibfnamefont {U.}~\bibnamefont
  {{Barres de Almeida}}}, \bibinfo {author} {\bibfnamefont {A.~R.}\
  \bibnamefont {{Bazer-Bachi}}}, \bibinfo {author} {\bibfnamefont
  {Y.}~\bibnamefont {{Becherini}}}, \bibinfo {author} {\bibfnamefont
  {B.}~\bibnamefont {{Behera}}}, \bibinfo {author} {\bibfnamefont
  {W.}~\bibnamefont {{Benbow}}}, \bibinfo {author} {\bibfnamefont
  {K.}~\bibnamefont {{Bernl{\"o}hr}}}, \bibinfo {author} {\bibfnamefont
  {C.}~\bibnamefont {{Boisson}}}, \bibinfo {author} {\bibfnamefont
  {A.}~\bibnamefont {{Bochow}}}, \ and\ \bibinfo {author} {\bibnamefont
  {et~al.}},\ }\href {\doibase 10.1103/PhysRevLett.101.261104} {\bibfield
  {journal} {\bibinfo  {journal} {Physical Review Letters}\ }\textbf {\bibinfo
  {volume} {101}},\ \bibinfo {eid} {261104} (\bibinfo {year} {2008})},\ \Eprint
  {http://arxiv.org/abs/0811.3894} {arXiv:0811.3894} \BibitemShut {NoStop}%
\bibitem [{\citenamefont {{HESS collaboration}}(2009)}]{Aharonian2009}%
  \BibitemOpen
  \bibfield  {author} {\bibinfo {author} {\bibnamefont {{HESS
  collaboration}}},\ }\href {\doibase 10.1051/0004-6361/200913323} {\bibfield
  {journal} {\bibinfo  {journal} {\aap}\ }\textbf {\bibinfo {volume} {508}},\
  \bibinfo {pages} {561} (\bibinfo {year} {2009})},\ \Eprint
  {http://arxiv.org/abs/0905.0105} {arXiv:0905.0105 [astro-ph.HE]} \BibitemShut
  {NoStop}%
\bibitem [{\citenamefont {{Staszak}}\ and\ \citenamefont {{for the VERITAS
  Collaboration}}(2015)}]{Staszak2015}%
  \BibitemOpen
  \bibfield  {author} {\bibinfo {author} {\bibfnamefont {D.}~\bibnamefont
  {{Staszak}}}\ and\ \bibinfo {author} {\bibnamefont {{for the VERITAS
  Collaboration}}},\ }\href@noop {} {\bibfield  {journal} {\bibinfo  {journal}
  {ArXiv e-prints}\ } (\bibinfo {year} {2015})},\ \Eprint
  {http://arxiv.org/abs/1508.06597} {arXiv:1508.06597 [astro-ph.HE]}
  \BibitemShut {NoStop}%
\bibitem [{\citenamefont {{Holder}}(2017)}]{Holder2017}%
  \BibitemOpen
  \bibfield  {author} {\bibinfo {author} {\bibfnamefont {J.}~\bibnamefont
  {{Holder}}},\ }in\ \href {\doibase 10.1063/1.4968898} {\emph {\bibinfo
  {booktitle} {6th International Symposium on High Energy Gamma-Ray
  Astronomy}}},\ \bibinfo {series} {American Institute of Physics Conference
  Series}, Vol.\ \bibinfo {volume} {1792}\ (\bibinfo {year} {2017})\ p.\
  \bibinfo {pages} {020013},\ \Eprint {http://arxiv.org/abs/1609.02881}
  {arXiv:1609.02881 [astro-ph.HE]} \BibitemShut {NoStop}%
\bibitem [{\citenamefont {Abdollahi}\ \emph {et~al.}(2017)\citenamefont
  {Abdollahi} \emph {et~al.}}]{Abdollahi2017}%
  \BibitemOpen
  \bibfield  {author} {\bibinfo {author} {\bibfnamefont {S.}~\bibnamefont
  {Abdollahi}} \emph {et~al.} (\bibinfo {collaboration} {Fermi-LAT}),\ }\href
  {\doibase 10.1103/PhysRevD.95.082007} {\bibfield  {journal} {\bibinfo
  {journal} {Phys. Rev.}\ }\textbf {\bibinfo {volume} {D95}},\ \bibinfo {pages}
  {082007} (\bibinfo {year} {2017})},\ \Eprint
  {http://arxiv.org/abs/1704.07195} {arXiv:1704.07195 [astro-ph.HE]}
  \BibitemShut {NoStop}%
\bibitem [{\citenamefont {{Meehan}}\ \emph {et~al.}(2017)\citenamefont
  {{Meehan}}, \citenamefont {{Vandenbroucke}},\ and\ \citenamefont {{for the
  Fermi-LAT Collaboration}}}]{Meehan2017}%
  \BibitemOpen
  \bibfield  {author} {\bibinfo {author} {\bibfnamefont {M.}~\bibnamefont
  {{Meehan}}}, \bibinfo {author} {\bibfnamefont {J.}~\bibnamefont
  {{Vandenbroucke}}}, \ and\ \bibinfo {author} {\bibnamefont {{for the
  Fermi-LAT Collaboration}}},\ }\href@noop {} {\bibfield  {journal} {\bibinfo
  {journal} {ArXiv e-prints}\ } (\bibinfo {year} {2017})},\ \Eprint
  {http://arxiv.org/abs/1708.07796} {arXiv:1708.07796 [astro-ph.HE]}
  \BibitemShut {NoStop}%
\bibitem [{\citenamefont {{AMS
  collaboration}}(2014{\natexlab{a}})}]{AMS02_lepton_sum}%
  \BibitemOpen
  \bibfield  {author} {\bibinfo {author} {\bibnamefont {{AMS collaboration}}},\
  }\href {\doibase 10.1103/PhysRevLett.113.221102} {\bibfield  {journal}
  {\bibinfo  {journal} {Physical Review Letters}\ }\textbf {\bibinfo {volume}
  {113}},\ \bibinfo {eid} {221102} (\bibinfo {year}
  {2014}{\natexlab{a}})}\BibitemShut {NoStop}%
\bibitem [{\citenamefont {{AMS
  collaboration}}(2014{\natexlab{b}})}]{AMS02_lepton}%
  \BibitemOpen
  \bibfield  {author} {\bibinfo {author} {\bibnamefont {{AMS collaboration}}},\
  }\href {\doibase 10.1103/PhysRevLett.113.121102} {\bibfield  {journal}
  {\bibinfo  {journal} {Physical Review Letters}\ }\textbf {\bibinfo {volume}
  {113}},\ \bibinfo {eid} {121102} (\bibinfo {year}
  {2014}{\natexlab{b}})}\BibitemShut {NoStop}%
\bibitem [{\citenamefont {{AMS
  collaboration}}(2013{\natexlab{a}})}]{AMS02_fraction01}%
  \BibitemOpen
  \bibfield  {author} {\bibinfo {author} {\bibnamefont {{AMS collaboration}}},\
  }\href {\doibase 10.1103/PhysRevLett.110.141102} {\bibfield  {journal}
  {\bibinfo  {journal} {Physical Review Letters}\ }\textbf {\bibinfo {volume}
  {110}},\ \bibinfo {eid} {141102} (\bibinfo {year}
  {2013}{\natexlab{a}})}\BibitemShut {NoStop}%
\bibitem [{\citenamefont {{AMS
  collaboration}}(2014{\natexlab{c}})}]{AMS02_fraction02}%
  \BibitemOpen
  \bibfield  {author} {\bibinfo {author} {\bibnamefont {{AMS collaboration}}},\
  }\href {\doibase 10.1103/PhysRevLett.113.121101} {\bibfield  {journal}
  {\bibinfo  {journal} {Physical Review Letters}\ }\textbf {\bibinfo {volume}
  {113}},\ \bibinfo {eid} {121101} (\bibinfo {year}
  {2014}{\natexlab{c}})}\BibitemShut {NoStop}%
\bibitem [{\citenamefont {collaboration}(2017{\natexlab{a}})}]{CALET2017}%
  \BibitemOpen
  \bibfield  {author} {\bibinfo {author} {\bibfnamefont {C.}~\bibnamefont
  {collaboration}} (\bibinfo {collaboration} {CALET Collaboration}),\ }\href
  {\doibase 10.1103/PhysRevLett.119.181101} {\bibfield  {journal} {\bibinfo
  {journal} {Phys. Rev. Lett.}\ }\textbf {\bibinfo {volume} {119}},\ \bibinfo
  {pages} {181101} (\bibinfo {year} {2017}{\natexlab{a}})}\BibitemShut
  {NoStop}%
\bibitem [{\citenamefont {Adriani}\ \emph {et~al.}(2018)\citenamefont
  {Adriani}, \citenamefont {Akaike}, \citenamefont {Asano}, \citenamefont
  {Asaoka}, \citenamefont {Bagliesi}, \citenamefont {Berti}, \citenamefont
  {Bigongiari}, \citenamefont {Binns}, \citenamefont {Bonechi}, \citenamefont
  {Bongi}, \citenamefont {Brogi}, \citenamefont {Buckley}, \citenamefont
  {Cannady}, \citenamefont {Castellini}, \citenamefont {Checchia},
  \citenamefont {Cherry}, \citenamefont {Collazuol}, \citenamefont {Di~Felice},
  \citenamefont {Ebisawa}, \citenamefont {Fuke}, \citenamefont {Guzik},
  \citenamefont {Hams}, \citenamefont {Hareyama}, \citenamefont {Hasebe},
  \citenamefont {Hibino}, \citenamefont {Ichimura}, \citenamefont {Ioka},
  \citenamefont {Ishizaki}, \citenamefont {Israel}, \citenamefont {Kasahara},
  \citenamefont {Kataoka}, \citenamefont {Kataoka}, \citenamefont {Katayose},
  \citenamefont {Kato}, \citenamefont {Kawanaka}, \citenamefont {Kawakubo},
  \citenamefont {Kohri}, \citenamefont {Krawczynski}, \citenamefont
  {Krizmanic}, \citenamefont {Lomtadze}, \citenamefont {Maestro}, \citenamefont
  {Marrocchesi}, \citenamefont {Messineo}, \citenamefont {Mitchell},
  \citenamefont {Miyake}, \citenamefont {Moiseev}, \citenamefont {Mori},
  \citenamefont {Mori}, \citenamefont {Mori}, \citenamefont {Motz},
  \citenamefont {Munakata}, \citenamefont {Murakami}, \citenamefont {Nakahira},
  \citenamefont {Nishimura}, \citenamefont {de~Nolfo}, \citenamefont {Okuno},
  \citenamefont {Ormes}, \citenamefont {Ozawa}, \citenamefont {Pacini},
  \citenamefont {Palma}, \citenamefont {Papini}, \citenamefont {Penacchioni},
  \citenamefont {Rauch}, \citenamefont {Ricciarini}, \citenamefont {Sakai},
  \citenamefont {Sakamoto}, \citenamefont {Sasaki}, \citenamefont {Shimizu},
  \citenamefont {Shiomi}, \citenamefont {Sparvoli}, \citenamefont
  {Spillantini}, \citenamefont {Stolzi}, \citenamefont {Suh}, \citenamefont
  {Sulaj}, \citenamefont {Takahashi}, \citenamefont {Takayanagi}, \citenamefont
  {Takita}, \citenamefont {Tamura}, \citenamefont {Tateyama}, \citenamefont
  {Terasawa}, \citenamefont {Tomida}, \citenamefont {Torii}, \citenamefont
  {Tsunesada}, \citenamefont {Uchihori}, \citenamefont {Ueno}, \citenamefont
  {Vannuccini}, \citenamefont {Wefel}, \citenamefont {Yamaoka}, \citenamefont
  {Yanagita}, \citenamefont {Yoshida},\ and\ \citenamefont
  {Yoshida}}]{CALET2018}%
  \BibitemOpen
  \bibfield  {author} {\bibinfo {author} {\bibfnamefont {O.}~\bibnamefont
  {Adriani}}, \bibinfo {author} {\bibfnamefont {Y.}~\bibnamefont {Akaike}},
  \bibinfo {author} {\bibfnamefont {K.}~\bibnamefont {Asano}}, \bibinfo
  {author} {\bibfnamefont {Y.}~\bibnamefont {Asaoka}}, \bibinfo {author}
  {\bibfnamefont {M.~G.}\ \bibnamefont {Bagliesi}}, \bibinfo {author}
  {\bibfnamefont {E.}~\bibnamefont {Berti}}, \bibinfo {author} {\bibfnamefont
  {G.}~\bibnamefont {Bigongiari}}, \bibinfo {author} {\bibfnamefont {W.~R.}\
  \bibnamefont {Binns}}, \bibinfo {author} {\bibfnamefont {S.}~\bibnamefont
  {Bonechi}}, \bibinfo {author} {\bibfnamefont {M.}~\bibnamefont {Bongi}},
  \bibinfo {author} {\bibfnamefont {P.}~\bibnamefont {Brogi}}, \bibinfo
  {author} {\bibfnamefont {J.~H.}\ \bibnamefont {Buckley}}, \bibinfo {author}
  {\bibfnamefont {N.}~\bibnamefont {Cannady}}, \bibinfo {author} {\bibfnamefont
  {G.}~\bibnamefont {Castellini}}, \bibinfo {author} {\bibfnamefont
  {C.}~\bibnamefont {Checchia}}, \bibinfo {author} {\bibfnamefont {M.~L.}\
  \bibnamefont {Cherry}}, \bibinfo {author} {\bibfnamefont {G.}~\bibnamefont
  {Collazuol}}, \bibinfo {author} {\bibfnamefont {V.}~\bibnamefont
  {Di~Felice}}, \bibinfo {author} {\bibfnamefont {K.}~\bibnamefont {Ebisawa}},
  \bibinfo {author} {\bibfnamefont {H.}~\bibnamefont {Fuke}}, \bibinfo {author}
  {\bibfnamefont {T.~G.}\ \bibnamefont {Guzik}}, \bibinfo {author}
  {\bibfnamefont {T.}~\bibnamefont {Hams}}, \bibinfo {author} {\bibfnamefont
  {M.}~\bibnamefont {Hareyama}}, \bibinfo {author} {\bibfnamefont
  {N.}~\bibnamefont {Hasebe}}, \bibinfo {author} {\bibfnamefont
  {K.}~\bibnamefont {Hibino}}, \bibinfo {author} {\bibfnamefont
  {M.}~\bibnamefont {Ichimura}}, \bibinfo {author} {\bibfnamefont
  {K.}~\bibnamefont {Ioka}}, \bibinfo {author} {\bibfnamefont {W.}~\bibnamefont
  {Ishizaki}}, \bibinfo {author} {\bibfnamefont {M.~H.}\ \bibnamefont
  {Israel}}, \bibinfo {author} {\bibfnamefont {K.}~\bibnamefont {Kasahara}},
  \bibinfo {author} {\bibfnamefont {J.}~\bibnamefont {Kataoka}}, \bibinfo
  {author} {\bibfnamefont {R.}~\bibnamefont {Kataoka}}, \bibinfo {author}
  {\bibfnamefont {Y.}~\bibnamefont {Katayose}}, \bibinfo {author}
  {\bibfnamefont {C.}~\bibnamefont {Kato}}, \bibinfo {author} {\bibfnamefont
  {N.}~\bibnamefont {Kawanaka}}, \bibinfo {author} {\bibfnamefont
  {Y.}~\bibnamefont {Kawakubo}}, \bibinfo {author} {\bibfnamefont
  {K.}~\bibnamefont {Kohri}}, \bibinfo {author} {\bibfnamefont {H.~S.}\
  \bibnamefont {Krawczynski}}, \bibinfo {author} {\bibfnamefont {J.~F.}\
  \bibnamefont {Krizmanic}}, \bibinfo {author} {\bibfnamefont {T.}~\bibnamefont
  {Lomtadze}}, \bibinfo {author} {\bibfnamefont {P.}~\bibnamefont {Maestro}},
  \bibinfo {author} {\bibfnamefont {P.~S.}\ \bibnamefont {Marrocchesi}},
  \bibinfo {author} {\bibfnamefont {A.~M.}\ \bibnamefont {Messineo}}, \bibinfo
  {author} {\bibfnamefont {J.~W.}\ \bibnamefont {Mitchell}}, \bibinfo {author}
  {\bibfnamefont {S.}~\bibnamefont {Miyake}}, \bibinfo {author} {\bibfnamefont
  {A.~A.}\ \bibnamefont {Moiseev}}, \bibinfo {author} {\bibfnamefont
  {K.}~\bibnamefont {Mori}}, \bibinfo {author} {\bibfnamefont {M.}~\bibnamefont
  {Mori}}, \bibinfo {author} {\bibfnamefont {N.}~\bibnamefont {Mori}}, \bibinfo
  {author} {\bibfnamefont {H.~M.}\ \bibnamefont {Motz}}, \bibinfo {author}
  {\bibfnamefont {K.}~\bibnamefont {Munakata}}, \bibinfo {author}
  {\bibfnamefont {H.}~\bibnamefont {Murakami}}, \bibinfo {author}
  {\bibfnamefont {S.}~\bibnamefont {Nakahira}}, \bibinfo {author}
  {\bibfnamefont {J.}~\bibnamefont {Nishimura}}, \bibinfo {author}
  {\bibfnamefont {G.~A.}\ \bibnamefont {de~Nolfo}}, \bibinfo {author}
  {\bibfnamefont {S.}~\bibnamefont {Okuno}}, \bibinfo {author} {\bibfnamefont
  {J.~F.}\ \bibnamefont {Ormes}}, \bibinfo {author} {\bibfnamefont
  {S.}~\bibnamefont {Ozawa}}, \bibinfo {author} {\bibfnamefont
  {L.}~\bibnamefont {Pacini}}, \bibinfo {author} {\bibfnamefont
  {F.}~\bibnamefont {Palma}}, \bibinfo {author} {\bibfnamefont
  {P.}~\bibnamefont {Papini}}, \bibinfo {author} {\bibfnamefont {A.~V.}\
  \bibnamefont {Penacchioni}}, \bibinfo {author} {\bibfnamefont {B.~F.}\
  \bibnamefont {Rauch}}, \bibinfo {author} {\bibfnamefont {S.~B.}\ \bibnamefont
  {Ricciarini}}, \bibinfo {author} {\bibfnamefont {K.}~\bibnamefont {Sakai}},
  \bibinfo {author} {\bibfnamefont {T.}~\bibnamefont {Sakamoto}}, \bibinfo
  {author} {\bibfnamefont {M.}~\bibnamefont {Sasaki}}, \bibinfo {author}
  {\bibfnamefont {Y.}~\bibnamefont {Shimizu}}, \bibinfo {author} {\bibfnamefont
  {A.}~\bibnamefont {Shiomi}}, \bibinfo {author} {\bibfnamefont
  {R.}~\bibnamefont {Sparvoli}}, \bibinfo {author} {\bibfnamefont
  {P.}~\bibnamefont {Spillantini}}, \bibinfo {author} {\bibfnamefont
  {F.}~\bibnamefont {Stolzi}}, \bibinfo {author} {\bibfnamefont {J.~E.}\
  \bibnamefont {Suh}}, \bibinfo {author} {\bibfnamefont {A.}~\bibnamefont
  {Sulaj}}, \bibinfo {author} {\bibfnamefont {I.}~\bibnamefont {Takahashi}},
  \bibinfo {author} {\bibfnamefont {M.}~\bibnamefont {Takayanagi}}, \bibinfo
  {author} {\bibfnamefont {M.}~\bibnamefont {Takita}}, \bibinfo {author}
  {\bibfnamefont {T.}~\bibnamefont {Tamura}}, \bibinfo {author} {\bibfnamefont
  {N.}~\bibnamefont {Tateyama}}, \bibinfo {author} {\bibfnamefont
  {T.}~\bibnamefont {Terasawa}}, \bibinfo {author} {\bibfnamefont
  {H.}~\bibnamefont {Tomida}}, \bibinfo {author} {\bibfnamefont
  {S.}~\bibnamefont {Torii}}, \bibinfo {author} {\bibfnamefont
  {Y.}~\bibnamefont {Tsunesada}}, \bibinfo {author} {\bibfnamefont
  {Y.}~\bibnamefont {Uchihori}}, \bibinfo {author} {\bibfnamefont
  {S.}~\bibnamefont {Ueno}}, \bibinfo {author} {\bibfnamefont {E.}~\bibnamefont
  {Vannuccini}}, \bibinfo {author} {\bibfnamefont {J.~P.}\ \bibnamefont
  {Wefel}}, \bibinfo {author} {\bibfnamefont {K.}~\bibnamefont {Yamaoka}},
  \bibinfo {author} {\bibfnamefont {S.}~\bibnamefont {Yanagita}}, \bibinfo
  {author} {\bibfnamefont {A.}~\bibnamefont {Yoshida}}, \ and\ \bibinfo
  {author} {\bibfnamefont {K.}~\bibnamefont {Yoshida}} (\bibinfo
  {collaboration} {CALET Collaboration}),\ }\href {\doibase
  10.1103/PhysRevLett.120.261102} {\bibfield  {journal} {\bibinfo  {journal}
  {Phys. Rev. Lett.}\ }\textbf {\bibinfo {volume} {120}},\ \bibinfo {pages}
  {261102} (\bibinfo {year} {2018})}\BibitemShut {NoStop}%
\bibitem [{\citenamefont {{Chang}}\ \emph {et~al.}(2008)\citenamefont
  {{Chang}}, \citenamefont {{Adams}}, \citenamefont {{Ahn}}, \citenamefont
  {{Bashindzhagyan}}, \citenamefont {{Christl}}, \citenamefont {{Ganel}},
  \citenamefont {{Guzik}}, \citenamefont {{Isbert}}, \citenamefont {{Kim}},
  \citenamefont {{Kuznetsov}}, \citenamefont {{Panasyuk}}, \citenamefont
  {{Panov}}, \citenamefont {{Schmidt}}, \citenamefont {{Seo}}, \citenamefont
  {{Sokolskaya}}, \citenamefont {{Watts}}, \citenamefont {{Wefel}},
  \citenamefont {{Wu}},\ and\ \citenamefont {{Zatsepin}}}]{Chang2008}%
  \BibitemOpen
  \bibfield  {author} {\bibinfo {author} {\bibfnamefont {J.}~\bibnamefont
  {{Chang}}}, \bibinfo {author} {\bibfnamefont {J.~H.}\ \bibnamefont
  {{Adams}}}, \bibinfo {author} {\bibfnamefont {H.~S.}\ \bibnamefont {{Ahn}}},
  \bibinfo {author} {\bibfnamefont {G.~L.}\ \bibnamefont {{Bashindzhagyan}}},
  \bibinfo {author} {\bibfnamefont {M.}~\bibnamefont {{Christl}}}, \bibinfo
  {author} {\bibfnamefont {O.}~\bibnamefont {{Ganel}}}, \bibinfo {author}
  {\bibfnamefont {T.~G.}\ \bibnamefont {{Guzik}}}, \bibinfo {author}
  {\bibfnamefont {J.}~\bibnamefont {{Isbert}}}, \bibinfo {author}
  {\bibfnamefont {K.~C.}\ \bibnamefont {{Kim}}}, \bibinfo {author}
  {\bibfnamefont {E.~N.}\ \bibnamefont {{Kuznetsov}}}, \bibinfo {author}
  {\bibfnamefont {M.~I.}\ \bibnamefont {{Panasyuk}}}, \bibinfo {author}
  {\bibfnamefont {A.~D.}\ \bibnamefont {{Panov}}}, \bibinfo {author}
  {\bibfnamefont {W.~K.~H.}\ \bibnamefont {{Schmidt}}}, \bibinfo {author}
  {\bibfnamefont {E.~S.}\ \bibnamefont {{Seo}}}, \bibinfo {author}
  {\bibfnamefont {N.~V.}\ \bibnamefont {{Sokolskaya}}}, \bibinfo {author}
  {\bibfnamefont {J.~W.}\ \bibnamefont {{Watts}}}, \bibinfo {author}
  {\bibfnamefont {J.~P.}\ \bibnamefont {{Wefel}}}, \bibinfo {author}
  {\bibfnamefont {J.}~\bibnamefont {{Wu}}}, \ and\ \bibinfo {author}
  {\bibfnamefont {V.~I.}\ \bibnamefont {{Zatsepin}}},\ }\href {\doibase
  10.1038/nature07477} {\bibfield  {journal} {\bibinfo  {journal} {\nat}\
  }\textbf {\bibinfo {volume} {456}},\ \bibinfo {pages} {362} (\bibinfo {year}
  {2008})}\BibitemShut {NoStop}%
\bibitem [{\citenamefont {{Fermi-LAT collaboration}}(2009)}]{Abdo2009}%
  \BibitemOpen
  \bibfield  {author} {\bibinfo {author} {\bibnamefont {{Fermi-LAT
  collaboration}}},\ }\href {\doibase 10.1103/PhysRevLett.102.181101}
  {\bibfield  {journal} {\bibinfo  {journal} {Physical Review Letters}\
  }\textbf {\bibinfo {volume} {102}},\ \bibinfo {eid} {181101} (\bibinfo {year}
  {2009})},\ \Eprint {http://arxiv.org/abs/0905.0025} {arXiv:0905.0025
  [astro-ph.HE]} \BibitemShut {NoStop}%
\bibitem [{\citenamefont {Aguilar}\ \emph {et~al.}(2014)\citenamefont {Aguilar}
  \emph {et~al.}}]{Aguilar2014}%
  \BibitemOpen
  \bibfield  {author} {\bibinfo {author} {\bibfnamefont {M.}~\bibnamefont
  {Aguilar}} \emph {et~al.} (\bibinfo {collaboration} {AMS}),\ }\href {\doibase
  10.1103/PhysRevLett.113.121102} {\bibfield  {journal} {\bibinfo  {journal}
  {Phys. Rev. Lett.}\ }\textbf {\bibinfo {volume} {113}},\ \bibinfo {pages}
  {121102} (\bibinfo {year} {2014})}\BibitemShut {NoStop}%
\bibitem [{\citenamefont {{Adriani}}\ \emph {et~al.}(2009)\citenamefont
  {{Adriani}}, \citenamefont {{Barbarino}}, \citenamefont {{Bazilevskaya}},
  \citenamefont {{Bellotti}}, \citenamefont {{Boezio}}, \citenamefont
  {{Bogomolov}}, \citenamefont {{Bonechi}}, \citenamefont {{Bongi}},
  \citenamefont {{Bonvicini}}, \citenamefont {{Bottai}},\ and\ \citenamefont
  {et~al.}}]{Adriani2009}%
  \BibitemOpen
  \bibfield  {author} {\bibinfo {author} {\bibfnamefont {O.}~\bibnamefont
  {{Adriani}}}, \bibinfo {author} {\bibfnamefont {G.~C.}\ \bibnamefont
  {{Barbarino}}}, \bibinfo {author} {\bibfnamefont {G.~A.}\ \bibnamefont
  {{Bazilevskaya}}}, \bibinfo {author} {\bibfnamefont {R.}~\bibnamefont
  {{Bellotti}}}, \bibinfo {author} {\bibfnamefont {M.}~\bibnamefont
  {{Boezio}}}, \bibinfo {author} {\bibfnamefont {E.~A.}\ \bibnamefont
  {{Bogomolov}}}, \bibinfo {author} {\bibfnamefont {L.}~\bibnamefont
  {{Bonechi}}}, \bibinfo {author} {\bibfnamefont {M.}~\bibnamefont {{Bongi}}},
  \bibinfo {author} {\bibfnamefont {V.}~\bibnamefont {{Bonvicini}}}, \bibinfo
  {author} {\bibfnamefont {S.}~\bibnamefont {{Bottai}}}, \ and\ \bibinfo
  {author} {\bibnamefont {et~al.}},\ }\href {\doibase 10.1038/nature07942}
  {\bibfield  {journal} {\bibinfo  {journal} {\nat}\ }\textbf {\bibinfo
  {volume} {458}},\ \bibinfo {pages} {607} (\bibinfo {year} {2009})},\ \Eprint
  {http://arxiv.org/abs/0810.4995} {arXiv:0810.4995} \BibitemShut {NoStop}%
\bibitem [{\citenamefont {{Fermi-LAT collaboration}}(2012)}]{Ackermann2012}%
  \BibitemOpen
  \bibfield  {author} {\bibinfo {author} {\bibnamefont {{Fermi-LAT
  collaboration}}},\ }\href {\doibase 10.1103/PhysRevLett.108.011103}
  {\bibfield  {journal} {\bibinfo  {journal} {Physical Review Letters}\
  }\textbf {\bibinfo {volume} {108}},\ \bibinfo {eid} {011103} (\bibinfo {year}
  {2012})},\ \Eprint {http://arxiv.org/abs/1109.0521} {arXiv:1109.0521
  [astro-ph.HE]} \BibitemShut {NoStop}%
\bibitem [{\citenamefont {Aguilar}\ \emph {et~al.}(2013)\citenamefont {Aguilar}
  \emph {et~al.}}]{Aguilar2013}%
  \BibitemOpen
  \bibfield  {author} {\bibinfo {author} {\bibfnamefont {M.}~\bibnamefont
  {Aguilar}} \emph {et~al.} (\bibinfo {collaboration} {AMS}),\ }\href {\doibase
  10.1103/PhysRevLett.110.141102} {\bibfield  {journal} {\bibinfo  {journal}
  {Phys. Rev. Lett.}\ }\textbf {\bibinfo {volume} {110}},\ \bibinfo {pages}
  {141102} (\bibinfo {year} {2013})}\BibitemShut {NoStop}%
\bibitem [{\citenamefont {Accardo}\ \emph {et~al.}(2014)\citenamefont {Accardo}
  \emph {et~al.}}]{Accardo2014}%
  \BibitemOpen
  \bibfield  {author} {\bibinfo {author} {\bibfnamefont {L.}~\bibnamefont
  {Accardo}} \emph {et~al.} (\bibinfo {collaboration} {AMS}),\ }\href {\doibase
  10.1103/PhysRevLett.113.121101} {\bibfield  {journal} {\bibinfo  {journal}
  {Phys. Rev. Lett.}\ }\textbf {\bibinfo {volume} {113}},\ \bibinfo {pages}
  {121101} (\bibinfo {year} {2014})}\BibitemShut {NoStop}%
\bibitem [{\citenamefont {Ambrosi}\ \emph {et~al.}(2017)\citenamefont {Ambrosi}
  \emph {et~al.}}]{DAMPE2017}%
  \BibitemOpen
  \bibfield  {author} {\bibinfo {author} {\bibfnamefont {G.}~\bibnamefont
  {Ambrosi}} \emph {et~al.} (\bibinfo {collaboration} {DAMPE collaboration}),\
  }\href {\doibase 10.1038/nature24475} {\bibfield  {journal} {\bibinfo
  {journal} {Nature}\ } (\bibinfo {year} {2017}),\ 10.1038/nature24475},\
  \Eprint {http://arxiv.org/abs/1711.10981} {arXiv:1711.10981 [astro-ph.HE]}
  \BibitemShut {NoStop}%
\bibitem [{\citenamefont {{Gu}}\ and\ \citenamefont {{He}}(2017)}]{Gu2017}%
  \BibitemOpen
  \bibfield  {author} {\bibinfo {author} {\bibfnamefont {P.-H.}\ \bibnamefont
  {{Gu}}}\ and\ \bibinfo {author} {\bibfnamefont {X.-G.}\ \bibnamefont
  {{He}}},\ }\href@noop {} {\bibfield  {journal} {\bibinfo  {journal} {ArXiv
  e-prints}\ } (\bibinfo {year} {2017})},\ \Eprint
  {http://arxiv.org/abs/1711.11000} {arXiv:1711.11000 [hep-ph]} \BibitemShut
  {NoStop}%
\bibitem [{\citenamefont {{Fang}}\ \emph {et~al.}(2017)\citenamefont {{Fang}},
  \citenamefont {{Bi}},\ and\ \citenamefont {{Yin}}}]{Fang2017}%
  \BibitemOpen
  \bibfield  {author} {\bibinfo {author} {\bibfnamefont {K.}~\bibnamefont
  {{Fang}}}, \bibinfo {author} {\bibfnamefont {X.-J.}\ \bibnamefont {{Bi}}}, \
  and\ \bibinfo {author} {\bibfnamefont {P.-F.}\ \bibnamefont {{Yin}}},\
  }\href@noop {} {\bibfield  {journal} {\bibinfo  {journal} {ArXiv e-prints}\ }
  (\bibinfo {year} {2017})},\ \Eprint {http://arxiv.org/abs/1711.10996}
  {arXiv:1711.10996 [astro-ph.HE]} \BibitemShut {NoStop}%
\bibitem [{\citenamefont {{Fan}}\ \emph {et~al.}(2017)\citenamefont {{Fan}},
  \citenamefont {{Huang}}, \citenamefont {{Spinrath}}, \citenamefont {{Sming
  Tsai}},\ and\ \citenamefont {{Yuan}}}]{Fan2017}%
  \BibitemOpen
  \bibfield  {author} {\bibinfo {author} {\bibfnamefont {Y.-Z.}\ \bibnamefont
  {{Fan}}}, \bibinfo {author} {\bibfnamefont {W.-C.}\ \bibnamefont {{Huang}}},
  \bibinfo {author} {\bibfnamefont {M.}~\bibnamefont {{Spinrath}}}, \bibinfo
  {author} {\bibfnamefont {Y.-L.}\ \bibnamefont {{Sming Tsai}}}, \ and\
  \bibinfo {author} {\bibfnamefont {Q.}~\bibnamefont {{Yuan}}},\ }\href@noop {}
  {\bibfield  {journal} {\bibinfo  {journal} {ArXiv e-prints}\ } (\bibinfo
  {year} {2017})},\ \Eprint {http://arxiv.org/abs/1711.10995} {arXiv:1711.10995
  [hep-ph]} \BibitemShut {NoStop}%
\bibitem [{\citenamefont {{Yuan}}\ \emph
  {et~al.}(2017{\natexlab{a}})\citenamefont {{Yuan}}, \citenamefont {{Feng}},
  \citenamefont {{Yin}}, \citenamefont {{Fan}}, \citenamefont {{Bi}},
  \citenamefont {{Cui}}, \citenamefont {{Dong}}, \citenamefont {{Guo}},
  \citenamefont {{Fang}}, \citenamefont {{Hu}}, \citenamefont {{Huang}},
  \citenamefont {{Lei}}, \citenamefont {{Li}}, \citenamefont {{Lin}},
  \citenamefont {{Liu}}, \citenamefont {{Ma}}, \citenamefont {{Peng}},
  \citenamefont {{Qiao}}, \citenamefont {{Shen}}, \citenamefont {{Su}},
  \citenamefont {{Wei}}, \citenamefont {{Xu}}, \citenamefont {{Yue}},
  \citenamefont {{Zang}}, \citenamefont {{Zhang}}, \citenamefont {{Zhang}},
  \citenamefont {{Zhang}}, \citenamefont {{Zhang}},\ and\ \citenamefont
  {{Zhang}}}]{Yuan2017_dampe}%
  \BibitemOpen
  \bibfield  {author} {\bibinfo {author} {\bibfnamefont {Q.}~\bibnamefont
  {{Yuan}}}, \bibinfo {author} {\bibfnamefont {L.}~\bibnamefont {{Feng}}},
  \bibinfo {author} {\bibfnamefont {P.-F.}\ \bibnamefont {{Yin}}}, \bibinfo
  {author} {\bibfnamefont {Y.-Z.}\ \bibnamefont {{Fan}}}, \bibinfo {author}
  {\bibfnamefont {X.-J.}\ \bibnamefont {{Bi}}}, \bibinfo {author}
  {\bibfnamefont {M.-Y.}\ \bibnamefont {{Cui}}}, \bibinfo {author}
  {\bibfnamefont {T.-K.}\ \bibnamefont {{Dong}}}, \bibinfo {author}
  {\bibfnamefont {Y.-Q.}\ \bibnamefont {{Guo}}}, \bibinfo {author}
  {\bibfnamefont {K.}~\bibnamefont {{Fang}}}, \bibinfo {author} {\bibfnamefont
  {H.-B.}\ \bibnamefont {{Hu}}}, \bibinfo {author} {\bibfnamefont
  {X.}~\bibnamefont {{Huang}}}, \bibinfo {author} {\bibfnamefont {S.-J.}\
  \bibnamefont {{Lei}}}, \bibinfo {author} {\bibfnamefont {X.}~\bibnamefont
  {{Li}}}, \bibinfo {author} {\bibfnamefont {S.-J.}\ \bibnamefont {{Lin}}},
  \bibinfo {author} {\bibfnamefont {H.}~\bibnamefont {{Liu}}}, \bibinfo
  {author} {\bibfnamefont {P.-X.}\ \bibnamefont {{Ma}}}, \bibinfo {author}
  {\bibfnamefont {W.-X.}\ \bibnamefont {{Peng}}}, \bibinfo {author}
  {\bibfnamefont {R.}~\bibnamefont {{Qiao}}}, \bibinfo {author} {\bibfnamefont
  {Z.-Q.}\ \bibnamefont {{Shen}}}, \bibinfo {author} {\bibfnamefont
  {M.}~\bibnamefont {{Su}}}, \bibinfo {author} {\bibfnamefont {Y.-F.}\
  \bibnamefont {{Wei}}}, \bibinfo {author} {\bibfnamefont {Z.-L.}\ \bibnamefont
  {{Xu}}}, \bibinfo {author} {\bibfnamefont {C.}~\bibnamefont {{Yue}}},
  \bibinfo {author} {\bibfnamefont {J.-J.}\ \bibnamefont {{Zang}}}, \bibinfo
  {author} {\bibfnamefont {C.}~\bibnamefont {{Zhang}}}, \bibinfo {author}
  {\bibfnamefont {X.}~\bibnamefont {{Zhang}}}, \bibinfo {author} {\bibfnamefont
  {Y.-P.}\ \bibnamefont {{Zhang}}}, \bibinfo {author} {\bibfnamefont {Y.-J.}\
  \bibnamefont {{Zhang}}}, \ and\ \bibinfo {author} {\bibfnamefont {Y.-L.}\
  \bibnamefont {{Zhang}}},\ }\href@noop {} {\bibfield  {journal} {\bibinfo
  {journal} {ArXiv e-prints}\ } (\bibinfo {year} {2017}{\natexlab{a}})},\
  \Eprint {http://arxiv.org/abs/1711.10989} {arXiv:1711.10989 [astro-ph.HE]}
  \BibitemShut {NoStop}%
\bibitem [{\citenamefont {{Duan}}\ \emph {et~al.}(2017)\citenamefont {{Duan}},
  \citenamefont {{Feng}}, \citenamefont {{Wang}}, \citenamefont {{Wu}},
  \citenamefont {{Yang}},\ and\ \citenamefont {{Zheng}}}]{Duan2017}%
  \BibitemOpen
  \bibfield  {author} {\bibinfo {author} {\bibfnamefont {G.~H.}\ \bibnamefont
  {{Duan}}}, \bibinfo {author} {\bibfnamefont {L.}~\bibnamefont {{Feng}}},
  \bibinfo {author} {\bibfnamefont {F.}~\bibnamefont {{Wang}}}, \bibinfo
  {author} {\bibfnamefont {L.}~\bibnamefont {{Wu}}}, \bibinfo {author}
  {\bibfnamefont {J.~M.}\ \bibnamefont {{Yang}}}, \ and\ \bibinfo {author}
  {\bibfnamefont {R.}~\bibnamefont {{Zheng}}},\ }\href@noop {} {\bibfield
  {journal} {\bibinfo  {journal} {ArXiv e-prints}\ } (\bibinfo {year}
  {2017})},\ \Eprint {http://arxiv.org/abs/1711.11012} {arXiv:1711.11012
  [hep-ph]} \BibitemShut {NoStop}%
\bibitem [{\citenamefont {Gu}(2017{\natexlab{a}})}]{Gu2017a}%
  \BibitemOpen
  \bibfield  {author} {\bibinfo {author} {\bibfnamefont {P.-H.}\ \bibnamefont
  {Gu}},\ }\href@noop {} {\bibfield  {journal} {\bibinfo  {journal} {ArXiv
  e-prints}\ } (\bibinfo {year} {2017}{\natexlab{a}})},\ \Eprint
  {http://arxiv.org/abs/1711.11333} {arXiv:1711.11333 [hep-ph]} \BibitemShut
  {NoStop}%
\bibitem [{\citenamefont {Chao}\ and\ \citenamefont {Yuan}(2017)}]{Chao2017}%
  \BibitemOpen
  \bibfield  {author} {\bibinfo {author} {\bibfnamefont {W.}~\bibnamefont
  {Chao}}\ and\ \bibinfo {author} {\bibfnamefont {Q.}~\bibnamefont {Yuan}},\
  }\href@noop {} {\bibfield  {journal} {\bibinfo  {journal} {ArXiv e-prints}\ }
  (\bibinfo {year} {2017})},\ \Eprint {http://arxiv.org/abs/1711.11182}
  {arXiv:1711.11182 [hep-ph]} \BibitemShut {NoStop}%
\bibitem [{\citenamefont {Tang}\ \emph {et~al.}(2017)\citenamefont {Tang},
  \citenamefont {Wu}, \citenamefont {Zhang},\ and\ \citenamefont
  {Zheng}}]{Tang2017}%
  \BibitemOpen
  \bibfield  {author} {\bibinfo {author} {\bibfnamefont {Y.-L.}\ \bibnamefont
  {Tang}}, \bibinfo {author} {\bibfnamefont {L.}~\bibnamefont {Wu}}, \bibinfo
  {author} {\bibfnamefont {M.}~\bibnamefont {Zhang}}, \ and\ \bibinfo {author}
  {\bibfnamefont {R.}~\bibnamefont {Zheng}},\ }\href@noop {} {\bibfield
  {journal} {\bibinfo  {journal} {ArXiv e-prints}\ } (\bibinfo {year}
  {2017})},\ \Eprint {http://arxiv.org/abs/1711.11058} {arXiv:1711.11058
  [hep-ph]} \BibitemShut {NoStop}%
\bibitem [{\citenamefont {Zu}\ \emph {et~al.}(2017)\citenamefont {Zu},
  \citenamefont {Zhang}, \citenamefont {Feng}, \citenamefont {Yuan},\ and\
  \citenamefont {Fan}}]{Zu2017}%
  \BibitemOpen
  \bibfield  {author} {\bibinfo {author} {\bibfnamefont {L.}~\bibnamefont
  {Zu}}, \bibinfo {author} {\bibfnamefont {C.}~\bibnamefont {Zhang}}, \bibinfo
  {author} {\bibfnamefont {L.}~\bibnamefont {Feng}}, \bibinfo {author}
  {\bibfnamefont {Q.}~\bibnamefont {Yuan}}, \ and\ \bibinfo {author}
  {\bibfnamefont {Y.-Z.}\ \bibnamefont {Fan}},\ }\href@noop {} {\bibfield
  {journal} {\bibinfo  {journal} {ArXiv e-prints}\ } (\bibinfo {year}
  {2017})},\ \Eprint {http://arxiv.org/abs/1711.11052} {arXiv:1711.11052
  [hep-ph]} \BibitemShut {NoStop}%
\bibitem [{\citenamefont {{Liu}}\ and\ \citenamefont {{Liu}}(2017)}]{Liu2017}%
  \BibitemOpen
  \bibfield  {author} {\bibinfo {author} {\bibfnamefont {X.}~\bibnamefont
  {{Liu}}}\ and\ \bibinfo {author} {\bibfnamefont {Z.}~\bibnamefont {{Liu}}},\
  }\href@noop {} {\bibfield  {journal} {\bibinfo  {journal} {ArXiv e-prints}\ }
  (\bibinfo {year} {2017})},\ \Eprint {http://arxiv.org/abs/1711.11579}
  {arXiv:1711.11579 [hep-ph]} \BibitemShut {NoStop}%
\bibitem [{\citenamefont {{Cao}}\ \emph {et~al.}(2017)\citenamefont {{Cao}},
  \citenamefont {{Feng}}, \citenamefont {{Guo}}, \citenamefont {{Shang}},
  \citenamefont {{Wang}},\ and\ \citenamefont {{Wu}}}]{Cao2017}%
  \BibitemOpen
  \bibfield  {author} {\bibinfo {author} {\bibfnamefont {J.}~\bibnamefont
  {{Cao}}}, \bibinfo {author} {\bibfnamefont {L.}~\bibnamefont {{Feng}}},
  \bibinfo {author} {\bibfnamefont {X.}~\bibnamefont {{Guo}}}, \bibinfo
  {author} {\bibfnamefont {L.}~\bibnamefont {{Shang}}}, \bibinfo {author}
  {\bibfnamefont {F.}~\bibnamefont {{Wang}}}, \ and\ \bibinfo {author}
  {\bibfnamefont {P.}~\bibnamefont {{Wu}}},\ }\href@noop {} {\bibfield
  {journal} {\bibinfo  {journal} {ArXiv e-prints}\ } (\bibinfo {year}
  {2017})},\ \Eprint {http://arxiv.org/abs/1711.11452} {arXiv:1711.11452
  [hep-ph]} \BibitemShut {NoStop}%
\bibitem [{\citenamefont {{Athron}}\ \emph {et~al.}(2017)\citenamefont
  {{Athron}}, \citenamefont {{Balazs}}, \citenamefont {{Fowlie}},\ and\
  \citenamefont {{Zhang}}}]{Athron2017}%
  \BibitemOpen
  \bibfield  {author} {\bibinfo {author} {\bibfnamefont {P.}~\bibnamefont
  {{Athron}}}, \bibinfo {author} {\bibfnamefont {C.}~\bibnamefont {{Balazs}}},
  \bibinfo {author} {\bibfnamefont {A.}~\bibnamefont {{Fowlie}}}, \ and\
  \bibinfo {author} {\bibfnamefont {Y.}~\bibnamefont {{Zhang}}},\ }\href@noop
  {} {\bibfield  {journal} {\bibinfo  {journal} {ArXiv e-prints}\ } (\bibinfo
  {year} {2017})},\ \Eprint {http://arxiv.org/abs/1711.11376} {arXiv:1711.11376
  [hep-ph]} \BibitemShut {NoStop}%
\bibitem [{\citenamefont {Chao}\ \emph {et~al.}(2017)\citenamefont {Chao},
  \citenamefont {Guo}, \citenamefont {Li},\ and\ \citenamefont
  {Shu}}]{Chao2017a}%
  \BibitemOpen
  \bibfield  {author} {\bibinfo {author} {\bibfnamefont {W.}~\bibnamefont
  {Chao}}, \bibinfo {author} {\bibfnamefont {H.-K.}\ \bibnamefont {Guo}},
  \bibinfo {author} {\bibfnamefont {H.-L.}\ \bibnamefont {Li}}, \ and\ \bibinfo
  {author} {\bibfnamefont {J.}~\bibnamefont {Shu}},\ }\href@noop {} {\bibfield
  {journal} {\bibinfo  {journal} {ArXiv e-prints}\ } (\bibinfo {year}
  {2017})},\ \Eprint {http://arxiv.org/abs/1712.00037} {arXiv:1712.00037
  [hep-ph]} \BibitemShut {NoStop}%
\bibitem [{\citenamefont {Gao}\ and\ \citenamefont {Ma}(2017)}]{Gao2017}%
  \BibitemOpen
  \bibfield  {author} {\bibinfo {author} {\bibfnamefont {Y.}~\bibnamefont
  {Gao}}\ and\ \bibinfo {author} {\bibfnamefont {Y.-Z.}\ \bibnamefont {Ma}},\
  }\href@noop {} {\bibfield  {journal} {\bibinfo  {journal} {ArXiv e-prints}\ }
  (\bibinfo {year} {2017})},\ \Eprint {http://arxiv.org/abs/1712.00370}
  {arXiv:1712.00370 [astro-ph.HE]} \BibitemShut {NoStop}%
\bibitem [{\citenamefont {{Niu}}\ \emph
  {et~al.}(2018{\natexlab{a}})\citenamefont {{Niu}}, \citenamefont {{Li}},
  \citenamefont {{Ding}}, \citenamefont {{Zhu}}, \citenamefont {{Xue}},\ and\
  \citenamefont {{Wang}}}]{Niu2017_dampe}%
  \BibitemOpen
  \bibfield  {author} {\bibinfo {author} {\bibfnamefont {J.-S.}\ \bibnamefont
  {{Niu}}}, \bibinfo {author} {\bibfnamefont {T.}~\bibnamefont {{Li}}},
  \bibinfo {author} {\bibfnamefont {R.}~\bibnamefont {{Ding}}}, \bibinfo
  {author} {\bibfnamefont {B.}~\bibnamefont {{Zhu}}}, \bibinfo {author}
  {\bibfnamefont {H.-F.}\ \bibnamefont {{Xue}}}, \ and\ \bibinfo {author}
  {\bibfnamefont {Y.}~\bibnamefont {{Wang}}},\ }\href {\doibase
  10.1103/PhysRevD.97.083012} {\bibfield  {journal} {\bibinfo  {journal}
  {\prd}\ }\textbf {\bibinfo {volume} {97}},\ \bibinfo {eid} {083012} (\bibinfo
  {year} {2018}{\natexlab{a}})},\ \Eprint {http://arxiv.org/abs/1712.00372}
  {arXiv:1712.00372 [astro-ph.HE]} \BibitemShut {NoStop}%
\bibitem [{\citenamefont {{Jin}}\ \emph {et~al.}(2017)\citenamefont {{Jin}},
  \citenamefont {{Yue}}, \citenamefont {{Zhang}},\ and\ \citenamefont
  {{Chen}}}]{Jin2017_dampe}%
  \BibitemOpen
  \bibfield  {author} {\bibinfo {author} {\bibfnamefont {H.-B.}\ \bibnamefont
  {{Jin}}}, \bibinfo {author} {\bibfnamefont {B.}~\bibnamefont {{Yue}}},
  \bibinfo {author} {\bibfnamefont {X.}~\bibnamefont {{Zhang}}}, \ and\
  \bibinfo {author} {\bibfnamefont {X.}~\bibnamefont {{Chen}}},\ }\href@noop {}
  {\bibfield  {journal} {\bibinfo  {journal} {ArXiv e-prints}\ } (\bibinfo
  {year} {2017})},\ \Eprint {http://arxiv.org/abs/1712.00362} {arXiv:1712.00362
  [astro-ph.HE]} \BibitemShut {NoStop}%
\bibitem [{\citenamefont {{Huang}}\ \emph {et~al.}(2017)\citenamefont
  {{Huang}}, \citenamefont {{Wu}}, \citenamefont {{Zhang}},\ and\ \citenamefont
  {{Zhou}}}]{Huang2017}%
  \BibitemOpen
  \bibfield  {author} {\bibinfo {author} {\bibfnamefont {X.-J.}\ \bibnamefont
  {{Huang}}}, \bibinfo {author} {\bibfnamefont {Y.-L.}\ \bibnamefont {{Wu}}},
  \bibinfo {author} {\bibfnamefont {W.-H.}\ \bibnamefont {{Zhang}}}, \ and\
  \bibinfo {author} {\bibfnamefont {Y.-F.}\ \bibnamefont {{Zhou}}},\
  }\href@noop {} {\bibfield  {journal} {\bibinfo  {journal} {ArXiv e-prints}\ }
  (\bibinfo {year} {2017})},\ \Eprint {http://arxiv.org/abs/1712.00005}
  {arXiv:1712.00005 [astro-ph.HE]} \BibitemShut {NoStop}%
\bibitem [{\citenamefont {Duan}\ \emph {et~al.}(2017)\citenamefont {Duan},
  \citenamefont {Feng}, \citenamefont {Wang}, \citenamefont {Wu}, \citenamefont
  {Yang},\ and\ \citenamefont {Zheng}}]{Duan2017a}%
  \BibitemOpen
  \bibfield  {author} {\bibinfo {author} {\bibfnamefont {G.~H.}\ \bibnamefont
  {Duan}}, \bibinfo {author} {\bibfnamefont {L.}~\bibnamefont {Feng}}, \bibinfo
  {author} {\bibfnamefont {F.}~\bibnamefont {Wang}}, \bibinfo {author}
  {\bibfnamefont {L.}~\bibnamefont {Wu}}, \bibinfo {author} {\bibfnamefont
  {J.~M.}\ \bibnamefont {Yang}}, \ and\ \bibinfo {author} {\bibfnamefont
  {R.}~\bibnamefont {Zheng}},\ }\href@noop {} {\bibfield  {journal} {\bibinfo
  {journal} {ArXiv e-prints}\ } (\bibinfo {year} {2017})},\ \Eprint
  {http://arxiv.org/abs/1711.11012} {arXiv:1711.11012 [hep-ph]} \BibitemShut
  {NoStop}%
\bibitem [{\citenamefont {Cao}\ \emph {et~al.}(2017{\natexlab{a}})\citenamefont
  {Cao}, \citenamefont {Feng}, \citenamefont {Guo}, \citenamefont {Shang},
  \citenamefont {Wang}, \citenamefont {Wu},\ and\ \citenamefont
  {Zu}}]{Cao2017a}%
  \BibitemOpen
  \bibfield  {author} {\bibinfo {author} {\bibfnamefont {J.}~\bibnamefont
  {Cao}}, \bibinfo {author} {\bibfnamefont {L.}~\bibnamefont {Feng}}, \bibinfo
  {author} {\bibfnamefont {X.}~\bibnamefont {Guo}}, \bibinfo {author}
  {\bibfnamefont {L.}~\bibnamefont {Shang}}, \bibinfo {author} {\bibfnamefont
  {F.}~\bibnamefont {Wang}}, \bibinfo {author} {\bibfnamefont {P.}~\bibnamefont
  {Wu}}, \ and\ \bibinfo {author} {\bibfnamefont {L.}~\bibnamefont {Zu}},\
  }\href@noop {} {\bibfield  {journal} {\bibinfo  {journal} {ArXiv e-prints}\ }
  (\bibinfo {year} {2017}{\natexlab{a}})},\ \Eprint
  {http://arxiv.org/abs/1712.01244} {arXiv:1712.01244 [hep-ph]} \BibitemShut
  {NoStop}%
\bibitem [{\citenamefont {Ghorbani}\ and\ \citenamefont
  {Ghorbani}(2017)}]{Ghorbani2017}%
  \BibitemOpen
  \bibfield  {author} {\bibinfo {author} {\bibfnamefont {K.}~\bibnamefont
  {Ghorbani}}\ and\ \bibinfo {author} {\bibfnamefont {P.~H.}\ \bibnamefont
  {Ghorbani}},\ }\href@noop {} {\bibfield  {journal} {\bibinfo  {journal}
  {ArXiv e-prints}\ } (\bibinfo {year} {2017})},\ \Eprint
  {http://arxiv.org/abs/1712.01239} {arXiv:1712.01239 [hep-ph]} \BibitemShut
  {NoStop}%
\bibitem [{\citenamefont {Nomura}\ and\ \citenamefont
  {Okada}(2017)}]{Nomura2017}%
  \BibitemOpen
  \bibfield  {author} {\bibinfo {author} {\bibfnamefont {T.}~\bibnamefont
  {Nomura}}\ and\ \bibinfo {author} {\bibfnamefont {H.}~\bibnamefont {Okada}},\
  }\href@noop {} {\bibfield  {journal} {\bibinfo  {journal} {ArXiv e-prints}\ }
  (\bibinfo {year} {2017})},\ \Eprint {http://arxiv.org/abs/1712.00941}
  {arXiv:1712.00941 [hep-ph]} \BibitemShut {NoStop}%
\bibitem [{\citenamefont {Gu}(2017{\natexlab{b}})}]{Gu2017b}%
  \BibitemOpen
  \bibfield  {author} {\bibinfo {author} {\bibfnamefont {P.-H.}\ \bibnamefont
  {Gu}},\ }\href@noop {} {\bibfield  {journal} {\bibinfo  {journal} {ArXiv
  e-prints}\ } (\bibinfo {year} {2017}{\natexlab{b}})},\ \Eprint
  {http://arxiv.org/abs/1712.00922} {arXiv:1712.00922 [hep-ph]} \BibitemShut
  {NoStop}%
\bibitem [{\citenamefont {Li}\ \emph {et~al.}(2017)\citenamefont {Li},
  \citenamefont {Okada},\ and\ \citenamefont {Shafi}}]{Li2017}%
  \BibitemOpen
  \bibfield  {author} {\bibinfo {author} {\bibfnamefont {T.}~\bibnamefont
  {Li}}, \bibinfo {author} {\bibfnamefont {N.}~\bibnamefont {Okada}}, \ and\
  \bibinfo {author} {\bibfnamefont {Q.}~\bibnamefont {Shafi}},\ }\href@noop {}
  {\bibfield  {journal} {\bibinfo  {journal} {ArXiv e-prints}\ } (\bibinfo
  {year} {2017})},\ \Eprint {http://arxiv.org/abs/1712.00869} {arXiv:1712.00869
  [hep-ph]} \BibitemShut {NoStop}%
\bibitem [{\citenamefont {Chen}\ \emph {et~al.}(2017)\citenamefont {Chen},
  \citenamefont {Chiang},\ and\ \citenamefont {Nomura}}]{Chen2017}%
  \BibitemOpen
  \bibfield  {author} {\bibinfo {author} {\bibfnamefont {C.-H.}\ \bibnamefont
  {Chen}}, \bibinfo {author} {\bibfnamefont {C.-W.}\ \bibnamefont {Chiang}}, \
  and\ \bibinfo {author} {\bibfnamefont {T.}~\bibnamefont {Nomura}},\
  }\href@noop {} {\bibfield  {journal} {\bibinfo  {journal} {ArXiv e-prints}\ }
  (\bibinfo {year} {2017})},\ \Eprint {http://arxiv.org/abs/1712.00793}
  {arXiv:1712.00793 [hep-ph]} \BibitemShut {NoStop}%
\bibitem [{\citenamefont {{Yang}}\ \emph {et~al.}(2017)\citenamefont {{Yang}},
  \citenamefont {{Su}},\ and\ \citenamefont {{Zhao}}}]{Yang2017}%
  \BibitemOpen
  \bibfield  {author} {\bibinfo {author} {\bibfnamefont {F.}~\bibnamefont
  {{Yang}}}, \bibinfo {author} {\bibfnamefont {M.}~\bibnamefont {{Su}}}, \ and\
  \bibinfo {author} {\bibfnamefont {Y.}~\bibnamefont {{Zhao}}},\ }\href@noop {}
  {\bibfield  {journal} {\bibinfo  {journal} {ArXiv e-prints}\ } (\bibinfo
  {year} {2017})},\ \Eprint {http://arxiv.org/abs/1712.01724} {arXiv:1712.01724
  [astro-ph.HE]} \BibitemShut {NoStop}%
\bibitem [{\citenamefont {Ding}\ \emph {et~al.}(2017)\citenamefont {Ding},
  \citenamefont {Han}, \citenamefont {Feng},\ and\ \citenamefont
  {Zhu}}]{Ding2017}%
  \BibitemOpen
  \bibfield  {author} {\bibinfo {author} {\bibfnamefont {R.}~\bibnamefont
  {Ding}}, \bibinfo {author} {\bibfnamefont {Z.-L.}\ \bibnamefont {Han}},
  \bibinfo {author} {\bibfnamefont {L.}~\bibnamefont {Feng}}, \ and\ \bibinfo
  {author} {\bibfnamefont {B.}~\bibnamefont {Zhu}},\ }\href@noop {} {\bibfield
  {journal} {\bibinfo  {journal} {ArXiv e-prints}\ } (\bibinfo {year}
  {2017})},\ \Eprint {http://arxiv.org/abs/1712.02021} {arXiv:1712.02021
  [hep-ph]} \BibitemShut {NoStop}%
\bibitem [{\citenamefont {{Ge}}\ and\ \citenamefont {{He}}(2017)}]{Ge2017}%
  \BibitemOpen
  \bibfield  {author} {\bibinfo {author} {\bibfnamefont {S.-F.}\ \bibnamefont
  {{Ge}}}\ and\ \bibinfo {author} {\bibfnamefont {H.-J.}\ \bibnamefont
  {{He}}},\ }\href@noop {} {\bibfield  {journal} {\bibinfo  {journal} {ArXiv
  e-prints}\ } (\bibinfo {year} {2017})},\ \Eprint
  {http://arxiv.org/abs/1712.02744} {arXiv:1712.02744 [astro-ph.HE]}
  \BibitemShut {NoStop}%
\bibitem [{\citenamefont {Okada}\ and\ \citenamefont {Seto}(2017)}]{Okada2017}%
  \BibitemOpen
  \bibfield  {author} {\bibinfo {author} {\bibfnamefont {N.}~\bibnamefont
  {Okada}}\ and\ \bibinfo {author} {\bibfnamefont {O.}~\bibnamefont {Seto}},\
  }\href@noop {} {\bibfield  {journal} {\bibinfo  {journal} {ArXiv e-prints}\ }
  (\bibinfo {year} {2017})},\ \Eprint {http://arxiv.org/abs/1712.03652}
  {arXiv:1712.03652 [hep-ph]} \BibitemShut {NoStop}%
\bibitem [{\citenamefont {Sui}\ and\ \citenamefont {Zhang}(2017)}]{Sui2017}%
  \BibitemOpen
  \bibfield  {author} {\bibinfo {author} {\bibfnamefont {Y.}~\bibnamefont
  {Sui}}\ and\ \bibinfo {author} {\bibfnamefont {Y.}~\bibnamefont {Zhang}},\
  }\href@noop {} {\bibfield  {journal} {\bibinfo  {journal} {ArXiv e-prints}\ }
  (\bibinfo {year} {2017})},\ \Eprint {http://arxiv.org/abs/1712.03642}
  {arXiv:1712.03642 [hep-ph]} \BibitemShut {NoStop}%
\bibitem [{\citenamefont {Cao}\ \emph {et~al.}(2017{\natexlab{b}})\citenamefont
  {Cao}, \citenamefont {Guo}, \citenamefont {Shang}, \citenamefont {Wang},\
  and\ \citenamefont {Wu}}]{Cao2017b}%
  \BibitemOpen
  \bibfield  {author} {\bibinfo {author} {\bibfnamefont {J.}~\bibnamefont
  {Cao}}, \bibinfo {author} {\bibfnamefont {X.}~\bibnamefont {Guo}}, \bibinfo
  {author} {\bibfnamefont {L.}~\bibnamefont {Shang}}, \bibinfo {author}
  {\bibfnamefont {F.}~\bibnamefont {Wang}}, \ and\ \bibinfo {author}
  {\bibfnamefont {P.}~\bibnamefont {Wu}},\ }\href@noop {} {\bibfield  {journal}
  {\bibinfo  {journal} {ArXiv e-prints}\ } (\bibinfo {year}
  {2017}{\natexlab{b}})},\ \Eprint {http://arxiv.org/abs/1712.05351}
  {arXiv:1712.05351 [hep-ph]} \BibitemShut {NoStop}%
\bibitem [{\citenamefont {Han}\ \emph {et~al.}(2017)\citenamefont {Han},
  \citenamefont {Wang},\ and\ \citenamefont {Ding}}]{Han2017}%
  \BibitemOpen
  \bibfield  {author} {\bibinfo {author} {\bibfnamefont {Z.-L.}\ \bibnamefont
  {Han}}, \bibinfo {author} {\bibfnamefont {W.}~\bibnamefont {Wang}}, \ and\
  \bibinfo {author} {\bibfnamefont {R.}~\bibnamefont {Ding}},\ }\href@noop {}
  {\bibfield  {journal} {\bibinfo  {journal} {ArXiv e-prints}\ } (\bibinfo
  {year} {2017})},\ \Eprint {http://arxiv.org/abs/1712.05722} {arXiv:1712.05722
  [hep-ph]} \BibitemShut {NoStop}%
\bibitem [{\citenamefont {{Cholis}}\ \emph {et~al.}(2017)\citenamefont
  {{Cholis}}, \citenamefont {{Karwal}},\ and\ \citenamefont
  {{Kamionkowski}}}]{Cholis2017}%
  \BibitemOpen
  \bibfield  {author} {\bibinfo {author} {\bibfnamefont {I.}~\bibnamefont
  {{Cholis}}}, \bibinfo {author} {\bibfnamefont {T.}~\bibnamefont {{Karwal}}},
  \ and\ \bibinfo {author} {\bibfnamefont {M.}~\bibnamefont {{Kamionkowski}}},\
  }\href@noop {} {\bibfield  {journal} {\bibinfo  {journal} {ArXiv e-prints}\ }
  (\bibinfo {year} {2017})},\ \Eprint {http://arxiv.org/abs/1712.00011}
  {arXiv:1712.00011 [astro-ph.HE]} \BibitemShut {NoStop}%
\bibitem [{\citenamefont {{Fowlie}}(2017)}]{Fowlie:2017fya}%
  \BibitemOpen
  \bibfield  {author} {\bibinfo {author} {\bibfnamefont {A.}~\bibnamefont
  {{Fowlie}}},\ }\href@noop {} {\bibfield  {journal} {\bibinfo  {journal}
  {ArXiv e-prints}\ } (\bibinfo {year} {2017})},\ \Eprint
  {http://arxiv.org/abs/1712.05089} {arXiv:1712.05089 [hep-ph]} \BibitemShut
  {NoStop}%
\bibitem [{\citenamefont {{Kachelrie{\ss}}}\ \emph {et~al.}(2015)\citenamefont
  {{Kachelrie{\ss}}}, \citenamefont {{Neronov}},\ and\ \citenamefont
  {{Semikoz}}}]{KNS2015}%
  \BibitemOpen
  \bibfield  {author} {\bibinfo {author} {\bibfnamefont {M.}~\bibnamefont
  {{Kachelrie{\ss}}}}, \bibinfo {author} {\bibfnamefont {A.}~\bibnamefont
  {{Neronov}}}, \ and\ \bibinfo {author} {\bibfnamefont {D.~V.}\ \bibnamefont
  {{Semikoz}}},\ }\href {\doibase 10.1103/PhysRevLett.115.181103} {\bibfield
  {journal} {\bibinfo  {journal} {Physical Review Letters}\ }\textbf {\bibinfo
  {volume} {115}},\ \bibinfo {eid} {181103} (\bibinfo {year} {2015})},\ \Eprint
  {http://arxiv.org/abs/1504.06472} {arXiv:1504.06472 [astro-ph.HE]}
  \BibitemShut {NoStop}%
\bibitem [{\citenamefont {{Kachelrie{\ss}}}\ \emph {et~al.}(2018)\citenamefont
  {{Kachelrie{\ss}}}, \citenamefont {{Neronov}},\ and\ \citenamefont
  {{Semikoz}}}]{KNS2018}%
  \BibitemOpen
  \bibfield  {author} {\bibinfo {author} {\bibfnamefont {M.}~\bibnamefont
  {{Kachelrie{\ss}}}}, \bibinfo {author} {\bibfnamefont {A.}~\bibnamefont
  {{Neronov}}}, \ and\ \bibinfo {author} {\bibfnamefont {D.~V.}\ \bibnamefont
  {{Semikoz}}},\ }\href {\doibase 10.1103/PhysRevD.97.063011} {\bibfield
  {journal} {\bibinfo  {journal} {\prd}\ }\textbf {\bibinfo {volume} {97}},\
  \bibinfo {eid} {063011} (\bibinfo {year} {2018})},\ \Eprint
  {http://arxiv.org/abs/1710.02321} {arXiv:1710.02321 [astro-ph.HE]}
  \BibitemShut {NoStop}%
\bibitem [{\citenamefont {{AMS collaboration}}(2013{\natexlab{b}})}]{AMS2013}%
  \BibitemOpen
  \bibfield  {author} {\bibinfo {author} {\bibnamefont {{AMS collaboration}}},\
  }\href {\doibase 10.1103/PhysRevLett.110.141102} {\bibfield  {journal}
  {\bibinfo  {journal} {Physical Review Letters}\ }\textbf {\bibinfo {volume}
  {110}},\ \bibinfo {eid} {141102} (\bibinfo {year}
  {2013}{\natexlab{b}})}\BibitemShut {NoStop}%
\bibitem [{\citenamefont {{HEAT collaboration}}(1997)}]{Barwick1997}%
  \BibitemOpen
  \bibfield  {author} {\bibinfo {author} {\bibnamefont {{HEAT
  collaboration}}},\ }\href {\doibase 10.1086/310706} {\bibfield  {journal}
  {\bibinfo  {journal} {\apjl}\ }\textbf {\bibinfo {volume} {482}},\ \bibinfo
  {pages} {L191} (\bibinfo {year} {1997})},\ \Eprint
  {http://arxiv.org/abs/astro-ph/9703192} {astro-ph/9703192} \BibitemShut
  {NoStop}%
\bibitem [{\citenamefont {{AMS-01 collaboration}}(2007)}]{AMS01}%
  \BibitemOpen
  \bibfield  {author} {\bibinfo {author} {\bibnamefont {{AMS-01
  collaboration}}},\ }\href {\doibase 10.1016/j.physletb.2007.01.024}
  {\bibfield  {journal} {\bibinfo  {journal} {Physics Letters B}\ }\textbf
  {\bibinfo {volume} {646}},\ \bibinfo {pages} {145} (\bibinfo {year}
  {2007})},\ \Eprint {http://arxiv.org/abs/astro-ph/0703154} {astro-ph/0703154}
  \BibitemShut {NoStop}%
\bibitem [{\citenamefont {{Evoli}}\ \emph {et~al.}(2008)\citenamefont
  {{Evoli}}, \citenamefont {{Gaggero}}, \citenamefont {{Grasso}},\ and\
  \citenamefont {{Maccione}}}]{Evoli2008}%
  \BibitemOpen
  \bibfield  {author} {\bibinfo {author} {\bibfnamefont {C.}~\bibnamefont
  {{Evoli}}}, \bibinfo {author} {\bibfnamefont {D.}~\bibnamefont {{Gaggero}}},
  \bibinfo {author} {\bibfnamefont {D.}~\bibnamefont {{Grasso}}}, \ and\
  \bibinfo {author} {\bibfnamefont {L.}~\bibnamefont {{Maccione}}},\ }\href
  {\doibase 10.1088/1475-7516/2008/10/018} {\bibfield  {journal} {\bibinfo
  {journal} {\jcap}\ }\textbf {\bibinfo {volume} {10}},\ \bibinfo {eid} {018}
  (\bibinfo {year} {2008})},\ \Eprint {http://arxiv.org/abs/0807.4730}
  {arXiv:0807.4730} \BibitemShut {NoStop}%
\bibitem [{\citenamefont {{J{\'o}hannesson}}\ \emph {et~al.}(2016)\citenamefont
  {{J{\'o}hannesson}}, \citenamefont {{Ruiz de Austri}}, \citenamefont
  {{Vincent}}, \citenamefont {{Moskalenko}}, \citenamefont {{Orlando}},
  \citenamefont {{Porter}}, \citenamefont {{Strong}}, \citenamefont {{Trotta}},
  \citenamefont {{Feroz}}, \citenamefont {{Graff}},\ and\ \citenamefont
  {{Hobson}}}]{Johannesson2016}%
  \BibitemOpen
  \bibfield  {author} {\bibinfo {author} {\bibfnamefont {G.}~\bibnamefont
  {{J{\'o}hannesson}}}, \bibinfo {author} {\bibfnamefont {R.}~\bibnamefont
  {{Ruiz de Austri}}}, \bibinfo {author} {\bibfnamefont {A.~C.}\ \bibnamefont
  {{Vincent}}}, \bibinfo {author} {\bibfnamefont {I.~V.}\ \bibnamefont
  {{Moskalenko}}}, \bibinfo {author} {\bibfnamefont {E.}~\bibnamefont
  {{Orlando}}}, \bibinfo {author} {\bibfnamefont {T.~A.}\ \bibnamefont
  {{Porter}}}, \bibinfo {author} {\bibfnamefont {A.~W.}\ \bibnamefont
  {{Strong}}}, \bibinfo {author} {\bibfnamefont {R.}~\bibnamefont {{Trotta}}},
  \bibinfo {author} {\bibfnamefont {F.}~\bibnamefont {{Feroz}}}, \bibinfo
  {author} {\bibfnamefont {P.}~\bibnamefont {{Graff}}}, \ and\ \bibinfo
  {author} {\bibfnamefont {M.~P.}\ \bibnamefont {{Hobson}}},\ }\href {\doibase
  10.3847/0004-637X/824/1/16} {\bibfield  {journal} {\bibinfo  {journal}
  {\apj}\ }\textbf {\bibinfo {volume} {824}},\ \bibinfo {eid} {16} (\bibinfo
  {year} {2016})},\ \Eprint {http://arxiv.org/abs/1602.02243} {arXiv:1602.02243
  [astro-ph.HE]} \BibitemShut {NoStop}%
\bibitem [{\citenamefont {{Niu}}\ and\ \citenamefont {{Li}}(2018)}]{Niu2017}%
  \BibitemOpen
  \bibfield  {author} {\bibinfo {author} {\bibfnamefont {J.-S.}\ \bibnamefont
  {{Niu}}}\ and\ \bibinfo {author} {\bibfnamefont {T.}~\bibnamefont {{Li}}},\
  }\href {\doibase 10.1103/PhysRevD.97.023015} {\bibfield  {journal} {\bibinfo
  {journal} {\prd}\ }\textbf {\bibinfo {volume} {97}},\ \bibinfo {eid} {023015}
  (\bibinfo {year} {2018})},\ \Eprint {http://arxiv.org/abs/1705.11089}
  {arXiv:1705.11089 [astro-ph.HE]} \BibitemShut {NoStop}%
\bibitem [{\citenamefont {{Yuan}}\ \emph
  {et~al.}(2017{\natexlab{b}})\citenamefont {{Yuan}}, \citenamefont {{Lin}},
  \citenamefont {{Fang}},\ and\ \citenamefont {{Bi}}}]{Yuan2017}%
  \BibitemOpen
  \bibfield  {author} {\bibinfo {author} {\bibfnamefont {Q.}~\bibnamefont
  {{Yuan}}}, \bibinfo {author} {\bibfnamefont {S.-J.}\ \bibnamefont {{Lin}}},
  \bibinfo {author} {\bibfnamefont {K.}~\bibnamefont {{Fang}}}, \ and\ \bibinfo
  {author} {\bibfnamefont {X.-J.}\ \bibnamefont {{Bi}}},\ }\href@noop {}
  {\bibfield  {journal} {\bibinfo  {journal} {ArXiv e-prints}\ } (\bibinfo
  {year} {2017}{\natexlab{b}})},\ \Eprint {http://arxiv.org/abs/1701.06149}
  {arXiv:1701.06149 [astro-ph.HE]} \BibitemShut {NoStop}%
\bibitem [{\citenamefont {{Yuan}}(2018)}]{Yuan2018}%
  \BibitemOpen
  \bibfield  {author} {\bibinfo {author} {\bibfnamefont {Q.}~\bibnamefont
  {{Yuan}}},\ }\href@noop {} {\bibfield  {journal} {\bibinfo  {journal} {ArXiv
  e-prints}\ } (\bibinfo {year} {2018})},\ \Eprint
  {http://arxiv.org/abs/1805.10649} {arXiv:1805.10649 [astro-ph.HE]}
  \BibitemShut {NoStop}%
\bibitem [{\citenamefont {{Zhu}}\ \emph {et~al.}(2018)\citenamefont {{Zhu}},
  \citenamefont {{Yuan}},\ and\ \citenamefont {{Wei}}}]{Zhu2018}%
  \BibitemOpen
  \bibfield  {author} {\bibinfo {author} {\bibfnamefont {C.-R.}\ \bibnamefont
  {{Zhu}}}, \bibinfo {author} {\bibfnamefont {Q.}~\bibnamefont {{Yuan}}}, \
  and\ \bibinfo {author} {\bibfnamefont {D.-M.}\ \bibnamefont {{Wei}}},\ }\href
  {\doibase 10.3847/1538-4357/aacff9} {\bibfield  {journal} {\bibinfo
  {journal} {\apj}\ }\textbf {\bibinfo {volume} {863}},\ \bibinfo {eid} {119}
  (\bibinfo {year} {2018})},\ \Eprint {http://arxiv.org/abs/1807.09470}
  {arXiv:1807.09470 [astro-ph.HE]} \BibitemShut {NoStop}%
\bibitem [{\citenamefont {{Panov}}\ \emph {et~al.}(2006)\citenamefont
  {{Panov}}, \citenamefont {{Adams}}, \citenamefont {{Ahn}}, \citenamefont
  {{Bashindzhagyan}}, \citenamefont {{Batkov}}, \citenamefont {{Chang}},
  \citenamefont {{Christl}}, \citenamefont {{Fazely}}, \citenamefont {{Ganel}},
  \citenamefont {{Gunashingha}}, \citenamefont {{Guzik}}, \citenamefont
  {{Isbert}}, \citenamefont {{Kim}}, \citenamefont {{Kouznetsov}},
  \citenamefont {{Panasyuk}}, \citenamefont {{Schmidt}}, \citenamefont {{Seo}},
  \citenamefont {{Sokolskaya}}, \citenamefont {{Watts}}, \citenamefont
  {{Wefel}}, \citenamefont {{Wu}},\ and\ \citenamefont
  {{Zatsepin}}}]{ATIC2006}%
  \BibitemOpen
  \bibfield  {author} {\bibinfo {author} {\bibfnamefont {A.~D.}\ \bibnamefont
  {{Panov}}}, \bibinfo {author} {\bibfnamefont {J.~H.}\ \bibnamefont
  {{Adams}}}, \bibinfo {author} {\bibfnamefont {H.~S.}\ \bibnamefont {{Ahn}}},
  \bibinfo {author} {\bibfnamefont {G.~L.}\ \bibnamefont {{Bashindzhagyan}}},
  \bibinfo {author} {\bibfnamefont {K.~E.}\ \bibnamefont {{Batkov}}}, \bibinfo
  {author} {\bibfnamefont {J.}~\bibnamefont {{Chang}}}, \bibinfo {author}
  {\bibfnamefont {M.}~\bibnamefont {{Christl}}}, \bibinfo {author}
  {\bibfnamefont {A.~R.}\ \bibnamefont {{Fazely}}}, \bibinfo {author}
  {\bibfnamefont {O.}~\bibnamefont {{Ganel}}}, \bibinfo {author} {\bibfnamefont
  {R.~M.}\ \bibnamefont {{Gunashingha}}}, \bibinfo {author} {\bibfnamefont
  {T.~G.}\ \bibnamefont {{Guzik}}}, \bibinfo {author} {\bibfnamefont
  {J.}~\bibnamefont {{Isbert}}}, \bibinfo {author} {\bibfnamefont {K.~C.}\
  \bibnamefont {{Kim}}}, \bibinfo {author} {\bibfnamefont {E.~N.}\ \bibnamefont
  {{Kouznetsov}}}, \bibinfo {author} {\bibfnamefont {M.~I.}\ \bibnamefont
  {{Panasyuk}}}, \bibinfo {author} {\bibfnamefont {W.~K.~H.}\ \bibnamefont
  {{Schmidt}}}, \bibinfo {author} {\bibfnamefont {E.~S.}\ \bibnamefont
  {{Seo}}}, \bibinfo {author} {\bibfnamefont {N.~V.}\ \bibnamefont
  {{Sokolskaya}}}, \bibinfo {author} {\bibfnamefont {J.~W.}\ \bibnamefont
  {{Watts}}}, \bibinfo {author} {\bibfnamefont {J.~P.}\ \bibnamefont
  {{Wefel}}}, \bibinfo {author} {\bibfnamefont {J.}~\bibnamefont {{Wu}}}, \
  and\ \bibinfo {author} {\bibfnamefont {V.~I.}\ \bibnamefont {{Zatsepin}}},\
  }\href@noop {} {\bibfield  {journal} {\bibinfo  {journal} {ArXiv Astrophysics
  e-prints}\ } (\bibinfo {year} {2006})},\ \Eprint
  {http://arxiv.org/abs/astro-ph/0612377} {astro-ph/0612377} \BibitemShut
  {NoStop}%
\bibitem [{\citenamefont {{Ahn}}\ \emph {et~al.}(2010)\citenamefont {{Ahn}},
  \citenamefont {{Allison}}, \citenamefont {{Bagliesi}}, \citenamefont
  {{Beatty}}, \citenamefont {{Bigongiari}}, \citenamefont {{Childers}},
  \citenamefont {{Conklin}}, \citenamefont {{Coutu}}, \citenamefont
  {{DuVernois}}, \citenamefont {{Ganel}}, \citenamefont {{Han}}, \citenamefont
  {{Jeon}}, \citenamefont {{Kim}}, \citenamefont {{Lee}}, \citenamefont
  {{Lutz}}, \citenamefont {{Maestro}}, \citenamefont {{Malinin}}, \citenamefont
  {{Marrocchesi}}, \citenamefont {{Minnick}}, \citenamefont {{Mognet}},
  \citenamefont {{Nam}}, \citenamefont {{Nam}}, \citenamefont {{Nutter}},
  \citenamefont {{Park}}, \citenamefont {{Park}}, \citenamefont {{Seo}},
  \citenamefont {{Sina}}, \citenamefont {{Wu}}, \citenamefont {{Yang}},
  \citenamefont {{Yoon}}, \citenamefont {{Zei}},\ and\ \citenamefont
  {{Zinn}}}]{CREAM2010}%
  \BibitemOpen
  \bibfield  {author} {\bibinfo {author} {\bibfnamefont {H.~S.}\ \bibnamefont
  {{Ahn}}}, \bibinfo {author} {\bibfnamefont {P.}~\bibnamefont {{Allison}}},
  \bibinfo {author} {\bibfnamefont {M.~G.}\ \bibnamefont {{Bagliesi}}},
  \bibinfo {author} {\bibfnamefont {J.~J.}\ \bibnamefont {{Beatty}}}, \bibinfo
  {author} {\bibfnamefont {G.}~\bibnamefont {{Bigongiari}}}, \bibinfo {author}
  {\bibfnamefont {J.~T.}\ \bibnamefont {{Childers}}}, \bibinfo {author}
  {\bibfnamefont {N.~B.}\ \bibnamefont {{Conklin}}}, \bibinfo {author}
  {\bibfnamefont {S.}~\bibnamefont {{Coutu}}}, \bibinfo {author} {\bibfnamefont
  {M.~A.}\ \bibnamefont {{DuVernois}}}, \bibinfo {author} {\bibfnamefont
  {O.}~\bibnamefont {{Ganel}}}, \bibinfo {author} {\bibfnamefont {J.~H.}\
  \bibnamefont {{Han}}}, \bibinfo {author} {\bibfnamefont {J.~A.}\ \bibnamefont
  {{Jeon}}}, \bibinfo {author} {\bibfnamefont {K.~C.}\ \bibnamefont {{Kim}}},
  \bibinfo {author} {\bibfnamefont {M.~H.}\ \bibnamefont {{Lee}}}, \bibinfo
  {author} {\bibfnamefont {L.}~\bibnamefont {{Lutz}}}, \bibinfo {author}
  {\bibfnamefont {P.}~\bibnamefont {{Maestro}}}, \bibinfo {author}
  {\bibfnamefont {A.}~\bibnamefont {{Malinin}}}, \bibinfo {author}
  {\bibfnamefont {P.~S.}\ \bibnamefont {{Marrocchesi}}}, \bibinfo {author}
  {\bibfnamefont {S.}~\bibnamefont {{Minnick}}}, \bibinfo {author}
  {\bibfnamefont {S.~I.}\ \bibnamefont {{Mognet}}}, \bibinfo {author}
  {\bibfnamefont {J.}~\bibnamefont {{Nam}}}, \bibinfo {author} {\bibfnamefont
  {S.}~\bibnamefont {{Nam}}}, \bibinfo {author} {\bibfnamefont {S.~L.}\
  \bibnamefont {{Nutter}}}, \bibinfo {author} {\bibfnamefont {I.~H.}\
  \bibnamefont {{Park}}}, \bibinfo {author} {\bibfnamefont {N.~H.}\
  \bibnamefont {{Park}}}, \bibinfo {author} {\bibfnamefont {E.~S.}\
  \bibnamefont {{Seo}}}, \bibinfo {author} {\bibfnamefont {R.}~\bibnamefont
  {{Sina}}}, \bibinfo {author} {\bibfnamefont {J.}~\bibnamefont {{Wu}}},
  \bibinfo {author} {\bibfnamefont {J.}~\bibnamefont {{Yang}}}, \bibinfo
  {author} {\bibfnamefont {Y.~S.}\ \bibnamefont {{Yoon}}}, \bibinfo {author}
  {\bibfnamefont {R.}~\bibnamefont {{Zei}}}, \ and\ \bibinfo {author}
  {\bibfnamefont {S.~Y.}\ \bibnamefont {{Zinn}}},\ }\href {\doibase
  10.1088/2041-8205/714/1/L89} {\bibfield  {journal} {\bibinfo  {journal}
  {\apjl}\ }\textbf {\bibinfo {volume} {714}},\ \bibinfo {pages} {L89}
  (\bibinfo {year} {2010})},\ \Eprint {http://arxiv.org/abs/1004.1123}
  {arXiv:1004.1123 [astro-ph.HE]} \BibitemShut {NoStop}%
\bibitem [{\citenamefont {{PAMELA
  collaboration}}(2011{\natexlab{a}})}]{PAMELA2011}%
  \BibitemOpen
  \bibfield  {author} {\bibinfo {author} {\bibnamefont {{PAMELA
  collaboration}}},\ }\href {\doibase 10.1126/science.1199172} {\bibfield
  {journal} {\bibinfo  {journal} {Science}\ }\textbf {\bibinfo {volume}
  {332}},\ \bibinfo {pages} {69} (\bibinfo {year} {2011}{\natexlab{a}})},\
  \Eprint {http://arxiv.org/abs/1103.4055} {arXiv:1103.4055 [astro-ph.HE]}
  \BibitemShut {NoStop}%
\bibitem [{\citenamefont {{AMS
  collaboration}}(2015{\natexlab{a}})}]{AMS02_proton}%
  \BibitemOpen
  \bibfield  {author} {\bibinfo {author} {\bibnamefont {{AMS collaboration}}},\
  }\href {\doibase 10.1103/PhysRevLett.114.171103} {\bibfield  {journal}
  {\bibinfo  {journal} {Physical Review Letters}\ }\textbf {\bibinfo {volume}
  {114}},\ \bibinfo {eid} {171103} (\bibinfo {year}
  {2015}{\natexlab{a}})}\BibitemShut {NoStop}%
\bibitem [{\citenamefont {{AMS
  collaboration}}(2015{\natexlab{b}})}]{AMS02_helium}%
  \BibitemOpen
  \bibfield  {author} {\bibinfo {author} {\bibnamefont {{AMS collaboration}}},\
  }\href {\doibase 10.1103/PhysRevLett.115.211101} {\bibfield  {journal}
  {\bibinfo  {journal} {Physical Review Letters}\ }\textbf {\bibinfo {volume}
  {115}},\ \bibinfo {eid} {211101} (\bibinfo {year}
  {2015}{\natexlab{b}})}\BibitemShut {NoStop}%
\bibitem [{\citenamefont {{PAMELA
  collaboration}}(2011{\natexlab{b}})}]{Adriani2011}%
  \BibitemOpen
  \bibfield  {author} {\bibinfo {author} {\bibnamefont {{PAMELA
  collaboration}}},\ }\href {\doibase 10.1126/science.1199172} {\bibfield
  {journal} {\bibinfo  {journal} {Science}\ }\textbf {\bibinfo {volume}
  {332}},\ \bibinfo {pages} {69} (\bibinfo {year} {2011}{\natexlab{b}})},\
  \Eprint {http://arxiv.org/abs/1103.4055} {arXiv:1103.4055 [astro-ph.HE]}
  \BibitemShut {NoStop}%
\bibitem [{\citenamefont {{Tan}}\ and\ \citenamefont {{Ng}}(1983)}]{Tan1983}%
  \BibitemOpen
  \bibfield  {author} {\bibinfo {author} {\bibfnamefont {L.~C.}\ \bibnamefont
  {{Tan}}}\ and\ \bibinfo {author} {\bibfnamefont {L.~K.}\ \bibnamefont
  {{Ng}}},\ }\href {\doibase 10.1088/0305-4616/9/10/015} {\bibfield  {journal}
  {\bibinfo  {journal} {Journal of Physics G Nuclear Physics}\ }\textbf
  {\bibinfo {volume} {9}},\ \bibinfo {pages} {1289} (\bibinfo {year}
  {1983})}\BibitemShut {NoStop}%
\bibitem [{\citenamefont {{Duperray}}\ \emph {et~al.}(2003)\citenamefont
  {{Duperray}}, \citenamefont {{Huang}}, \citenamefont {{Protasov}},\ and\
  \citenamefont {{Bu{\'e}nerd}}}]{Duperray2003}%
  \BibitemOpen
  \bibfield  {author} {\bibinfo {author} {\bibfnamefont {R.~P.}\ \bibnamefont
  {{Duperray}}}, \bibinfo {author} {\bibfnamefont {C.-Y.}\ \bibnamefont
  {{Huang}}}, \bibinfo {author} {\bibfnamefont {K.~V.}\ \bibnamefont
  {{Protasov}}}, \ and\ \bibinfo {author} {\bibfnamefont {M.}~\bibnamefont
  {{Bu{\'e}nerd}}},\ }\href {\doibase 10.1103/PhysRevD.68.094017} {\bibfield
  {journal} {\bibinfo  {journal} {\prd}\ }\textbf {\bibinfo {volume} {68}},\
  \bibinfo {eid} {094017} (\bibinfo {year} {2003})},\ \Eprint
  {http://arxiv.org/abs/astro-ph/0305274} {astro-ph/0305274} \BibitemShut
  {NoStop}%
\bibitem [{\citenamefont {{Kappl}}\ and\ \citenamefont
  {{Winkler}}(2014)}]{Kappl2014}%
  \BibitemOpen
  \bibfield  {author} {\bibinfo {author} {\bibfnamefont {R.}~\bibnamefont
  {{Kappl}}}\ and\ \bibinfo {author} {\bibfnamefont {M.~W.}\ \bibnamefont
  {{Winkler}}},\ }\href {\doibase 10.1088/1475-7516/2014/09/051} {\bibfield
  {journal} {\bibinfo  {journal} {\jcap}\ }\textbf {\bibinfo {volume} {9}},\
  \bibinfo {eid} {051} (\bibinfo {year} {2014})},\ \Eprint
  {http://arxiv.org/abs/1408.0299} {arXiv:1408.0299 [hep-ph]} \BibitemShut
  {NoStop}%
\bibitem [{\citenamefont {{di Mauro}}\ \emph {et~al.}(2014)\citenamefont {{di
  Mauro}}, \citenamefont {{Donato}}, \citenamefont {{Goudelis}},\ and\
  \citenamefont {{Serpico}}}]{diMauro2014}%
  \BibitemOpen
  \bibfield  {author} {\bibinfo {author} {\bibfnamefont {M.}~\bibnamefont {{di
  Mauro}}}, \bibinfo {author} {\bibfnamefont {F.}~\bibnamefont {{Donato}}},
  \bibinfo {author} {\bibfnamefont {A.}~\bibnamefont {{Goudelis}}}, \ and\
  \bibinfo {author} {\bibfnamefont {P.~D.}\ \bibnamefont {{Serpico}}},\ }\href
  {\doibase 10.1103/PhysRevD.90.085017} {\bibfield  {journal} {\bibinfo
  {journal} {\prd}\ }\textbf {\bibinfo {volume} {90}},\ \bibinfo {eid} {085017}
  (\bibinfo {year} {2014})},\ \Eprint {http://arxiv.org/abs/1408.0288}
  {arXiv:1408.0288 [hep-ph]} \BibitemShut {NoStop}%
\bibitem [{\citenamefont {Lin}\ \emph {et~al.}(2015)\citenamefont {Lin},
  \citenamefont {Yuan},\ and\ \citenamefont {Bi}}]{Lin2015}%
  \BibitemOpen
  \bibfield  {author} {\bibinfo {author} {\bibfnamefont {S.-J.}\ \bibnamefont
  {Lin}}, \bibinfo {author} {\bibfnamefont {Q.}~\bibnamefont {Yuan}}, \ and\
  \bibinfo {author} {\bibfnamefont {X.-J.}\ \bibnamefont {Bi}},\ }\href
  {\doibase 10.1103/PhysRevD.91.063508} {\bibfield  {journal} {\bibinfo
  {journal} {Physical Review D}\ }\textbf {\bibinfo {volume} {91}},\ \bibinfo
  {pages} {063508} (\bibinfo {year} {2015})},\ \Eprint
  {http://arxiv.org/abs/1409.6248} {arXiv:1409.6248 [astro-ph.HE]} \BibitemShut
  {NoStop}%
\bibitem [{\citenamefont {{Delahaye}}\ \emph {et~al.}(2009)\citenamefont
  {{Delahaye}}, \citenamefont {{Lineros}}, \citenamefont {{Donato}},
  \citenamefont {{Fornengo}}, \citenamefont {{Lavalle}}, \citenamefont
  {{Salati}},\ and\ \citenamefont {{Taillet}}}]{Delahaye2009}%
  \BibitemOpen
  \bibfield  {author} {\bibinfo {author} {\bibfnamefont {T.}~\bibnamefont
  {{Delahaye}}}, \bibinfo {author} {\bibfnamefont {R.}~\bibnamefont
  {{Lineros}}}, \bibinfo {author} {\bibfnamefont {F.}~\bibnamefont {{Donato}}},
  \bibinfo {author} {\bibfnamefont {N.}~\bibnamefont {{Fornengo}}}, \bibinfo
  {author} {\bibfnamefont {J.}~\bibnamefont {{Lavalle}}}, \bibinfo {author}
  {\bibfnamefont {P.}~\bibnamefont {{Salati}}}, \ and\ \bibinfo {author}
  {\bibfnamefont {R.}~\bibnamefont {{Taillet}}},\ }\href {\doibase
  10.1051/0004-6361/200811130} {\bibfield  {journal} {\bibinfo  {journal}
  {\aap}\ }\textbf {\bibinfo {volume} {501}},\ \bibinfo {pages} {821} (\bibinfo
  {year} {2009})},\ \Eprint {http://arxiv.org/abs/0809.5268} {arXiv:0809.5268}
  \BibitemShut {NoStop}%
\bibitem [{\citenamefont {{Mori}}(2009)}]{Mori2009}%
  \BibitemOpen
  \bibfield  {author} {\bibinfo {author} {\bibfnamefont {M.}~\bibnamefont
  {{Mori}}},\ }\href {\doibase 10.1016/j.astropartphys.2009.03.004} {\bibfield
  {journal} {\bibinfo  {journal} {Astroparticle Physics}\ }\textbf {\bibinfo
  {volume} {31}},\ \bibinfo {pages} {341} (\bibinfo {year} {2009})},\ \Eprint
  {http://arxiv.org/abs/0903.3260} {arXiv:0903.3260 [astro-ph.HE]} \BibitemShut
  {NoStop}%
\bibitem [{\citenamefont {{Gleeson}}\ and\ \citenamefont
  {{Axford}}(1968)}]{Gleeson1968}%
  \BibitemOpen
  \bibfield  {author} {\bibinfo {author} {\bibfnamefont {L.~J.}\ \bibnamefont
  {{Gleeson}}}\ and\ \bibinfo {author} {\bibfnamefont {W.~I.}\ \bibnamefont
  {{Axford}}},\ }\href {\doibase 10.1086/149822} {\bibfield  {journal}
  {\bibinfo  {journal} {\apj}\ }\textbf {\bibinfo {volume} {154}},\ \bibinfo
  {pages} {1011} (\bibinfo {year} {1968})}\BibitemShut {NoStop}%
\bibitem [{\citenamefont {{Navarro}}\ \emph {et~al.}(2004)\citenamefont
  {{Navarro}}, \citenamefont {{Hayashi}}, \citenamefont {{Power}},
  \citenamefont {{Jenkins}}, \citenamefont {{Frenk}}, \citenamefont {{White}},
  \citenamefont {{Springel}}, \citenamefont {{Stadel}},\ and\ \citenamefont
  {{Quinn}}}]{Navarro2004}%
  \BibitemOpen
  \bibfield  {author} {\bibinfo {author} {\bibfnamefont {J.~F.}\ \bibnamefont
  {{Navarro}}}, \bibinfo {author} {\bibfnamefont {E.}~\bibnamefont
  {{Hayashi}}}, \bibinfo {author} {\bibfnamefont {C.}~\bibnamefont {{Power}}},
  \bibinfo {author} {\bibfnamefont {A.~R.}\ \bibnamefont {{Jenkins}}}, \bibinfo
  {author} {\bibfnamefont {C.~S.}\ \bibnamefont {{Frenk}}}, \bibinfo {author}
  {\bibfnamefont {S.~D.~M.}\ \bibnamefont {{White}}}, \bibinfo {author}
  {\bibfnamefont {V.}~\bibnamefont {{Springel}}}, \bibinfo {author}
  {\bibfnamefont {J.}~\bibnamefont {{Stadel}}}, \ and\ \bibinfo {author}
  {\bibfnamefont {T.~R.}\ \bibnamefont {{Quinn}}},\ }\href {\doibase
  10.1111/j.1365-2966.2004.07586.x} {\bibfield  {journal} {\bibinfo  {journal}
  {\mnras}\ }\textbf {\bibinfo {volume} {349}},\ \bibinfo {pages} {1039}
  (\bibinfo {year} {2004})},\ \Eprint {http://arxiv.org/abs/astro-ph/0311231}
  {astro-ph/0311231} \BibitemShut {NoStop}%
\bibitem [{\citenamefont {{Merritt}}\ \emph {et~al.}(2006)\citenamefont
  {{Merritt}}, \citenamefont {{Graham}}, \citenamefont {{Moore}}, \citenamefont
  {{Diemand}},\ and\ \citenamefont {{Terzi{\'c}}}}]{Merritt2006}%
  \BibitemOpen
  \bibfield  {author} {\bibinfo {author} {\bibfnamefont {D.}~\bibnamefont
  {{Merritt}}}, \bibinfo {author} {\bibfnamefont {A.~W.}\ \bibnamefont
  {{Graham}}}, \bibinfo {author} {\bibfnamefont {B.}~\bibnamefont {{Moore}}},
  \bibinfo {author} {\bibfnamefont {J.}~\bibnamefont {{Diemand}}}, \ and\
  \bibinfo {author} {\bibfnamefont {B.}~\bibnamefont {{Terzi{\'c}}}},\ }\href
  {\doibase 10.1086/508988} {\bibfield  {journal} {\bibinfo  {journal} {\aj}\
  }\textbf {\bibinfo {volume} {132}},\ \bibinfo {pages} {2685} (\bibinfo {year}
  {2006})},\ \Eprint {http://arxiv.org/abs/astro-ph/0509417} {astro-ph/0509417}
  \BibitemShut {NoStop}%
\bibitem [{\citenamefont {{Einasto}}(2009)}]{Einasto2009}%
  \BibitemOpen
  \bibfield  {author} {\bibinfo {author} {\bibfnamefont {J.}~\bibnamefont
  {{Einasto}}},\ }\href@noop {} {\bibfield  {journal} {\bibinfo  {journal}
  {ArXiv e-prints}\ } (\bibinfo {year} {2009})},\ \Eprint
  {http://arxiv.org/abs/0901.0632} {arXiv:0901.0632 [astro-ph.CO]} \BibitemShut
  {NoStop}%
\bibitem [{\citenamefont {{Navarro}}\ \emph {et~al.}(2010)\citenamefont
  {{Navarro}}, \citenamefont {{Ludlow}}, \citenamefont {{Springel}},
  \citenamefont {{Wang}}, \citenamefont {{Vogelsberger}}, \citenamefont
  {{White}}, \citenamefont {{Jenkins}}, \citenamefont {{Frenk}},\ and\
  \citenamefont {{Helmi}}}]{Navarro2010}%
  \BibitemOpen
  \bibfield  {author} {\bibinfo {author} {\bibfnamefont {J.~F.}\ \bibnamefont
  {{Navarro}}}, \bibinfo {author} {\bibfnamefont {A.}~\bibnamefont {{Ludlow}}},
  \bibinfo {author} {\bibfnamefont {V.}~\bibnamefont {{Springel}}}, \bibinfo
  {author} {\bibfnamefont {J.}~\bibnamefont {{Wang}}}, \bibinfo {author}
  {\bibfnamefont {M.}~\bibnamefont {{Vogelsberger}}}, \bibinfo {author}
  {\bibfnamefont {S.~D.~M.}\ \bibnamefont {{White}}}, \bibinfo {author}
  {\bibfnamefont {A.}~\bibnamefont {{Jenkins}}}, \bibinfo {author}
  {\bibfnamefont {C.~S.}\ \bibnamefont {{Frenk}}}, \ and\ \bibinfo {author}
  {\bibfnamefont {A.}~\bibnamefont {{Helmi}}},\ }\href {\doibase
  10.1111/j.1365-2966.2009.15878.x} {\bibfield  {journal} {\bibinfo  {journal}
  {\mnras}\ }\textbf {\bibinfo {volume} {402}},\ \bibinfo {pages} {21}
  (\bibinfo {year} {2010})},\ \Eprint {http://arxiv.org/abs/0810.1522}
  {arXiv:0810.1522} \BibitemShut {NoStop}%
\bibitem [{\citenamefont {{Catena}}\ and\ \citenamefont
  {{Ullio}}(2010)}]{Catena2010}%
  \BibitemOpen
  \bibfield  {author} {\bibinfo {author} {\bibfnamefont {R.}~\bibnamefont
  {{Catena}}}\ and\ \bibinfo {author} {\bibfnamefont {P.}~\bibnamefont
  {{Ullio}}},\ }\href {\doibase 10.1088/1475-7516/2010/08/004} {\bibfield
  {journal} {\bibinfo  {journal} {\jcap}\ }\textbf {\bibinfo {volume} {8}},\
  \bibinfo {eid} {004} (\bibinfo {year} {2010})},\ \Eprint
  {http://arxiv.org/abs/0907.0018} {arXiv:0907.0018} \BibitemShut {NoStop}%
\bibitem [{\citenamefont {{Weber}}\ and\ \citenamefont {{de
  Boer}}(2010)}]{Weber2010}%
  \BibitemOpen
  \bibfield  {author} {\bibinfo {author} {\bibfnamefont {M.}~\bibnamefont
  {{Weber}}}\ and\ \bibinfo {author} {\bibfnamefont {W.}~\bibnamefont {{de
  Boer}}},\ }\href {\doibase 10.1051/0004-6361/200913381} {\bibfield  {journal}
  {\bibinfo  {journal} {\aap}\ }\textbf {\bibinfo {volume} {509}},\ \bibinfo
  {eid} {A25} (\bibinfo {year} {2010})},\ \Eprint
  {http://arxiv.org/abs/0910.4272} {arXiv:0910.4272 [astro-ph.CO]} \BibitemShut
  {NoStop}%
\bibitem [{\citenamefont {{Salucci}}\ \emph {et~al.}(2010)\citenamefont
  {{Salucci}}, \citenamefont {{Nesti}}, \citenamefont {{Gentile}},\ and\
  \citenamefont {{Frigerio Martins}}}]{Salucci2010}%
  \BibitemOpen
  \bibfield  {author} {\bibinfo {author} {\bibfnamefont {P.}~\bibnamefont
  {{Salucci}}}, \bibinfo {author} {\bibfnamefont {F.}~\bibnamefont {{Nesti}}},
  \bibinfo {author} {\bibfnamefont {G.}~\bibnamefont {{Gentile}}}, \ and\
  \bibinfo {author} {\bibfnamefont {C.}~\bibnamefont {{Frigerio Martins}}},\
  }\href {\doibase 10.1051/0004-6361/201014385} {\bibfield  {journal} {\bibinfo
   {journal} {\aap}\ }\textbf {\bibinfo {volume} {523}},\ \bibinfo {eid} {A83}
  (\bibinfo {year} {2010})},\ \Eprint {http://arxiv.org/abs/1003.3101}
  {arXiv:1003.3101} \BibitemShut {NoStop}%
\bibitem [{\citenamefont {{Pato}}\ \emph {et~al.}(2010)\citenamefont {{Pato}},
  \citenamefont {{Agertz}}, \citenamefont {{Bertone}}, \citenamefont
  {{Moore}},\ and\ \citenamefont {{Teyssier}}}]{Pato2010}%
  \BibitemOpen
  \bibfield  {author} {\bibinfo {author} {\bibfnamefont {M.}~\bibnamefont
  {{Pato}}}, \bibinfo {author} {\bibfnamefont {O.}~\bibnamefont {{Agertz}}},
  \bibinfo {author} {\bibfnamefont {G.}~\bibnamefont {{Bertone}}}, \bibinfo
  {author} {\bibfnamefont {B.}~\bibnamefont {{Moore}}}, \ and\ \bibinfo
  {author} {\bibfnamefont {R.}~\bibnamefont {{Teyssier}}},\ }\href {\doibase
  10.1103/PhysRevD.82.023531} {\bibfield  {journal} {\bibinfo  {journal}
  {\prd}\ }\textbf {\bibinfo {volume} {82}},\ \bibinfo {eid} {023531} (\bibinfo
  {year} {2010})},\ \Eprint {http://arxiv.org/abs/1006.1322} {arXiv:1006.1322
  [astro-ph.HE]} \BibitemShut {NoStop}%
\bibitem [{\citenamefont {{Iocco}}\ \emph {et~al.}(2011)\citenamefont
  {{Iocco}}, \citenamefont {{Pato}}, \citenamefont {{Bertone}},\ and\
  \citenamefont {{Jetzer}}}]{Iocco2011}%
  \BibitemOpen
  \bibfield  {author} {\bibinfo {author} {\bibfnamefont {F.}~\bibnamefont
  {{Iocco}}}, \bibinfo {author} {\bibfnamefont {M.}~\bibnamefont {{Pato}}},
  \bibinfo {author} {\bibfnamefont {G.}~\bibnamefont {{Bertone}}}, \ and\
  \bibinfo {author} {\bibfnamefont {P.}~\bibnamefont {{Jetzer}}},\ }\href
  {\doibase 10.1088/1475-7516/2011/11/029} {\bibfield  {journal} {\bibinfo
  {journal} {\jcap}\ }\textbf {\bibinfo {volume} {11}},\ \bibinfo {eid} {029}
  (\bibinfo {year} {2011})},\ \Eprint {http://arxiv.org/abs/1107.5810}
  {arXiv:1107.5810 [astro-ph.GA]} \BibitemShut {NoStop}%
\bibitem [{\citenamefont {{AMS collaboration}}(2016)}]{AMS02_pbar_proton}%
  \BibitemOpen
  \bibfield  {author} {\bibinfo {author} {\bibnamefont {{AMS collaboration}}},\
  }\href {\doibase 10.1103/PhysRevLett.117.091103} {\bibfield  {journal}
  {\bibinfo  {journal} {Physical Review Letters}\ }\textbf {\bibinfo {volume}
  {117}},\ \bibinfo {eid} {091103} (\bibinfo {year} {2016})}\BibitemShut
  {NoStop}%
\bibitem [{\citenamefont {{Niu}}\ \emph
  {et~al.}(2018{\natexlab{b}})\citenamefont {{Niu}}, \citenamefont {{Li}},\
  and\ \citenamefont {{Xue}}}]{Niu2018}%
  \BibitemOpen
  \bibfield  {author} {\bibinfo {author} {\bibfnamefont {J.-S.}\ \bibnamefont
  {{Niu}}}, \bibinfo {author} {\bibfnamefont {T.}~\bibnamefont {{Li}}}, \ and\
  \bibinfo {author} {\bibfnamefont {H.-F.}\ \bibnamefont {{Xue}}},\ }\href@noop
  {} {\bibfield  {journal} {\bibinfo  {journal} {ArXiv e-prints}\ } (\bibinfo
  {year} {2018}{\natexlab{b}})},\ \Eprint {http://arxiv.org/abs/1810.09301}
  {arXiv:1810.09301 [astro-ph.HE]} \BibitemShut {NoStop}%
\bibitem [{\citenamefont {collaboration}(2017{\natexlab{b}})}]{Fermi2017}%
  \BibitemOpen
  \bibfield  {author} {\bibinfo {author} {\bibfnamefont {F.-L.}\ \bibnamefont
  {collaboration}} (\bibinfo {collaboration} {The Fermi-LAT Collaboration}),\
  }\href {\doibase 10.1103/PhysRevD.95.082007} {\bibfield  {journal} {\bibinfo
  {journal} {Phys. Rev. D}\ }\textbf {\bibinfo {volume} {95}},\ \bibinfo
  {pages} {082007} (\bibinfo {year} {2017}{\natexlab{b}})}\BibitemShut
  {NoStop}%
\bibitem [{\citenamefont {{Profumo}}(2012)}]{Profumo2012}%
  \BibitemOpen
  \bibfield  {author} {\bibinfo {author} {\bibfnamefont {S.}~\bibnamefont
  {{Profumo}}},\ }\href {\doibase 10.2478/s11534-011-0099-z} {\bibfield
  {journal} {\bibinfo  {journal} {Central European Journal of Physics}\
  }\textbf {\bibinfo {volume} {10}},\ \bibinfo {pages} {1} (\bibinfo {year}
  {2012})},\ \Eprint {http://arxiv.org/abs/0812.4457} {arXiv:0812.4457}
  \BibitemShut {NoStop}%
\bibitem [{\citenamefont {{Reynolds}}(1988)}]{Reynolds1988}%
  \BibitemOpen
  \bibfield  {author} {\bibinfo {author} {\bibfnamefont {S.~P.}\ \bibnamefont
  {{Reynolds}}},\ }\href {\doibase 10.1086/166243} {\bibfield  {journal}
  {\bibinfo  {journal} {\apj}\ }\textbf {\bibinfo {volume} {327}},\ \bibinfo
  {pages} {853} (\bibinfo {year} {1988})}\BibitemShut {NoStop}%
\bibitem [{\citenamefont {{Thompson}}\ \emph {et~al.}(1994)\citenamefont
  {{Thompson}}, \citenamefont {{Arzoumanian}}, \citenamefont {{Bertsch}},
  \citenamefont {{Brazier}}, \citenamefont {{Chiang}}, \citenamefont
  {{D'Amico}}, \citenamefont {{Dingus}}, \citenamefont {{Esposito}},
  \citenamefont {{Fierro}}, \citenamefont {{Fichtel}}, \citenamefont
  {{Hartman}}, \citenamefont {{Hunter}}, \citenamefont {{Johnston}},
  \citenamefont {{Kanbach}}, \citenamefont {{Kaspi}}, \citenamefont
  {{Kniffen}}, \citenamefont {{Lin}}, \citenamefont {{Lyne}}, \citenamefont
  {{Manchester}}, \citenamefont {{Mattox}}, \citenamefont
  {{Mayer-Hasselwander}}, \citenamefont {{Michelson}}, \citenamefont {{von
  Montigny}}, \citenamefont {{Nel}}, \citenamefont {{Nice}}, \citenamefont
  {{Nolan}}, \citenamefont {{Ramanamurthy}}, \citenamefont {{Shemar}},
  \citenamefont {{Schneid}}, \citenamefont {{Sreekumar}},\ and\ \citenamefont
  {{Taylor}}}]{Thompson1994}%
  \BibitemOpen
  \bibfield  {author} {\bibinfo {author} {\bibfnamefont {D.~J.}\ \bibnamefont
  {{Thompson}}}, \bibinfo {author} {\bibfnamefont {Z.}~\bibnamefont
  {{Arzoumanian}}}, \bibinfo {author} {\bibfnamefont {D.~L.}\ \bibnamefont
  {{Bertsch}}}, \bibinfo {author} {\bibfnamefont {K.~T.~S.}\ \bibnamefont
  {{Brazier}}}, \bibinfo {author} {\bibfnamefont {J.}~\bibnamefont {{Chiang}}},
  \bibinfo {author} {\bibfnamefont {N.}~\bibnamefont {{D'Amico}}}, \bibinfo
  {author} {\bibfnamefont {B.~L.}\ \bibnamefont {{Dingus}}}, \bibinfo {author}
  {\bibfnamefont {J.~A.}\ \bibnamefont {{Esposito}}}, \bibinfo {author}
  {\bibfnamefont {J.~M.}\ \bibnamefont {{Fierro}}}, \bibinfo {author}
  {\bibfnamefont {C.~E.}\ \bibnamefont {{Fichtel}}}, \bibinfo {author}
  {\bibfnamefont {R.~C.}\ \bibnamefont {{Hartman}}}, \bibinfo {author}
  {\bibfnamefont {S.~D.}\ \bibnamefont {{Hunter}}}, \bibinfo {author}
  {\bibfnamefont {S.}~\bibnamefont {{Johnston}}}, \bibinfo {author}
  {\bibfnamefont {G.}~\bibnamefont {{Kanbach}}}, \bibinfo {author}
  {\bibfnamefont {V.~M.}\ \bibnamefont {{Kaspi}}}, \bibinfo {author}
  {\bibfnamefont {D.~A.}\ \bibnamefont {{Kniffen}}}, \bibinfo {author}
  {\bibfnamefont {Y.~C.}\ \bibnamefont {{Lin}}}, \bibinfo {author}
  {\bibfnamefont {A.~G.}\ \bibnamefont {{Lyne}}}, \bibinfo {author}
  {\bibfnamefont {R.~N.}\ \bibnamefont {{Manchester}}}, \bibinfo {author}
  {\bibfnamefont {J.~R.}\ \bibnamefont {{Mattox}}}, \bibinfo {author}
  {\bibfnamefont {H.~A.}\ \bibnamefont {{Mayer-Hasselwander}}}, \bibinfo
  {author} {\bibfnamefont {P.~F.}\ \bibnamefont {{Michelson}}}, \bibinfo
  {author} {\bibfnamefont {C.}~\bibnamefont {{von Montigny}}}, \bibinfo
  {author} {\bibfnamefont {H.~I.}\ \bibnamefont {{Nel}}}, \bibinfo {author}
  {\bibfnamefont {D.~J.}\ \bibnamefont {{Nice}}}, \bibinfo {author}
  {\bibfnamefont {P.~L.}\ \bibnamefont {{Nolan}}}, \bibinfo {author}
  {\bibfnamefont {P.~V.}\ \bibnamefont {{Ramanamurthy}}}, \bibinfo {author}
  {\bibfnamefont {S.~L.}\ \bibnamefont {{Shemar}}}, \bibinfo {author}
  {\bibfnamefont {E.~J.}\ \bibnamefont {{Schneid}}}, \bibinfo {author}
  {\bibfnamefont {P.}~\bibnamefont {{Sreekumar}}}, \ and\ \bibinfo {author}
  {\bibfnamefont {J.~H.}\ \bibnamefont {{Taylor}}},\ }\href {\doibase
  10.1086/174895} {\bibfield  {journal} {\bibinfo  {journal} {\apj}\ }\textbf
  {\bibinfo {volume} {436}},\ \bibinfo {pages} {229} (\bibinfo {year}
  {1994})}\BibitemShut {NoStop}%
\bibitem [{\citenamefont {{Fierro}}\ \emph {et~al.}(1995)\citenamefont
  {{Fierro}}, \citenamefont {{Arzoumanian}}, \citenamefont {{Bailes}},
  \citenamefont {{Bell}}, \citenamefont {{Bertsch}}, \citenamefont {{Brazier}},
  \citenamefont {{Chiang}}, \citenamefont {{D'Amico}}, \citenamefont
  {{Dingus}}, \citenamefont {{Esposito}}, \citenamefont {{Fichtel}},
  \citenamefont {{Hartman}}, \citenamefont {{Hunter}}, \citenamefont
  {{Johnston}}, \citenamefont {{Kanbach}}, \citenamefont {{Kaspi}},
  \citenamefont {{Kniffen}}, \citenamefont {{Lin}}, \citenamefont {{Lyne}},
  \citenamefont {{Manchester}}, \citenamefont {{Mattox}}, \citenamefont
  {{Mayer-Hasselwander}}, \citenamefont {{Michelson}}, \citenamefont {{von
  Montigny}}, \citenamefont {{Nel}}, \citenamefont {{Nice}}, \citenamefont
  {{Nolan}}, \citenamefont {{Schneid}}, \citenamefont {{Shriver}},
  \citenamefont {{Sreekumar}}, \citenamefont {{Taylor}}, \citenamefont
  {{Thompson}},\ and\ \citenamefont {{Willis}}}]{Fierro1995}%
  \BibitemOpen
  \bibfield  {author} {\bibinfo {author} {\bibfnamefont {J.~M.}\ \bibnamefont
  {{Fierro}}}, \bibinfo {author} {\bibfnamefont {Z.}~\bibnamefont
  {{Arzoumanian}}}, \bibinfo {author} {\bibfnamefont {M.}~\bibnamefont
  {{Bailes}}}, \bibinfo {author} {\bibfnamefont {J.~F.}\ \bibnamefont
  {{Bell}}}, \bibinfo {author} {\bibfnamefont {D.~L.}\ \bibnamefont
  {{Bertsch}}}, \bibinfo {author} {\bibfnamefont {K.~T.~S.}\ \bibnamefont
  {{Brazier}}}, \bibinfo {author} {\bibfnamefont {J.}~\bibnamefont {{Chiang}}},
  \bibinfo {author} {\bibfnamefont {N.}~\bibnamefont {{D'Amico}}}, \bibinfo
  {author} {\bibfnamefont {B.~L.}\ \bibnamefont {{Dingus}}}, \bibinfo {author}
  {\bibfnamefont {J.~A.}\ \bibnamefont {{Esposito}}}, \bibinfo {author}
  {\bibfnamefont {C.~E.}\ \bibnamefont {{Fichtel}}}, \bibinfo {author}
  {\bibfnamefont {R.~C.}\ \bibnamefont {{Hartman}}}, \bibinfo {author}
  {\bibfnamefont {S.~D.}\ \bibnamefont {{Hunter}}}, \bibinfo {author}
  {\bibfnamefont {S.}~\bibnamefont {{Johnston}}}, \bibinfo {author}
  {\bibfnamefont {G.}~\bibnamefont {{Kanbach}}}, \bibinfo {author}
  {\bibfnamefont {V.~M.}\ \bibnamefont {{Kaspi}}}, \bibinfo {author}
  {\bibfnamefont {D.~A.}\ \bibnamefont {{Kniffen}}}, \bibinfo {author}
  {\bibfnamefont {Y.~C.}\ \bibnamefont {{Lin}}}, \bibinfo {author}
  {\bibfnamefont {A.~G.}\ \bibnamefont {{Lyne}}}, \bibinfo {author}
  {\bibfnamefont {R.~N.}\ \bibnamefont {{Manchester}}}, \bibinfo {author}
  {\bibfnamefont {J.~R.}\ \bibnamefont {{Mattox}}}, \bibinfo {author}
  {\bibfnamefont {H.~A.}\ \bibnamefont {{Mayer-Hasselwander}}}, \bibinfo
  {author} {\bibfnamefont {P.~F.}\ \bibnamefont {{Michelson}}}, \bibinfo
  {author} {\bibfnamefont {C.}~\bibnamefont {{von Montigny}}}, \bibinfo
  {author} {\bibfnamefont {H.~I.}\ \bibnamefont {{Nel}}}, \bibinfo {author}
  {\bibfnamefont {D.}~\bibnamefont {{Nice}}}, \bibinfo {author} {\bibfnamefont
  {P.~L.}\ \bibnamefont {{Nolan}}}, \bibinfo {author} {\bibfnamefont {E.~J.}\
  \bibnamefont {{Schneid}}}, \bibinfo {author} {\bibfnamefont {S.~K.}\
  \bibnamefont {{Shriver}}}, \bibinfo {author} {\bibfnamefont {P.}~\bibnamefont
  {{Sreekumar}}}, \bibinfo {author} {\bibfnamefont {J.~H.}\ \bibnamefont
  {{Taylor}}}, \bibinfo {author} {\bibfnamefont {D.~J.}\ \bibnamefont
  {{Thompson}}}, \ and\ \bibinfo {author} {\bibfnamefont {T.~D.}\ \bibnamefont
  {{Willis}}},\ }\href {\doibase 10.1086/175919} {\bibfield  {journal}
  {\bibinfo  {journal} {\apj}\ }\textbf {\bibinfo {volume} {447}},\ \bibinfo
  {pages} {807} (\bibinfo {year} {1995})}\BibitemShut {NoStop}%
\bibitem [{\citenamefont {{Jungman}}\ \emph {et~al.}(1996)\citenamefont
  {{Jungman}}, \citenamefont {{Kamionkowski}},\ and\ \citenamefont
  {{Griest}}}]{Jungman1996}%
  \BibitemOpen
  \bibfield  {author} {\bibinfo {author} {\bibfnamefont {G.}~\bibnamefont
  {{Jungman}}}, \bibinfo {author} {\bibfnamefont {M.}~\bibnamefont
  {{Kamionkowski}}}, \ and\ \bibinfo {author} {\bibfnamefont {K.}~\bibnamefont
  {{Griest}}},\ }\href {\doibase 10.1016/0370-1573(95)00058-5} {\bibfield
  {journal} {\bibinfo  {journal} {\physrep}\ }\textbf {\bibinfo {volume}
  {267}},\ \bibinfo {pages} {195} (\bibinfo {year} {1996})},\ \Eprint
  {http://arxiv.org/abs/hep-ph/9506380} {hep-ph/9506380} \BibitemShut {NoStop}%
\bibitem [{\citenamefont {{Fermi-LAT collaboration}}(2011)}]{Ackermann:2011wa}%
  \BibitemOpen
  \bibfield  {author} {\bibinfo {author} {\bibnamefont {{Fermi-LAT
  collaboration}}},\ }\href {\doibase 10.1103/PhysRevLett.107.241302}
  {\bibfield  {journal} {\bibinfo  {journal} {Physical Review Letters}\
  }\textbf {\bibinfo {volume} {107}},\ \bibinfo {eid} {241302} (\bibinfo {year}
  {2011})},\ \Eprint {http://arxiv.org/abs/1108.3546} {arXiv:1108.3546
  [astro-ph.HE]} \BibitemShut {NoStop}%
\bibitem [{\citenamefont {{Geringer-Sameth}}\ and\ \citenamefont
  {{Koushiappas}}(2011)}]{GeringerSameth:2011iw}%
  \BibitemOpen
  \bibfield  {author} {\bibinfo {author} {\bibfnamefont {A.}~\bibnamefont
  {{Geringer-Sameth}}}\ and\ \bibinfo {author} {\bibfnamefont {S.~M.}\
  \bibnamefont {{Koushiappas}}},\ }\href@noop {} {\bibfield  {journal}
  {\bibinfo  {journal} {ArXiv e-prints}\ } (\bibinfo {year} {2011})},\ \Eprint
  {http://arxiv.org/abs/1108.2914} {arXiv:1108.2914 [astro-ph.CO]} \BibitemShut
  {NoStop}%
\bibitem [{\citenamefont {{Sming Tsai}}\ \emph {et~al.}(2012)\citenamefont
  {{Sming Tsai}}, \citenamefont {{Yuan}},\ and\ \citenamefont
  {{Huang}}}]{Tsai:2012cs}%
  \BibitemOpen
  \bibfield  {author} {\bibinfo {author} {\bibfnamefont {Y.-L.}\ \bibnamefont
  {{Sming Tsai}}}, \bibinfo {author} {\bibfnamefont {Q.}~\bibnamefont
  {{Yuan}}}, \ and\ \bibinfo {author} {\bibfnamefont {X.}~\bibnamefont
  {{Huang}}},\ }\href@noop {} {\bibfield  {journal} {\bibinfo  {journal} {ArXiv
  e-prints}\ } (\bibinfo {year} {2012})},\ \Eprint
  {http://arxiv.org/abs/1212.3990} {arXiv:1212.3990 [astro-ph.HE]} \BibitemShut
  {NoStop}%
\bibitem [{\citenamefont {{Fermi-LAT
  collaboration}}(2015)}]{Ackermann:2015zua}%
  \BibitemOpen
  \bibfield  {author} {\bibinfo {author} {\bibnamefont {{Fermi-LAT
  collaboration}}},\ }\href@noop {} {\bibfield  {journal} {\bibinfo  {journal}
  {ArXiv e-prints}\ } (\bibinfo {year} {2015})},\ \Eprint
  {http://arxiv.org/abs/1503.02641} {arXiv:1503.02641 [astro-ph.HE]}
  \BibitemShut {NoStop}%
\bibitem [{\citenamefont {{Li}}\ \emph {et~al.}(2015)\citenamefont {{Li}},
  \citenamefont {{Liang}}, \citenamefont {{Duan}}, \citenamefont {{Shen}},
  \citenamefont {{Huang}}, \citenamefont {{Li}}, \citenamefont {{Fan}},
  \citenamefont {{Liao}}, \citenamefont {{Feng}},\ and\ \citenamefont
  {{Chang}}}]{Li:2015kag}%
  \BibitemOpen
  \bibfield  {author} {\bibinfo {author} {\bibfnamefont {S.}~\bibnamefont
  {{Li}}}, \bibinfo {author} {\bibfnamefont {Y.-F.}\ \bibnamefont {{Liang}}},
  \bibinfo {author} {\bibfnamefont {K.-K.}\ \bibnamefont {{Duan}}}, \bibinfo
  {author} {\bibfnamefont {Z.-Q.}\ \bibnamefont {{Shen}}}, \bibinfo {author}
  {\bibfnamefont {X.}~\bibnamefont {{Huang}}}, \bibinfo {author} {\bibfnamefont
  {X.}~\bibnamefont {{Li}}}, \bibinfo {author} {\bibfnamefont {Y.-Z.}\
  \bibnamefont {{Fan}}}, \bibinfo {author} {\bibfnamefont {N.-H.}\ \bibnamefont
  {{Liao}}}, \bibinfo {author} {\bibfnamefont {L.}~\bibnamefont {{Feng}}}, \
  and\ \bibinfo {author} {\bibfnamefont {J.}~\bibnamefont {{Chang}}},\
  }\href@noop {} {\bibfield  {journal} {\bibinfo  {journal} {ArXiv e-prints}\ }
  (\bibinfo {year} {2015})},\ \Eprint {http://arxiv.org/abs/1511.09252}
  {arXiv:1511.09252 [astro-ph.HE]} \BibitemShut {NoStop}%
\bibitem [{\citenamefont {{Profumo}}\ \emph {et~al.}(2017)\citenamefont
  {{Profumo}}, \citenamefont {{Queiroz}}, \citenamefont {{Silk}},\ and\
  \citenamefont {{Siqueira}}}]{Profumo:2017obk}%
  \BibitemOpen
  \bibfield  {author} {\bibinfo {author} {\bibfnamefont {S.}~\bibnamefont
  {{Profumo}}}, \bibinfo {author} {\bibfnamefont {F.~S.}\ \bibnamefont
  {{Queiroz}}}, \bibinfo {author} {\bibfnamefont {J.}~\bibnamefont {{Silk}}}, \
  and\ \bibinfo {author} {\bibfnamefont {C.}~\bibnamefont {{Siqueira}}},\
  }\href@noop {} {\bibfield  {journal} {\bibinfo  {journal} {ArXiv e-prints}\ }
  (\bibinfo {year} {2017})},\ \Eprint {http://arxiv.org/abs/1711.03133}
  {arXiv:1711.03133 [hep-ph]} \BibitemShut {NoStop}%
\bibitem [{\citenamefont {{Planck collaboration}}(2015)}]{Ade:2015xua}%
  \BibitemOpen
  \bibfield  {author} {\bibinfo {author} {\bibnamefont {{Planck
  collaboration}}},\ }\href@noop {} {\bibfield  {journal} {\bibinfo  {journal}
  {ArXiv e-prints}\ } (\bibinfo {year} {2015})},\ \Eprint
  {http://arxiv.org/abs/1502.01589} {arXiv:1502.01589} \BibitemShut {NoStop}%
\bibitem [{\citenamefont {{Feldman}}\ \emph {et~al.}(2008)\citenamefont
  {{Feldman}}, \citenamefont {{Liu}},\ and\ \citenamefont
  {{Nath}}}]{Feldman:2008xs}%
  \BibitemOpen
  \bibfield  {author} {\bibinfo {author} {\bibfnamefont {D.}~\bibnamefont
  {{Feldman}}}, \bibinfo {author} {\bibfnamefont {Z.}~\bibnamefont {{Liu}}}, \
  and\ \bibinfo {author} {\bibfnamefont {P.}~\bibnamefont {{Nath}}},\
  }\href@noop {} {\bibfield  {journal} {\bibinfo  {journal} {ArXiv e-prints}\ }
  (\bibinfo {year} {2008})},\ \Eprint {http://arxiv.org/abs/0810.5762}
  {arXiv:0810.5762 [hep-ph]} \BibitemShut {NoStop}%
\bibitem [{\citenamefont {{Ibe}}\ \emph {et~al.}(2008)\citenamefont {{Ibe}},
  \citenamefont {{Murayama}},\ and\ \citenamefont {{Yanagida}}}]{Ibe:2008ye}%
  \BibitemOpen
  \bibfield  {author} {\bibinfo {author} {\bibfnamefont {M.}~\bibnamefont
  {{Ibe}}}, \bibinfo {author} {\bibfnamefont {H.}~\bibnamefont {{Murayama}}}, \
  and\ \bibinfo {author} {\bibfnamefont {T.~T.}\ \bibnamefont {{Yanagida}}},\
  }\href@noop {} {\bibfield  {journal} {\bibinfo  {journal} {ArXiv e-prints}\ }
  (\bibinfo {year} {2008})},\ \Eprint {http://arxiv.org/abs/0812.0072}
  {arXiv:0812.0072 [hep-ph]} \BibitemShut {NoStop}%
\bibitem [{\citenamefont {{Guo}}\ and\ \citenamefont
  {{Wu}}(2009)}]{Guo:2009aj}%
  \BibitemOpen
  \bibfield  {author} {\bibinfo {author} {\bibfnamefont {W.-L.}\ \bibnamefont
  {{Guo}}}\ and\ \bibinfo {author} {\bibfnamefont {Y.-L.}\ \bibnamefont
  {{Wu}}},\ }\href@noop {} {\bibfield  {journal} {\bibinfo  {journal} {ArXiv
  e-prints}\ } (\bibinfo {year} {2009})},\ \Eprint
  {http://arxiv.org/abs/0901.1450} {arXiv:0901.1450 [hep-ph]} \BibitemShut
  {NoStop}%
\bibitem [{\citenamefont {{Bi}}\ \emph {et~al.}(2009)\citenamefont {{Bi}},
  \citenamefont {{He}},\ and\ \citenamefont {{Yuan}}}]{Bi:2009uj}%
  \BibitemOpen
  \bibfield  {author} {\bibinfo {author} {\bibfnamefont {X.-J.}\ \bibnamefont
  {{Bi}}}, \bibinfo {author} {\bibfnamefont {X.-G.}\ \bibnamefont {{He}}}, \
  and\ \bibinfo {author} {\bibfnamefont {Q.}~\bibnamefont {{Yuan}}},\
  }\href@noop {} {\bibfield  {journal} {\bibinfo  {journal} {ArXiv e-prints}\ }
  (\bibinfo {year} {2009})},\ \Eprint {http://arxiv.org/abs/0903.0122}
  {arXiv:0903.0122 [hep-ph]} \BibitemShut {NoStop}%
\bibitem [{\citenamefont {{Bi}}\ \emph {et~al.}(2011)\citenamefont {{Bi}},
  \citenamefont {{Yin}},\ and\ \citenamefont {{Yuan}}}]{Bi:2011qm}%
  \BibitemOpen
  \bibfield  {author} {\bibinfo {author} {\bibfnamefont {X.-J.}\ \bibnamefont
  {{Bi}}}, \bibinfo {author} {\bibfnamefont {P.-F.}\ \bibnamefont {{Yin}}}, \
  and\ \bibinfo {author} {\bibfnamefont {Q.}~\bibnamefont {{Yuan}}},\
  }\href@noop {} {\bibfield  {journal} {\bibinfo  {journal} {ArXiv e-prints}\ }
  (\bibinfo {year} {2011})},\ \Eprint {http://arxiv.org/abs/1106.6027}
  {arXiv:1106.6027 [hep-ph]} \BibitemShut {NoStop}%
\bibitem [{\citenamefont {{Hisano}}\ \emph {et~al.}(2011)\citenamefont
  {{Hisano}}, \citenamefont {{Kawasaki}}, \citenamefont {{Kohri}},
  \citenamefont {{Moroi}}, \citenamefont {{Nakayama}},\ and\ \citenamefont
  {{Sekiguchi}}}]{Hisano:2011dc}%
  \BibitemOpen
  \bibfield  {author} {\bibinfo {author} {\bibfnamefont {J.}~\bibnamefont
  {{Hisano}}}, \bibinfo {author} {\bibfnamefont {M.}~\bibnamefont
  {{Kawasaki}}}, \bibinfo {author} {\bibfnamefont {K.}~\bibnamefont {{Kohri}}},
  \bibinfo {author} {\bibfnamefont {T.}~\bibnamefont {{Moroi}}}, \bibinfo
  {author} {\bibfnamefont {K.}~\bibnamefont {{Nakayama}}}, \ and\ \bibinfo
  {author} {\bibfnamefont {T.}~\bibnamefont {{Sekiguchi}}},\ }\href@noop {}
  {\bibfield  {journal} {\bibinfo  {journal} {ArXiv e-prints}\ } (\bibinfo
  {year} {2011})},\ \Eprint {http://arxiv.org/abs/1102.4658} {arXiv:1102.4658
  [hep-ph]} \BibitemShut {NoStop}%
\bibitem [{\citenamefont {{Bai}}\ \emph {et~al.}(2017)\citenamefont {{Bai}},
  \citenamefont {{Berger}},\ and\ \citenamefont {{Lu}}}]{Bai:2017fav}%
  \BibitemOpen
  \bibfield  {author} {\bibinfo {author} {\bibfnamefont {Y.}~\bibnamefont
  {{Bai}}}, \bibinfo {author} {\bibfnamefont {J.}~\bibnamefont {{Berger}}}, \
  and\ \bibinfo {author} {\bibfnamefont {S.}~\bibnamefont {{Lu}}},\ }\href@noop
  {} {\bibfield  {journal} {\bibinfo  {journal} {ArXiv e-prints}\ } (\bibinfo
  {year} {2017})},\ \Eprint {http://arxiv.org/abs/1706.09974} {arXiv:1706.09974
  [hep-ph]} \BibitemShut {NoStop}%
\bibitem [{\citenamefont {{Xiang}}\ \emph {et~al.}(2017)\citenamefont
  {{Xiang}}, \citenamefont {{Bi}}, \citenamefont {{Lin}},\ and\ \citenamefont
  {{Yin}}}]{Xiang:2017jou}%
  \BibitemOpen
  \bibfield  {author} {\bibinfo {author} {\bibfnamefont {Q.-F.}\ \bibnamefont
  {{Xiang}}}, \bibinfo {author} {\bibfnamefont {X.-J.}\ \bibnamefont {{Bi}}},
  \bibinfo {author} {\bibfnamefont {S.-J.}\ \bibnamefont {{Lin}}}, \ and\
  \bibinfo {author} {\bibfnamefont {P.-F.}\ \bibnamefont {{Yin}}},\ }\href@noop
  {} {\bibfield  {journal} {\bibinfo  {journal} {ArXiv e-prints}\ } (\bibinfo
  {year} {2017})},\ \Eprint {http://arxiv.org/abs/1707.09313} {arXiv:1707.09313
  [astro-ph.HE]} \BibitemShut {NoStop}%
\bibitem [{\citenamefont {{Niu}}\ \emph
  {et~al.}(2018{\natexlab{c}})\citenamefont {{Niu}}, \citenamefont {{Li}},
  \citenamefont {{Zong}}, \citenamefont {{Xue}},\ and\ \citenamefont
  {{Wang}}}]{Niu2017_DAV}%
  \BibitemOpen
  \bibfield  {author} {\bibinfo {author} {\bibfnamefont {J.-S.}\ \bibnamefont
  {{Niu}}}, \bibinfo {author} {\bibfnamefont {T.}~\bibnamefont {{Li}}},
  \bibinfo {author} {\bibfnamefont {W.}~\bibnamefont {{Zong}}}, \bibinfo
  {author} {\bibfnamefont {H.-F.}\ \bibnamefont {{Xue}}}, \ and\ \bibinfo
  {author} {\bibfnamefont {Y.}~\bibnamefont {{Wang}}},\ }\href {\doibase
  10.1103/PhysRevD.98.103023} {\bibfield  {journal} {\bibinfo  {journal}
  {\prd}\ }\textbf {\bibinfo {volume} {98}},\ \bibinfo {eid} {103023} (\bibinfo
  {year} {2018}{\natexlab{c}})},\ \Eprint {http://arxiv.org/abs/1709.08804}
  {arXiv:1709.08804 [astro-ph.HE]} \BibitemShut {NoStop}%
\bibitem [{\citenamefont {{Maurin}}\ \emph {et~al.}(2014)\citenamefont
  {{Maurin}}, \citenamefont {{Melot}},\ and\ \citenamefont
  {{Taillet}}}]{Maurin2014}%
  \BibitemOpen
  \bibfield  {author} {\bibinfo {author} {\bibfnamefont {D.}~\bibnamefont
  {{Maurin}}}, \bibinfo {author} {\bibfnamefont {F.}~\bibnamefont {{Melot}}}, \
  and\ \bibinfo {author} {\bibfnamefont {R.}~\bibnamefont {{Taillet}}},\ }\href
  {\doibase 10.1051/0004-6361/201321344} {\bibfield  {journal} {\bibinfo
  {journal} {\aap}\ }\textbf {\bibinfo {volume} {569}},\ \bibinfo {eid} {A32}
  (\bibinfo {year} {2014})},\ \Eprint {http://arxiv.org/abs/1302.5525}
  {arXiv:1302.5525 [astro-ph.HE]} \BibitemShut {NoStop}%
\end{thebibliography}

%

\clearpage
\section*{Appendix}

\begin{table*}[!htbp]
\begin{center}
\begin{tabular}{lllll}
  \hline\hline
ID  &Prior & Best-fit &Posterior mean and   &Posterior 95\%    \\
    &range&value  &Standard deviation & range  \\
\hline
$D_{0}\ (10^{28}\cm^{2}\s^{-1})\ ^{a}$
    &[1, 20]  &14.32  &14.24$\pm$0.21     &[13.55, 14.83]   \\

$\delta$
  &[0.1, 1.0] &0.322  &0.318$\pm$0.006     &[0.305, 0.336]     \\

$z_h\ (\kpc)$
  &[0.5, 30.0]  &25.88  &25.61$\pm$0.94     &[23.86, 27.35]     \\

$v_{A}\ (\km/\s)$
  &[0, 80]  &41.33  &40.82$\pm$0.64     &[39.04, 42.67]    \\

$N_{\p}\ ^{b}$
  &[1, 8] &4.47   &4.47$\pm$0.01    &[4.44, 4.49]  \\

$R_{\p1}\ (\GV)$
  &[1, 30]  &25.90  &25.60$\pm$0.45      &[24.62, 27.14]     \\
  
$R_{\p2}\ (\GV)$
  &[60, 1000]  &454.80  &466.64$\pm$14.92      &[416.01, 502.10]     \\
  
$\nu_{\p1}$
  &[1.0, 4.0] &2.190  &2.197$\pm$0.013     &[2.159, 2.227]    \\

$\nu_{\p2}$
  &[1.0, 4.0] &2.461  &2.456$\pm$0.007     &[2.445, 2.479]    \\

$\nu_{\p3}$
  &[1.0, 4.0] &2.336  &2.351$\pm$0.012     &[2.318, 2.365]    \\
  
$R_{\He1}\ (\GV)$
  &[1, 30]  &11.89  &12.09$\pm$0.19      &[11.58, 12.72]    \\
  
$R_{\He2}\ (\GV)$
  &[60, 1000]  &247.27  &246.63$\pm$10.39      &[220.06, 279.23]    \\

$\nu_{\He1}$
  &[1.0, 4.0] &2.184  &2.191$\pm$0.013     &[2.155, 2.220]     \\

$\nu_{\He2}$
  &[1.0, 4.0] &2.417  &2.422$\pm$0.007     &[2.404, 2.434]     \\

$\nu_{\He3}$
  &[1.0, 4.0] &2.205  &2.215$\pm$0.014     &[2.185, 2.245]     \\

\hline
$\phinuc\ (\GV)$
    &[0, 1.5] &0.73   &0.74$\pm$0.02    &[0.70, 0.77]   \\
$\phipbar\ (\GV)$
    &[0, 1.5] &0.31   &0.27$\pm$0.03    &[0.20, 0.36]   \\

$c_{\He}$
  &[0.1, 10.0] &4.056  &3.81$\pm$0.11     &[3.61, 4.18]   \\
$c_{\pbar}$
  &[0.1, 10.0] &1.382  &1.36$\pm$0.03     &[1.28, 1.44]   \\

\hline

$\log (N_{\e})\ ^{c}$
  &[-4, 0] &-1.9479   &-1.938$\pm$0.008    &[-1.957, -1.923]  \\

$\log (R_{\e} / \GV)$
  &[0, 3]  &1.636  &1.65$\pm$0.02      &[1.59, 1.71]     \\
  
$\nu_{\e1}$
  &[1.0, 4.0] &2.433  &2.57$\pm$0.06     &[2.37, 2.69]    \\

$\nu_{\e2}$
  &[1.0, 4.0] &2.392  &2.39$\pm$0.01     &[2.36, 2.42]    \\
  
$\log (N_{\psr})\ ^{d}$
  &[-8, -4]  &-6.14  &-6.14$\pm$0.02      &[-6.17, -6.11]    \\
  
$\nu_{\psr}$
  &[0, 3.0] &0.63  &0.67$\pm$0.03     &[0.55, 0.73]     \\

$\log (R_{c} / \GV))$
  &[2, 5]  &2.84  &2.83$\pm$0.03      &[2.76, 2.91]    \\

\hline
$\phipos\ (\GV)$
    &[0, 1.5] &1.395   &1.37$\pm$0.02    &[1.34, 1.42]   \\

$\cpos$
  &[0.1, 10.0] &5.22  &5.10$\pm$0.11     &[4.99, 5.28]   \\

  \hline\hline
\end{tabular}
\end{center}
\scriptsize{$^{a}$ Here $D_{0}$ is defined at the reference rigidity $R = 4 \GV$.}\\
\scriptsize{$^{b}$ $N_{\p}$ is the post-propagated normalization flux of protons at 100 GeV in unit $10^{-2}\m^{-2}\s^{-1}\sr^{-1}\GeV^{-1}$.}\\
\scriptsize{$^{c}$ $N_{\e}$ is the  post-propagated normalization flux of electrons at 25 GeV in unit $\m^{-2}\s^{-1}\sr^{-1}\GeV^{-1}$.}\\
\scriptsize{$^{d}$ $N_{\psr}$ is the post-propagated normalization flux of electrons at 300 GeV in unit $\m^{-2}\s^{-1}\sr^{-1}\GeV^{-1}$.}\\
\caption{
Constraints on the parameters of DAMPE CREs spectrum, pulsar scenario. The prior interval, best-fit value, statistic mean, standard deviation and the allowed range at $95\%$ CL are listed for parameters. For best fit values, we have $\chi^{2}/d.o.f. = 243.13/299$.}
\label{tab:params_psr_dampe}
\end{table*}

\begin{table*}[!htbp]
\begin{center}
\begin{tabular}{lllll}
  \hline\hline
ID  &Prior & Best-fit &Posterior mean and   &Posterior 95\%    \\
    &range&value  &Standard deviation & range  \\
\hline
$D_{0}\ (10^{28}\cm^{2}\s^{-1})\ ^{a}$
    &[1, 20]  &16.74  &14.40$\pm$1.13     &[12.50, 17.36]   \\

$\delta$
  &[0.1, 1.0] &0.289  &0.314$\pm$0.014     &[0.279, 0.342]     \\

$z_h\ (\kpc)$
  &[0.5, 30.0]  &24.14  &24.73$\pm$2.07     &[19.58, 28.86]     \\

$v_{A}\ (\km/\s)$
  &[0, 80]  &46.73  &41.76$\pm$2.84     &[38.01, 48.69]    \\

$N_{\p}\ ^{b}$
  &[1, 8] &4.46   &4.45$\pm$0.02    &[4.42, 4.49]  \\

$R_{\p1}\ (\GV)$
  &[1, 30]  &28.34  &26.80$\pm$1.61      &[25.12, 29.90]     \\
  
$R_{\p2}\ (\GV)$
  &[60, 1000]  &596.36  &486.39$\pm$78.40      &[428.26, 712.11]     \\
  
$\nu_{\p1}$
  &[1.0, 4.0] &2.223  &2.181$\pm$0.026     &[2.125, 2.245]    \\

$\nu_{\p2}$
  &[1.0, 4.0] &2.483  &2.463$\pm$0.013     &[2.435, 2.495]    \\

$\nu_{\p3}$
  &[1.0, 4.0] &2.352  &2.342$\pm$0.016     &[2.307, 2.371]    \\
  
$R_{\He1}\ (\GV)$
  &[1, 30]  &13.00  &12.30$\pm$0.60      &[11.41, 13.51]    \\
  
$R_{\He2}\ (\GV)$
  &[60, 1000]  &255.16  &239.77$\pm$32.42      &[190.45, 297.17]    \\

$\nu_{\He1}$
  &[1.0, 4.0] &2.204  &2.172$\pm$0.022     &[2.127, 2.225]     \\

$\nu_{\He2}$
  &[1.0, 4.0] &2.436  &2.418$\pm$0.013     &[2.391, 2.449]     \\

$\nu_{\He3}$
  &[1.0, 4.0] &2.244  &2.232$\pm$0.024     &[2.197, 2.276]     \\

\hline
$\phinuc\ (\GV)$
    &[0, 1.5] &0.72   &0.71$\pm$0.03    &[0.65, 0.75]   \\
$\phipbar\ (\GV)$
    &[0, 1.5] &0.06   &0.21$\pm$0.09    &[0.01, 0.33]   \\

$c_{\He}$
  &[0.1, 10.0] &4.10  &4.09$\pm$0.31     &[3.72, 4.79]   \\
$c_{\pbar}$
  &[0.1, 10.0] &1.438  &1.35$\pm$0.07     &[1.20, 1.49]   \\

\hline

$\log (N_{\e})\ ^{c}$
  &[-4, 0] &-1.993   &-1.988$\pm$0.012    &[-2.004, -1.978]  \\

$\log (R_{\e} / \GV)$
  &[0, 3]  &2.207  &1.74$\pm$0.28      &[1.12, 2.47]     \\
  
$\nu_{\e1}$
  &[1.0, 4.0] &2.504  &2.55$\pm$0.05     &[2.47, 2.65]    \\

$\nu_{\e2}$
  &[1.0, 4.0] &2.475  &2.45$\pm$0.03     &[2.38, 2.52]    \\
  
$\log (N_{\psr})\ ^{d}$
  &[-8, -4]  &-6.17  &-6.18$\pm$0.03      &[-6.25, -6.14]    \\
  
$\nu_{\psr}$
  &[0, 3.0] &0.69  &0.67$\pm$0.14     &[0.30, 0.97]     \\

$\log (R_{c} / \GV))$
  &[2, 5]  &2.84  &2.80$\pm$0.11      &[2.54, 3.05]    \\

\hline
$\phipos\ (\GV)$
    &[0, 1.5] &1.44   &1.41$\pm$0.04    &[1.37, 1.49]   \\

$\cpos$
  &[0.1, 10.0] &5.45  &5.24$\pm$0.25     &[5.00, 5.48]   \\

  \hline\hline
\end{tabular}
\end{center}
\scriptsize{$^{a}$ Here $D_{0}$ is defined at the reference rigidity $R = 4 \GV$.}\\
\scriptsize{$^{b}$ $N_{\p}$ is the post-propagated normalization flux of protons at 100 GeV in unit $10^{-2}\m^{-2}\s^{-1}\sr^{-1}\GeV^{-1}$.}\\
\scriptsize{$^{c}$ $N_{\e}$ is the  post-propagated normalization flux of electrons at 25 GeV in unit $\m^{-2}\s^{-1}\sr^{-1}\GeV^{-1}$.}\\
\scriptsize{$^{d}$ $N_{\psr}$ is the post-propagated normalization flux of electrons at 300 GeV in unit $\m^{-2}\s^{-1}\sr^{-1}\GeV^{-1}$.}\\
\caption{
Constraints on the parameters of CALET CREs spectrum, pulsar scenario. The prior interval, best-fit value, statistic mean, standard deviation and the allowed range at $95\%$ CL are listed for parameters. For best fit values, we have $\chi^{2}/d.o.f. = 229.98/301$.}
\label{tab:params_psr_calet}
\end{table*}

\begin{table*}[!htbp]
\begin{center}
\begin{tabular}{lllll}
  \hline\hline
ID  &Prior & Best-fit &Posterior mean and   &Posterior 95\%    \\
    &range&value  &Standard deviation & range  \\
\hline
$D_{0}\ (10^{28}\cm^{2}\s^{-1})\ ^{1}$
    &[1, 20]  &15.12  &15.16$\pm$0.13     &[14.87, 15.62]   \\

$\delta$
  &[0.1, 1.0] &0.334  &0.334$\pm$0.003     &[0.323, 0.342]     \\

$z_h\ (\kpc)$
  &[0.5, 30.0]  &28.57  &28.59$\pm$0.20     &[28.13, 29.15]     \\

$v_{A}\ (\km/\s)$
  &[0, 80]  &42.81  &42.92$\pm$0.64     &[41.72, 45.02]    \\

$N_{\p}\ ^{b}$
  &[1, 8] &4.49   &4.49$\pm$0.01    &[4.46, 4.52]  \\

$R_{\p1}\ (\GV)$
  &[1, 30]  &25.02  &25.03$\pm$0.22      &[24.59, 25.68]     \\
  
$R_{\p2}\ (\GV)$
  &[60, 1000]  &461.99  &463.00$\pm$5.36      &[449.86, 478.37]     \\
  
$\nu_{\p1}$
  &[1.0, 4.0] &2.186  &2.187$\pm$0.009     &[2.166, 2.209]    \\

$\nu_{\p2}$
  &[1.0, 4.0] &2.463  &2.462$\pm$0.006     &[2.447, 2.474]    \\

$\nu_{\p3}$
  &[1.0, 4.0] &2.324  &2.326$\pm$0.010     &[2.308, 2.345]    \\
  
$R_{\He1}\ (\GV)$
  &[1, 30]  &10.91  &10.90$\pm$0.13      &[10.52, 11.22]    \\
  
$R_{\He2}\ (\GV)$
  &[60, 1000]  &239.35  &238.80$\pm$8.65      &[215.60, 262.77]    \\

$\nu_{\He1}$
  &[1.0, 4.0] &2.172  &2.170$\pm$0.010     &[2.146, 2.192]     \\

$\nu_{\He2}$
  &[1.0, 4.0] &2.410  &2.409$\pm$0.006     &[2.395, 2.421]     \\

$\nu_{\He3}$
  &[1.0, 4.0] &2.205  &2.205$\pm$0.010     &[2.180, 2.231]     \\

\hline
$\phinuc\ (\GV)$
    &[0, 1.5] &0.77   &0.77$\pm$0.01    &[0.73, 0.79]   \\
$\phipbar\ (\GV)$
    &[0, 1.5] &0.43   &0.42$\pm$0.02    &[0.36, 0.46]   \\

$c_{\He}$
  &[0.1, 10.0] &4.50  &4.53$\pm$0.14     &[3.66, 4.66]   \\
$c_{\pbar}$
  &[0.1, 10.0] &1.56  &1.56$\pm$0.02     &[1.51, 1.61]   \\

\hline

$\log (N_{\e})\ ^{c}$
  &[-4, 0] &-1.933   &-1.934$\pm$0.006    &[-1.945, -1.923]  \\

$\log (R_{\e} /\GV)$
  &[0, 3]  &1.676  &1.680$\pm$0.025      &[1.63, 1.76]     \\
  
$\nu_{\e1}$
  &[1.0, 4.0] &2.55  &2.55$\pm$0.02     &[2.51, 2.60]    \\

$\nu_{\e2}$
  &[1.0, 4.0] &2.37  &2.37$\pm$0.01     &[2.35, 2.39]    \\
  
$\log (\Mdm / \GeV)$
  &[1, 6]  &3.275  &3.270$\pm$0.005      &[3.258, 3.275]    \\
  
$\log (\sigv)\ ^{d}$
  &[-28, -18] &-22.39  &-22.40$\pm$0.03     &[-22.48, -22.36]     \\

$\etae$
  &[0, 1]  &0.465  &0.460$\pm$0.011      &[0.438, 0.476]    \\
  
$\etamu$
  &[0, 1]  &0.510  &0.513$\pm$0.009      &[0.499, 0.530]    \\
  
$\etatau$
  &[0, 1]  &0.025  &0.027$\pm$0.007      &[0.016, 0.037]    \\

\hline
$\phipos\ (\GV)$
    &[0, 1.5] &1.36   &1.36$\pm$0.01    &[1.33, 1.41]   \\

$\cpos$
  &[0.1, 10.0] &5.19  &5.19$\pm$0.09     &[5.09, 5.33]   \\

  \hline\hline
\end{tabular}
\end{center}
\scriptsize{$^{a}$ Here $D_{0}$ is defined at the reference rigidity $R = 4 \GV$.}\\
\scriptsize{$^{b}$ $N_{\p}$ is the post-propagated normalization flux of protons at 100 GeV in unit $10^{-2}\m^{-2}\s^{-1}\sr^{-1}\GeV^{-1}$.}\\
\scriptsize{$^{c}$ $N_{\e}$ is the  post-propagated normalization flux of electrons at 25 GeV in unit $\m^{-2}\s^{-1}\sr^{-1}\GeV^{-1}$.}\\
\scriptsize{$^d$ $(\sigv)$ is in unit $\cm^{3} \s^{-1}$}\\
\caption{
Constraints on the parameters of DAMPE CREs spectrum, DM scenario. The prior interval, best-fit value, statistic mean, standard deviation and the allowed range at $95\%$ CL are listed for parameters. For best fit values, we have $\chi^{2}/d.o.f. = 262.94/297$.}
\label{tab:params_dm_dampe}
\end{table*}

\begin{table*}[!htbp]
\begin{center}
\begin{tabular}{lllll}
  \hline\hline
ID  &Prior & Best-fit &Posterior mean and   &Posterior 95\%    \\
    &range&value  &Standard deviation & range  \\
\hline
$D_{0}\ (10^{28}\cm^{2}\s^{-1})\ ^{a}$
    &[1, 20]  &16.68  &16.00$\pm$0.75     &[14.72, 18.22]   \\

$\delta$
  &[0.1, 1.0] &0.304  &0.307$\pm$0.009     &[0.285, 0.327]     \\

$z_h\ (\kpc)$
  &[0.5, 30.0]  &25.93  &27.64$\pm$1.72     &[22.52, 28.79]     \\

$v_{A}\ (\km/\s)$
  &[0, 80]  &47.07  &45.92$\pm$1.93     &[43.02, 51.38]    \\

$N_{\p}\ ^{b}$
  &[1, 8] &4.47   &4.45$\pm$0.02    &[4.44, 4.49]  \\

$R_{\p1}\ (\GV)$
  &[1, 30]  &27.80  &27.09$\pm$0.93      &[26.12, 29.56]     \\
  
$R_{\p2}\ (\GV)$
  &[60, 1000]  &569.43  &592.75$\pm$40.33      &[505.98, 713.39]     \\
  
$\nu_{\p1}$
  &[1.0, 4.0] &2.201  &2.186$\pm$0.019     &[2.152, 2.230]    \\

$\nu_{\p2}$
  &[1.0, 4.0] &2.471  &2.458$\pm$0.010     &[2.4457, 2.485]    \\

$\nu_{\p3}$
  &[1.0, 4.0] &2.349  &2.363$\pm$0.029     &[2.312, 2.376]    \\
  
$R_{\He1}\ (\GV)$
  &[1, 30]  &11.77  &11.46$\pm$0.32      &[11.19, 12.18]    \\
  
$R_{\He2}\ (\GV)$
  &[60, 1000]  &225.19  &268.93$\pm$29.10      &[220.29, 343.72]    \\

$\nu_{\He1}$
  &[1.0, 4.0] &2.178  &2.160$\pm$0.015     &[2.134, 2.198]     \\

$\nu_{\He2}$
  &[1.0, 4.0] &2.421  &2.414$\pm$0.010     &[2.398, 2.430]     \\

$\nu_{\He3}$
  &[1.0, 4.0] &2.234  &2.221$\pm$0.026     &[2.185, 2.256]     \\

\hline
$\phinuc\ (\GV)$
    &[0, 1.5] &0.72   &0.69$\pm$0.02    &[0.66, 0.73]   \\
$\phipbar\ (\GV)$
    &[0, 1.5] &0.15   &0.17$\pm$0.05    &[0.03, 0.25]   \\

$c_{\He}$
  &[0.1, 10.0] &4.89  &4.96$\pm$0.21     &[4.56, 5.58]   \\
$c_{\pbar}$
  &[0.1, 10.0] &1.48  &1.44$\pm$0.05     &[1.40, 1.56]   \\

\hline

$\log (N_{\e})\ ^{c}$
  &[-4, 0] &-1.990   &-1.994$\pm$0.011    &[-2.006, -1.980]  \\

$\log (R_{\e} /\GV)$
  &[0, 3]  &1.789  &1.764$\pm$0.036      &[1.71, 1.87]     \\
  
$\nu_{\e1}$
  &[1.0, 4.0] &2.54  &2.57$\pm$0.05     &[2.47, 2.61]    \\

$\nu_{\e2}$
  &[1.0, 4.0] &2.43  &2.39$\pm$0.03     &[2.39, 2.47]    \\
  
$\log (\Mdm / \GeV)$
  &[1, 6]  &3.109  &3.109$\pm$0.003      &[3.103, 3.116]    \\
  
$\log (\sigv)\ ^{d}$
  &[-28, -18] &-22.25  &-22.21$\pm$0.11     &[-22.41, -22.21]     \\

$\etae$
  &[0, 1]  &0.930  &0.918$\pm$0.032      &[0.861, 0.967]    \\
  
$\etamu$
  &[0, 1]  &0.042  &0.039$\pm$0.022      &[0.008, 0.078]    \\
  
$\etatau$
  &[0, 1]  &0.028  &0.043$\pm$0.027      &[0.006, 0.090]    \\

\hline
$\phipos\ (\GV)$
    &[0, 1.5] &1.42   &1.44$\pm$0.02    &[1.40, 1.49]   \\

$\cpos$
  &[0.1, 10.0] &5.22  &5.30$\pm$0.16     &[5.08, 5.46]   \\

  \hline\hline
\end{tabular}
\end{center}
\scriptsize{$^{a}$ Here $D_{0}$ is defined at the reference rigidity $R = 4 \GV$.}\\
\scriptsize{$^{b}$ $N_{\p}$ is the post-propagated normalization flux of protons at 100 GeV in unit $10^{-2}\m^{-2}\s^{-1}\sr^{-1}\GeV^{-1}$.}\\
\scriptsize{$^{c}$ $N_{\e}$ is the  post-propagated normalization flux of electrons at 25 GeV in unit $\m^{-2}\s^{-1}\sr^{-1}\GeV^{-1}$.}\\
\scriptsize{$^d$ $(\sigv)$ is in unit $\cm^{3} \s^{-1}$}\\
\caption{
Constraints on the parameters of CALET CREs spectrum, pulsar scenario. The prior interval, best-fit value, statistic mean, standard deviation and the allowed range at $95\%$ CL are listed for parameters. For best fit values, we have $\chi^{2}/d.o.f. = 265.03/299$.}
\label{tab:params_dm_calet}
\end{table*}

\begin{figure*}[!htbp]
  \centering
  \includegraphics[width=0.48\textwidth]{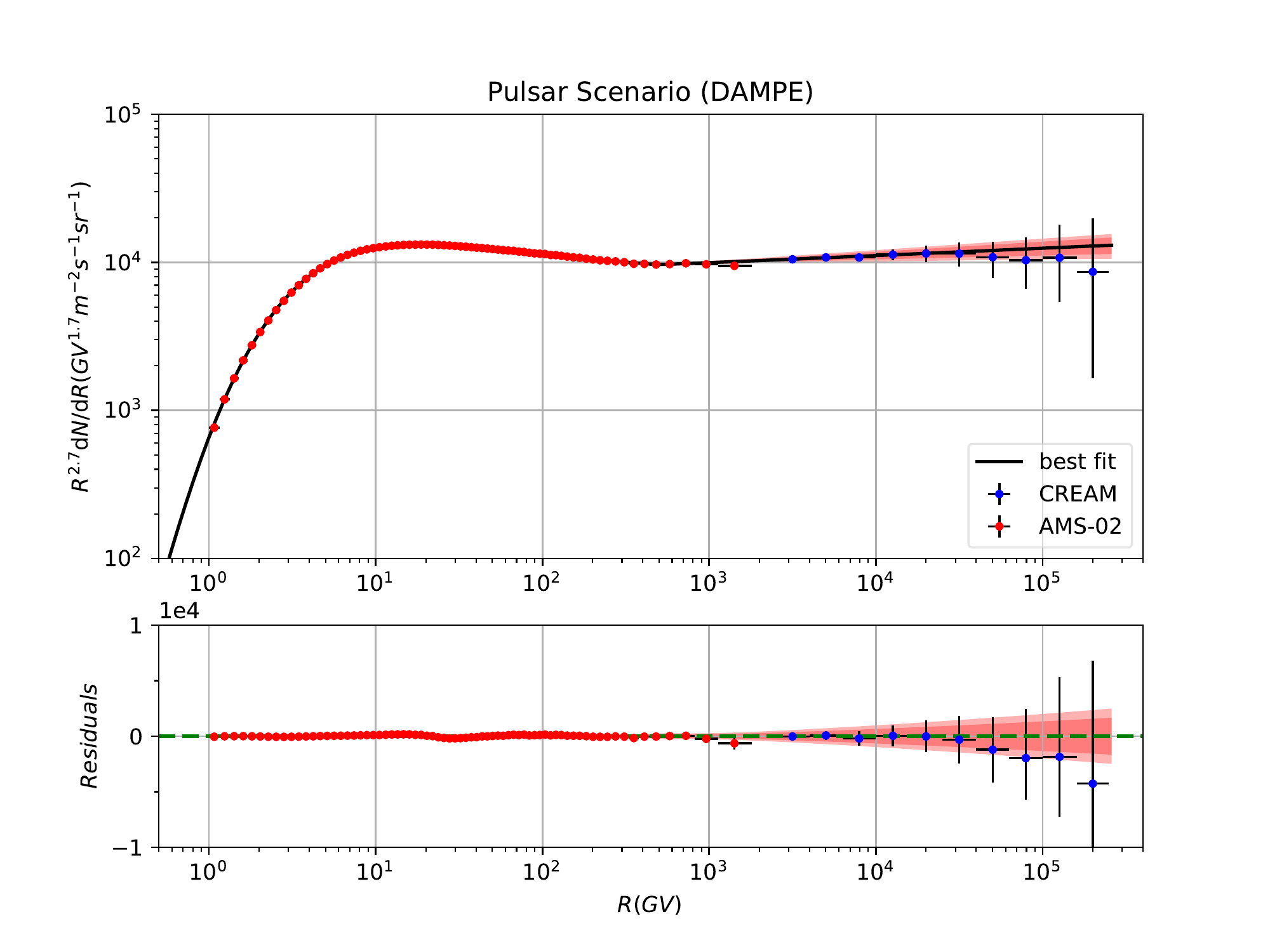}
  \includegraphics[width=0.48\textwidth]{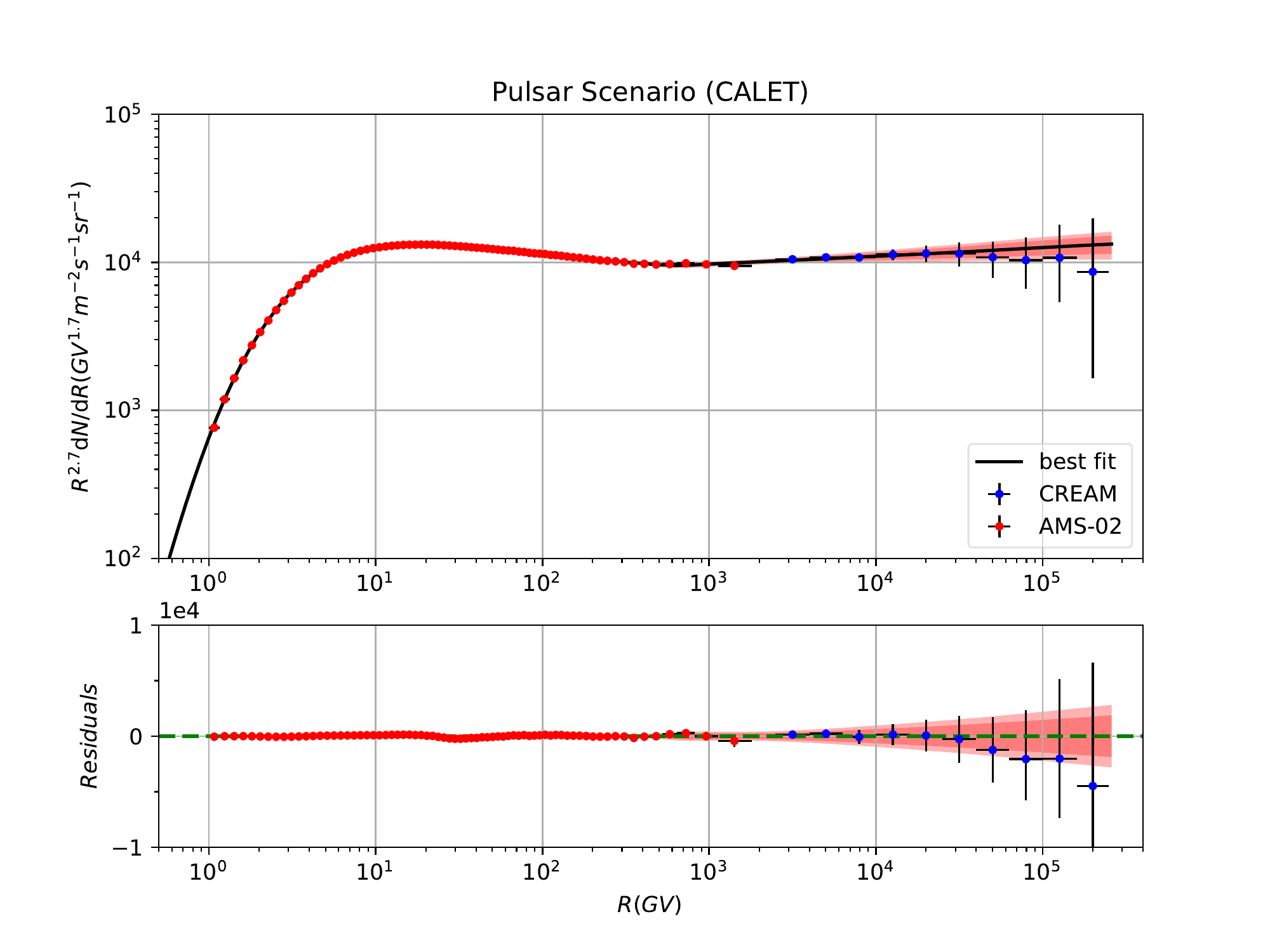}
  \includegraphics[width=0.48\textwidth]{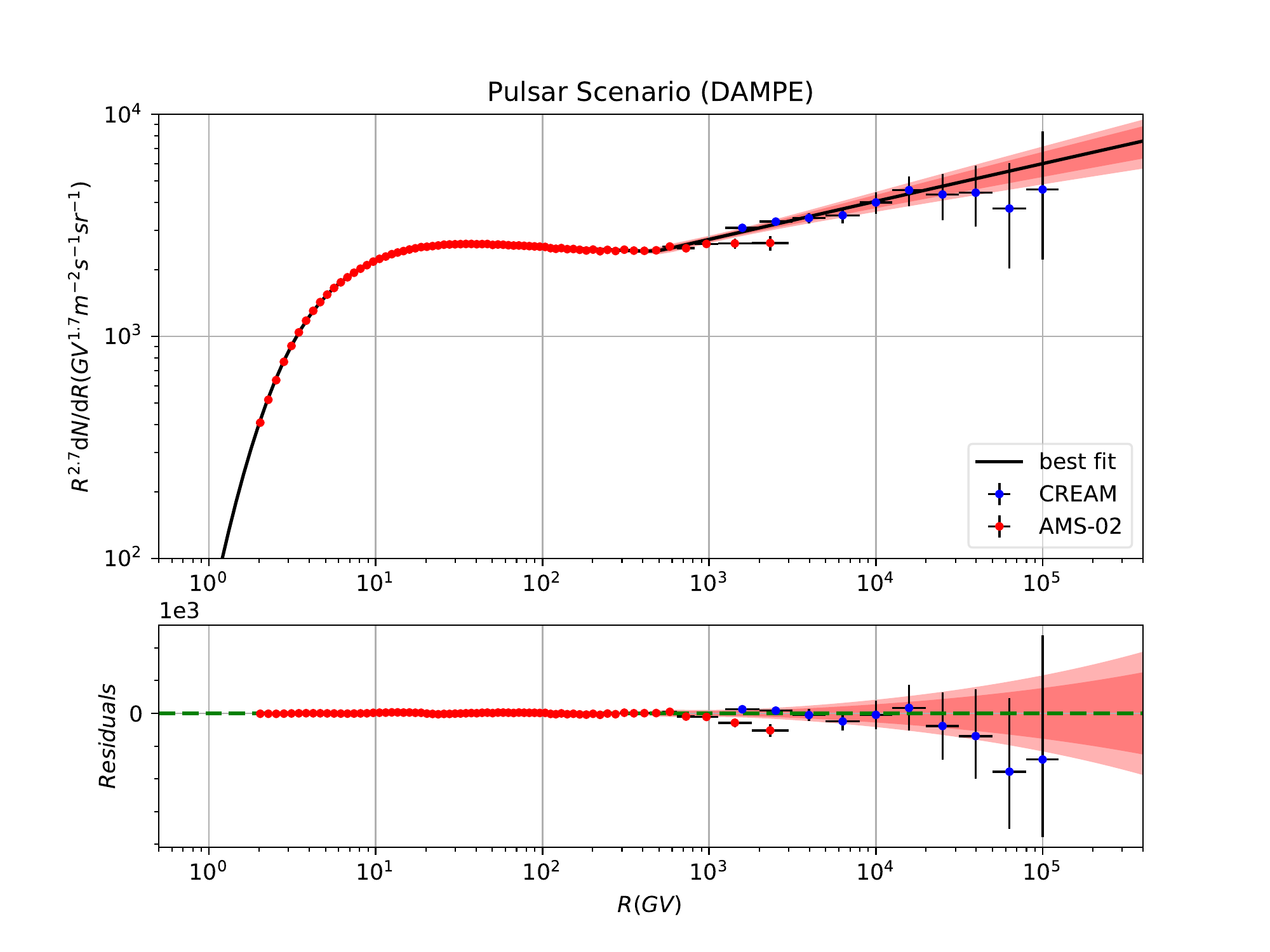}
  \includegraphics[width=0.48\textwidth]{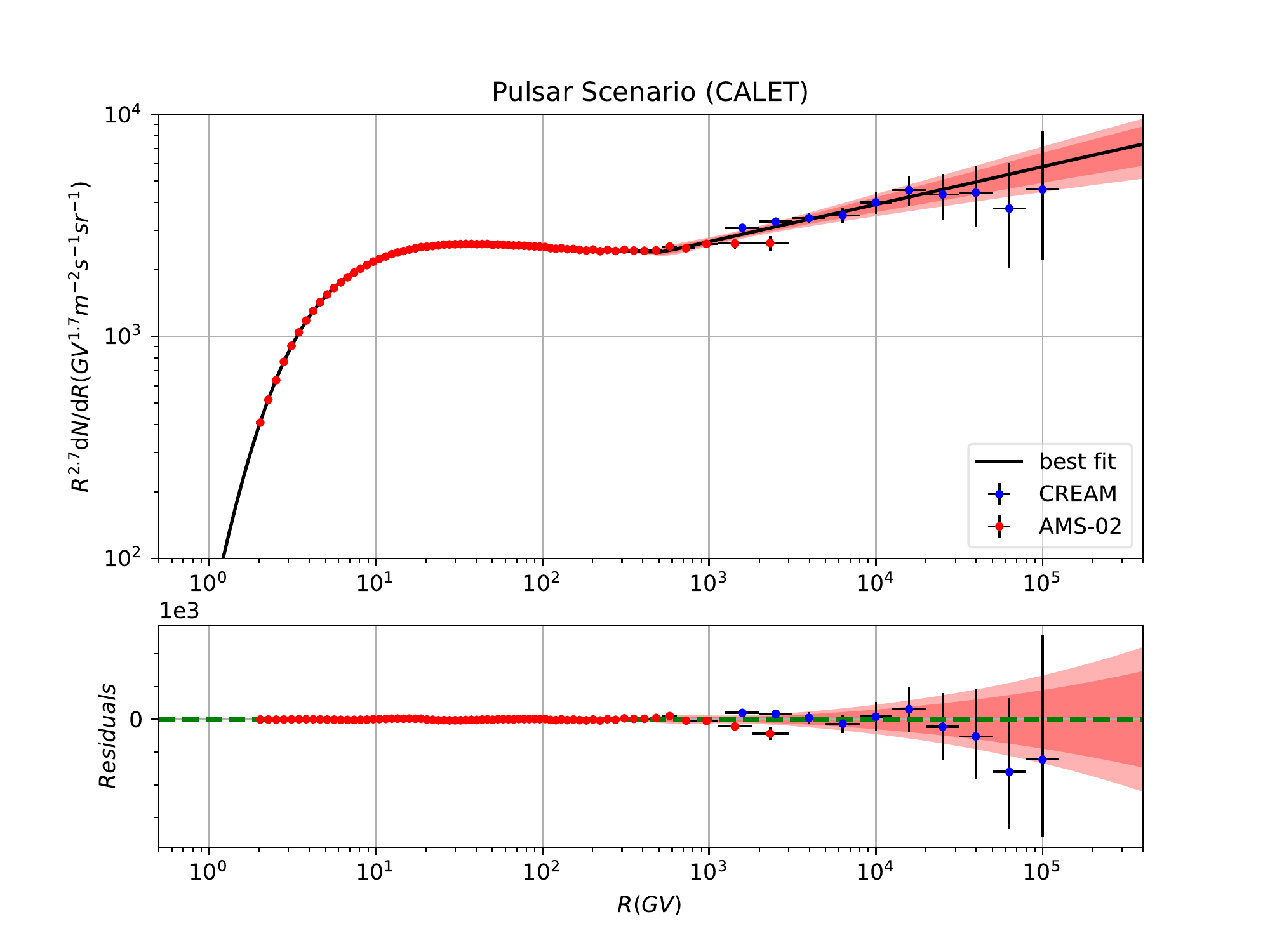}
  \includegraphics[width=0.48\textwidth]{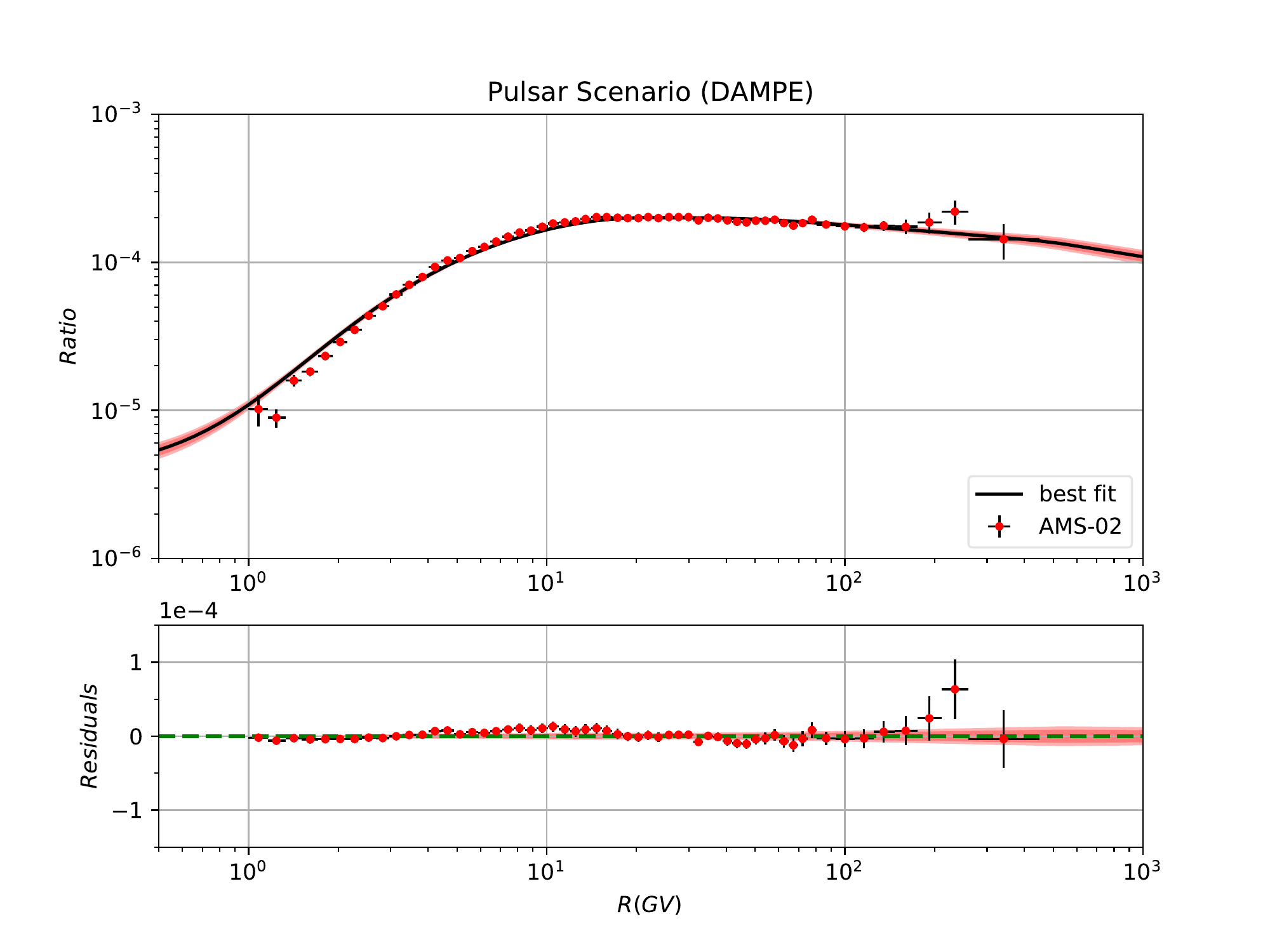}
  \includegraphics[width=0.48\textwidth]{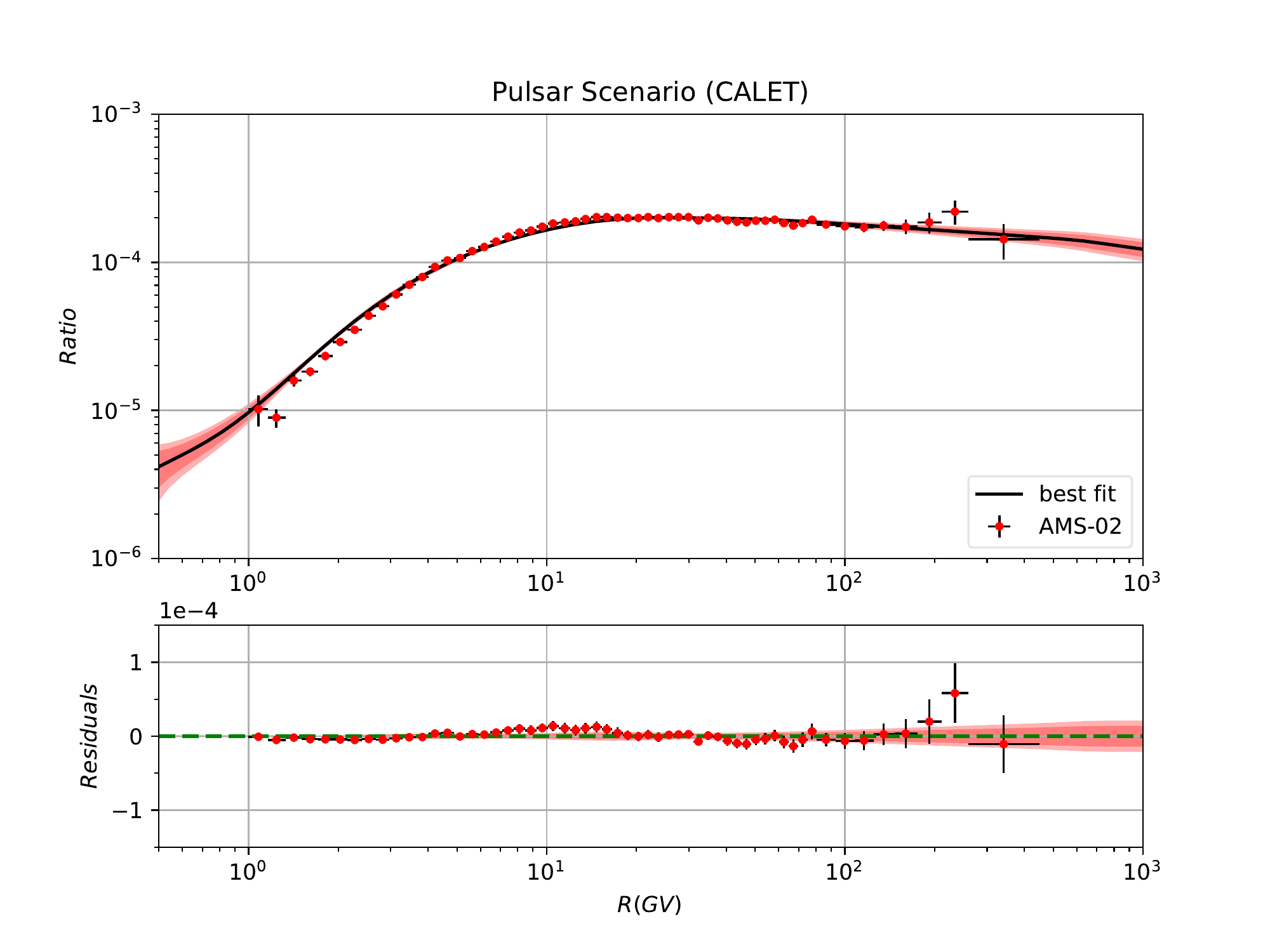}
  \caption{The global fitting results and the corresponding residuals to the proton flux, helium flux, $\pbarp$ ratio, and positron flux for 2 scenarios. The $2\sigma$ (deep red) and $3\sigma$ (light red) bound are also showed in the figures.}
\label{fig:nuclei_psr_results}
\end{figure*}

\begin{figure*}[!htbp]
  \centering
  \includegraphics[width=0.48\textwidth]{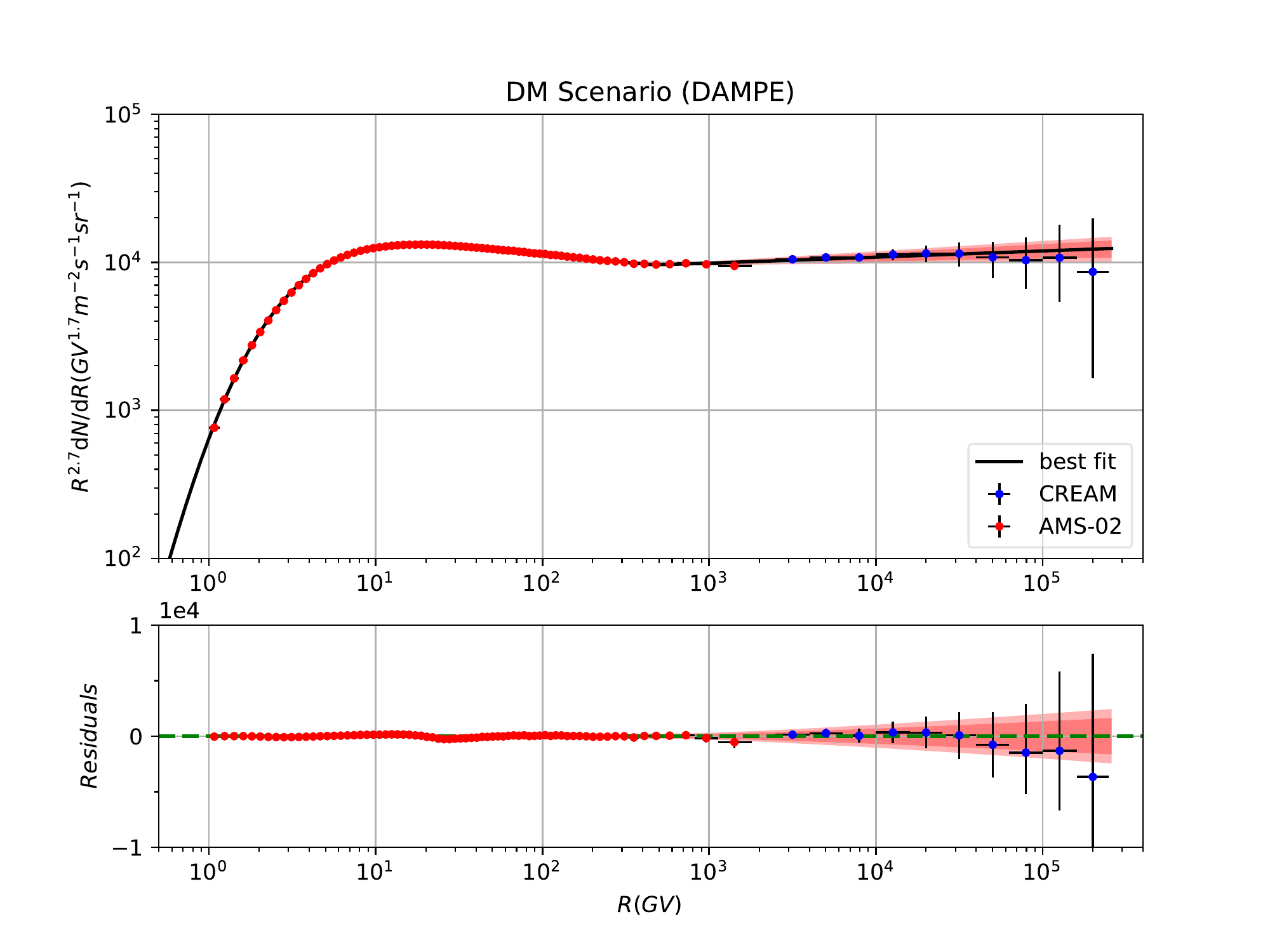}
  \includegraphics[width=0.48\textwidth]{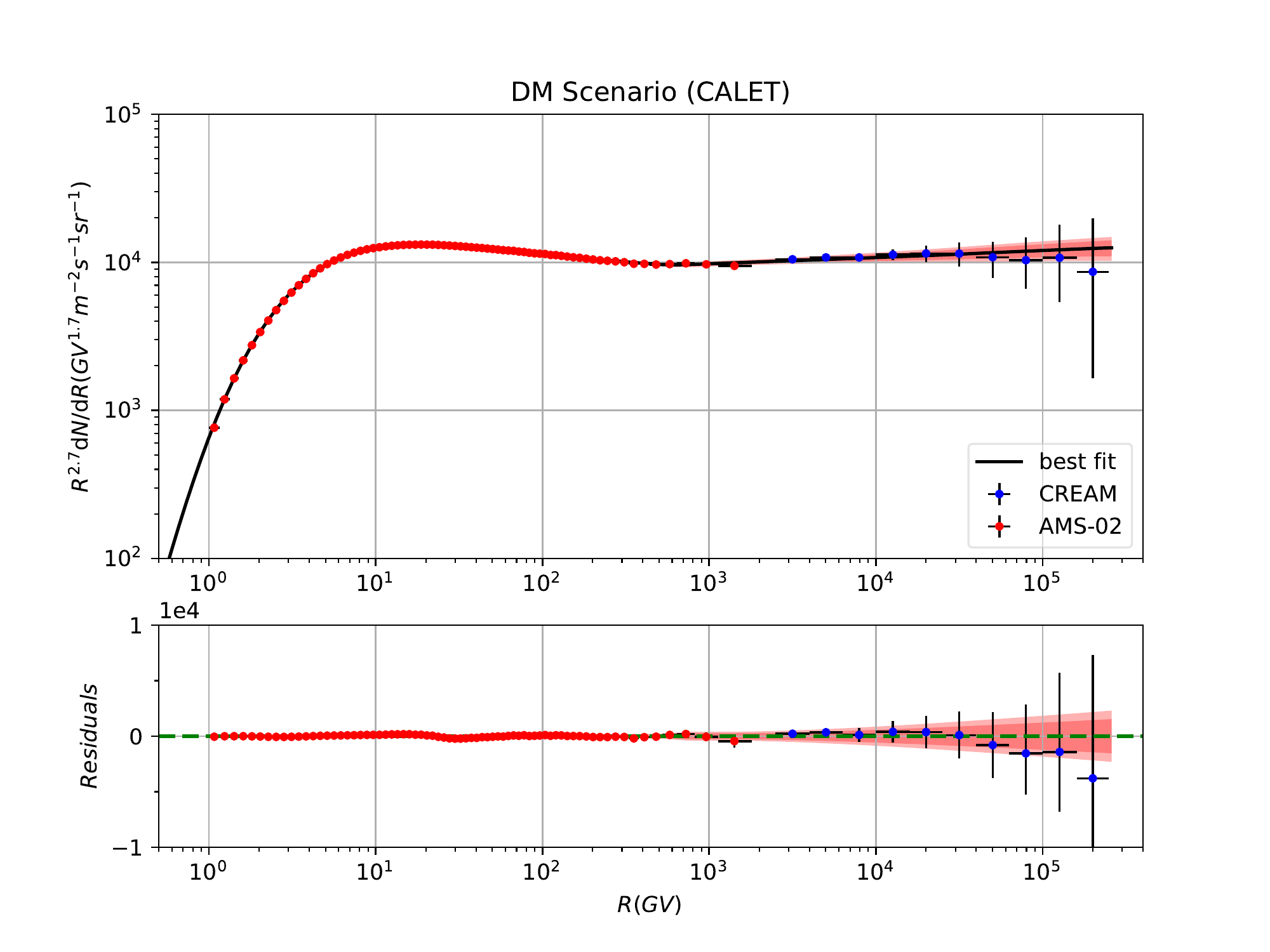}
  \includegraphics[width=0.48\textwidth]{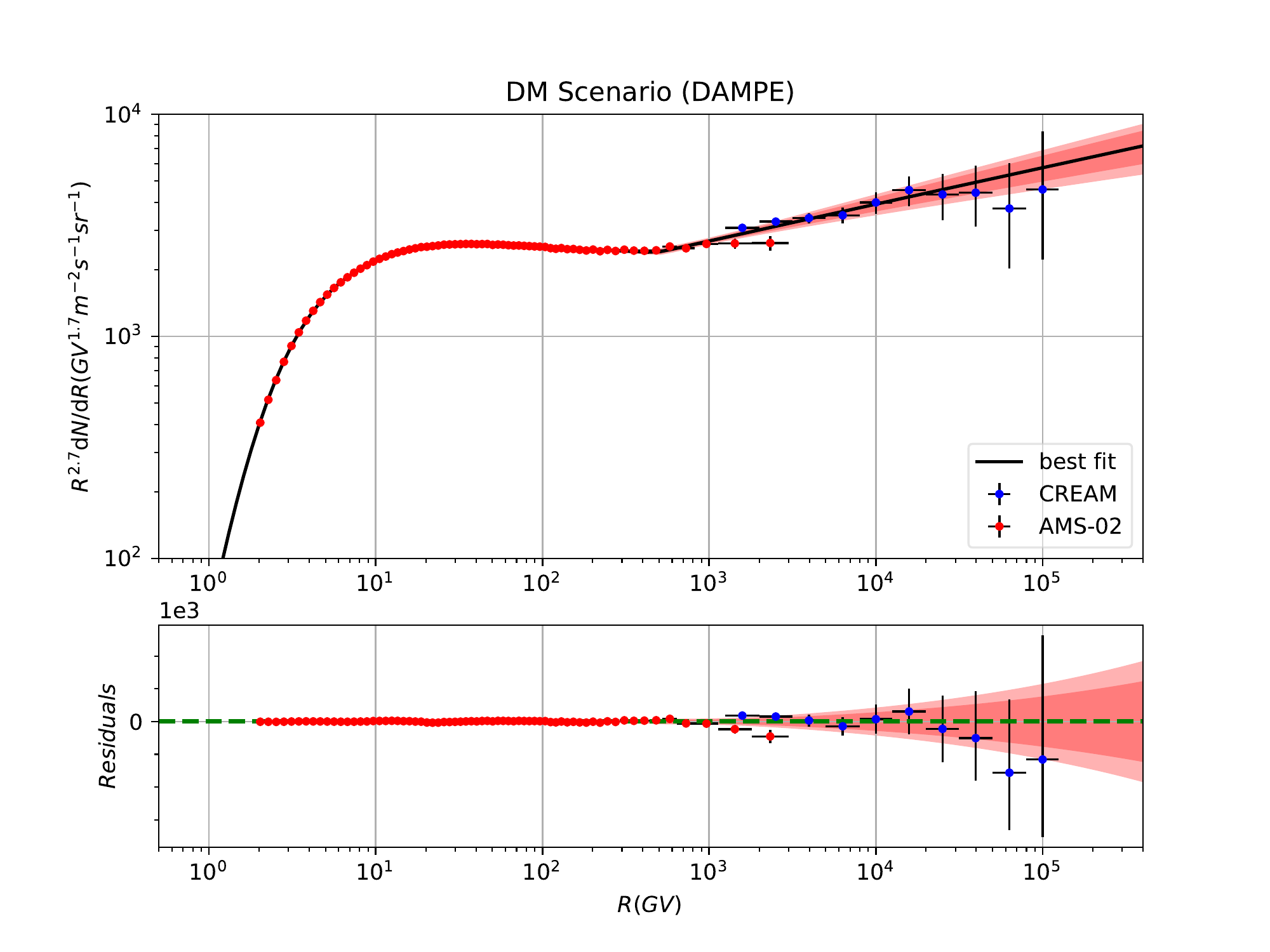}
  \includegraphics[width=0.48\textwidth]{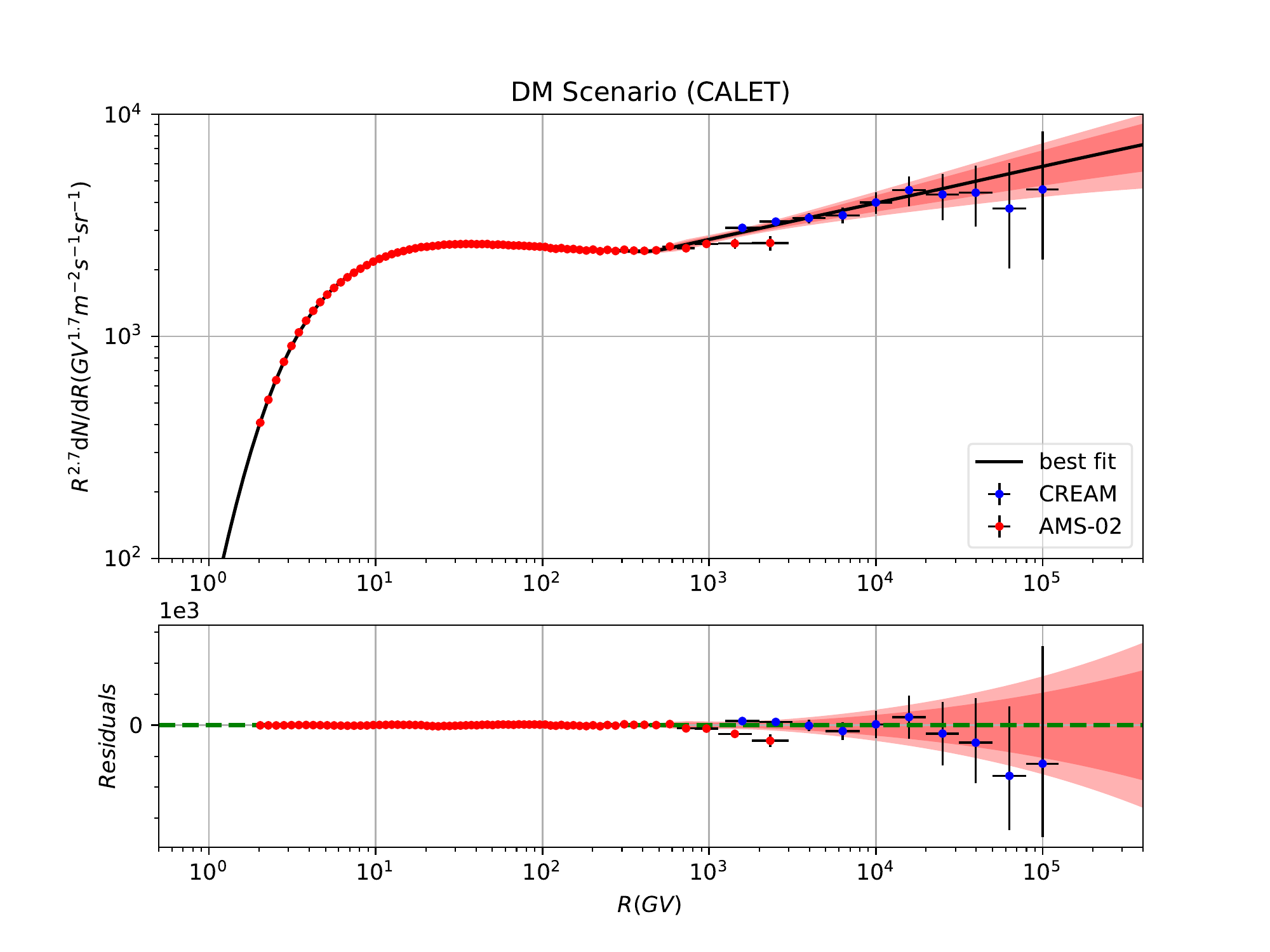}
  \includegraphics[width=0.48\textwidth]{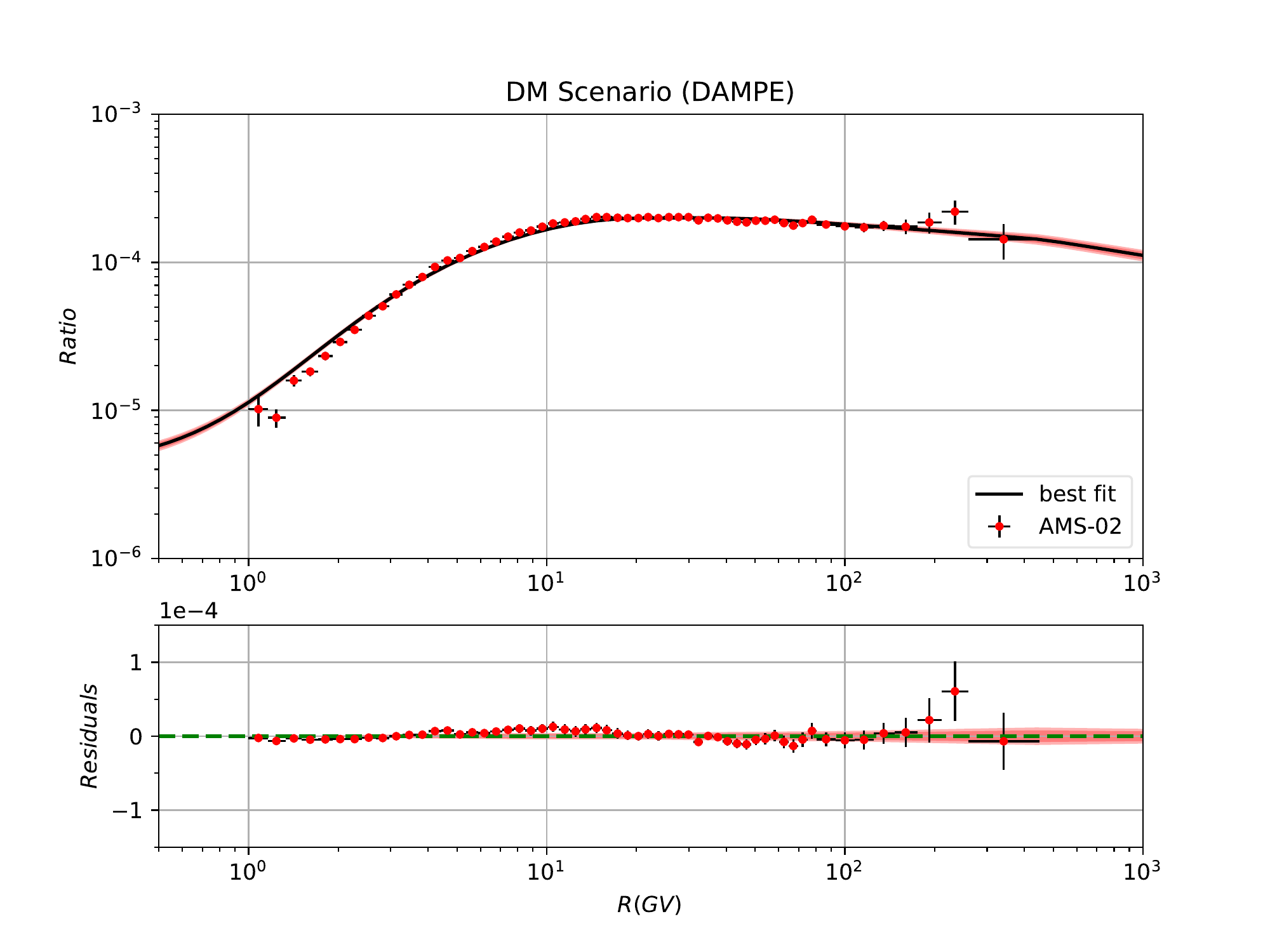}
  \includegraphics[width=0.48\textwidth]{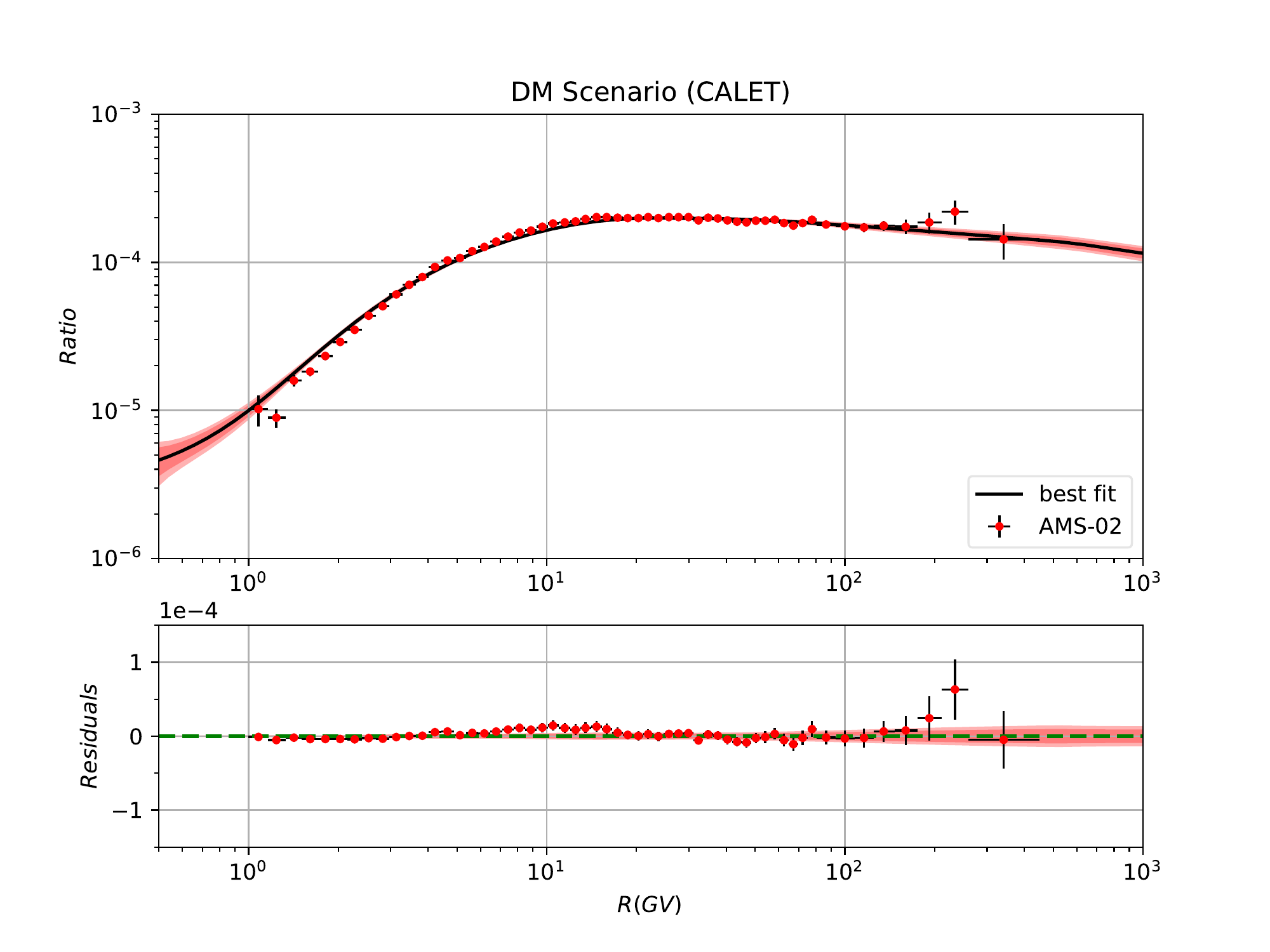}
  \caption{The global fitting results and the corresponding residuals to the proton flux, helium flux, $\pbarp$ ratio, and positron flux for 2 scenarios. The $2\sigma$ (deep red) and $3\sigma$ (light red) bound are also showed in the figures.}
\label{fig:nuclei_dm_results}
\end{figure*}

\end{document}